\documentclass[12pt]{article}
\usepackage{amssymb,amsmath}
\makeatletter

\usepackage{arydshln}

\newcount\comment@nesting

\newenvironment{comment}{%
    \begingroup
    \global\comment@nesting=0
    \def\do##1{\catcode`##1=12\relax}%
    \dospecials
    \catcode`\^^M=\active
    \comment@processline
}{%
    \endgroup
}

\begingroup
\xdef\comment@begincomment{\string\\begin\string\{comment\string\}}
\xdef\comment@endcomment{\string\\end\string\{comment\string\}}
\endgroup

\begingroup
\catcode`\^^M=\active
\def\@temp{\endgroup\def\comment@processline##1^^M}%
\@temp{%
    \def\comment@curline{#1}%
    \let\@next=\comment@processline
    \ifx\comment@curline\comment@endcomment
        \ifnum\comment@nesting=0
            \def\@next{\end{comment}}%
        \else
            \global\advance\comment@nesting by -1
        \fi
    \else
        \ifx\comment@curline\comment@begincomment
            \global\advance\comment@nesting by 1
        \fi
    \fi
    \@next
}

\makeatother

\usepackage[dvipdfmx]{graphicx}
\usepackage{bm}
\usepackage{color}
\usepackage{here}
\usepackage{enumerate}
\usepackage{here}
\usepackage[vcentermath]{youngtab}
\usepackage{subfigure}
\usepackage{tikz}
\usepackage{hyperref}

\newcommand{\Zb}{\mathbb{Z}}

\newcommand{\Nb}{\mathbb{N}}
\newcommand{\Acal}{\mathcal{A}}
\newcommand{\Ccal}{\mathcal{C}}

\newcommand{\Mcal}{\mathcal{M}}
\newcommand{\Ncal}{\mathcal{N}}

\DeclareMathOperator*{\Tr}{{\rm Tr}}

\newcommand{\II}{\mathbb{II}}

\numberwithin{equation}{section}

\allowdisplaybreaks[1]

\definecolor{mygreen}{rgb}{0,0.714,0.286}

\begin{document}

\thispagestyle{empty}
\begin{flushright}

\end{flushright}
\vskip1.5cm
\begin{center}
{\Large \bf 
Chern-Simons theories with defects,
\\
Rogers-Ramanujan type functions 
\\
\vspace*{0.35cm}
and eta-products
}

\vskip1.5cm
Tadashi Okazaki\footnote{tokazaki@seu.edu.cn}

\bigskip
{\it Shing-Tung Yau Center of Southeast University,\\
Yifu Architecture Building, No.2 Sipailou, Xuanwu district, \\
Nanjing, Jiangsu, 210096, China
}

\bigskip
and
\\
\bigskip
Douglas J. Smith\footnote{douglas.smith@durham.ac.uk}

\bigskip
{\it Department of Mathematical Sciences, Durham University,\\
Upper Mountjoy, Stockton Road, Durham DH1 3LE, UK}

\end{center}

\vskip1cm
\begin{abstract}
We study the line defect half-indices of 3d $\mathcal{N}=2$ supersymmetric Chern-Simons (CS) theories with (special) unitary, symplectic, orthogonal and exceptional gauge groups. We find that they have several beautiful infinite product $q$-series expressions in terms of Ramanujan's general theta function. 
For the theories with fundamental chiral multiplets, the pairs of the Neumann half-indices and the one-point functions of the fundamental Wilson lines form a basis for the line defect indices in terms of Rogers-Ramanujan type functions which correspond to Rogers-Ramanujan type identities relating infinite series and infinite products. Furthermore, the theories with an adjoint chiral admit expressions as eta-products. In particular, for the $SU(N)_{-2N}$ CS theory, there is a one-to-one correspondence between the BPS boundary local operators and the $N$-core partitions. 
\end{abstract}
\newpage
\setcounter{tocdepth}{3}
\tableofcontents

\section{Introduction and conclusion}
\label{sec_intro}

3d $\mathcal{N}=2$ supersymmetric field theories can preserve $\mathcal{N}=(0,2)$ supersymmetry at a boundary \cite{Gadde:2013wq, Okazaki:2013kaa}  
so that they can be viewed as 3d/2d bulk-boundary systems characterized by the BPS boundary conditions. 
The spectra of the BPS local operators in the bulk-boundary system can be examined by the half-indices \cite{Gadde:2013wq, Gadde:2013sca, Yoshida:2014ssa, Dimofte:2017tpi} 
which can be viewed as certain supersymmetric partition functions on $HS^2\times S^1$ where $HS^2$ is a hemisphere and $S^1$ is a circle. 
They can be decorated as ``line defect half-indices'' (which we also refer to as line operator $1$-point functions) by introducing BPS line defect operators wrapping $S^1$. 
The configurations are realized as 3d/2d/1d bulk-boundary-line systems which further generalize the bulk-boundary systems. 
They encode more general spectra of BPS local operators sitting at the junctions of lines and boundary which may or may not be gauge invariant \cite{Dimofte:2017tpi}\footnote{Note that we can calculate half-indices in the presence of line operators which may not be gauge invariant. What is being counted in such expressions is the gauge invariant combinations of the line operator with local operators.}.

Recently, we have found in \cite{Okazaki:2024paq} that 
the line defect half-indices for the 3d $\mathcal{N}=2$ supersymmetric $SU(N)$ Chern-Simons (CS) theories 
with Neumann boundary conditions for gauge fields can be described by the celebrated $q$-series functions, instances of Ramanujan's general theta function defined by 
\cite{MR1117903}
\begin{align}
\label{r_theta}
f(a,b)
&=\sum_{m\in \mathbb{Z}}
a^{\frac{m(m+1)}{2}}b^{\frac{m(m-1)}{2}}
=(-a;ab)_{\infty}(-b;ab)_{\infty}(ab;ab)_{\infty}, 
\end{align}
where $|ab|<1$. 
Ramanujan's general theta function can be related in various ways to the Jacobi theta functions defined by
\begin{align}
\label{theta12}
\vartheta_1(z;\tau)
&=\sum_{n\in \mathbb{Z}}(-1)^{n-\frac12} q^{\frac12 (n+\frac12)^2}x^{n+\frac12}, 
& \vartheta_2 (z;\tau)
&=\sum_{n\in \mathbb{Z}} q^{\frac12 (n+\frac12)^2}x^{n+\frac12}, \\
\label{theta3}
\vartheta_3(z;\tau)
&=\sum_{n\in \mathbb{Z}} q^{\frac{n^2}{2}} x^n,
& \vartheta_4(z;\tau)
&=\sum_{n\in \mathbb{Z}} (-1)^n q^{\frac{n^2}{2}} x^n, 
\end{align}
with $q=e^{2\pi i\tau}$, $x=e^{2\pi iz}$, $\tau\in \mathbb{H}$ and $z\in \mathbb{C}$. 

For $|q|<1$ the following notations are introduced
\begin{align}
\label{r_theta_phi}
\varphi(q)&:=f(q,q)=\sum_{n\in \mathbb{Z}}q^{n^2}=\frac{(-q;-q)_{\infty}}{(q;-q)_{\infty}}, \\
\label{r_theta_psi}
\psi(q)&:=f(q,q^3)=\sum_{n=0}^{\infty}q^{\frac{n(n+1)}{2}}=\frac{(q^2;q^2)_{\infty}}{(q;q^2)_{\infty}}, \\
\label{r_theta_f}
f(-q)&:=f(-q,-q^2)=\sum_{n\in \mathbb{Z}}(-1)^n q^{\frac{n(3n-1)}{2}}=(q;q)_{\infty}, \\
\label{r_theta_chi}
\chi(q)&:=(-q;q^2)_{\infty}. 
\end{align}

In this paper we examine spectra of the BPS boundary local operators in the presence of Wilson line operators 
for 3d $\mathcal{N}=2$ supersymmetric CS theories of special unitary, symplectic, orthogonal and exceptional gauge groups 
coupled to chiral multiplets transforming in the fundamental and adjoint representations by analyzing the line defect half-indices. 

We recall that the Rogers-Ramanujan functions are defined by
\begin{align}
\label{RR_identity1}
G(q)&=\sum_{n=0}^{\infty}
\frac{q^{n^2}}{(1-q)(1-q^2)\cdots (1-q^n)}
\nonumber\\
&
=\prod_{n=1}^{\infty}\frac{1}{(1-q^{5n-1})(1-q^{5n-4})}
=\frac{f(-q^2,-q^3)}{f(-q)}
, \\
\label{RR_identity2}
H(q)&=\sum_{n=0}^{\infty}
\frac{q^{n(n+1)}}{(1-q)(1-q^2)\cdots (1-q^n)}
\nonumber\\
&=\prod_{n=1}^{\infty}\frac{1}{(1-q^{5n-2})(1-q^{5n-3})}
=\frac{f(-q,-q^4)}{f(-q)} \; .
\end{align}
The product-series equalities (\ref{RR_identity1}) and (\ref{RR_identity2}) are known as the Rogers-Ramanujan identities (e.g. see \cite{MR1634067, MR3752624}) and the existence of both infinite sum and infinite product expressions is the key feature of Rogers-Ramanujan functions. 
These identities are of the greatest significance in the theory of partitions and number theory \cite{MR1634067, MR3752624}. 
The function (\ref{RR_identity1}) is the generating function for partitions of $n$ into parts with minimal difference $2$ with all parts greater than $0$ 
or equivalently that for partitions of $n$ of the form $5k+1$ and $5k+4$. 
The function (\ref{RR_identity2}) is the generating function for partitions of $n$ into parts with minimal difference $2$ with all parts greater than $1$ 
or equivalently that for partitions of $n$ of the form $5k+2$ and $5k+3$.

In previous work \cite{Okazaki:2024paq} we highlighted the physical interpretation of the Rogers-Ramanujan identities and generalizations. For background on Rogers-Ramanujan identities see for example \cite{MR1634067, MR3752624}. In this article we explore other generalizations and identify their physical interpretation as half-indices. We denote such generalizations of Rogers-Ramanujan functions as {\em Rogers-Ramanujan type functions}, emphasising that the key feature is that they are $q$-series with both infinite sum and infinite product representations. It should be noted that in some cases infinite sum expressions can be interpreted as $q$-series arising from half-indices where we have Dirichlet boundary conditions for the vector multiplets. Such sums can also arise through direct evaluation of half-indices where we have Neumann boundary conditions for the vector multiplets. The Rogers-Ramanujan identities may then have an interpretation as giving a closed form infinite product expression which can be understood as an index or half-index for free chiral multiplets, a phenomenon known as a confining (or s-confining) duality. See \cite{Amariti:2015kha,Nii:2016jzi,Pasquetti:2019uop,Pasquetti:2019tix,Nii:2019dwi,Benvenuti:2020gvy,Benvenuti:2021nwt,Bajeot:2022lah,Amariti:2022wae,Okazaki:2023hiv,Amariti:2023wts, Okazaki:2023kpq, Okazaki:2023hiv} for such examples in 3d $\Ncal = 2$ theories. Confining dualities are useful to construct a sequence of dual theories and to find an essential or basic part of dualities. In addition, where these arise from Neumann half-indices we have an integral representation which is not always known in the mathematical literature. Such integrals can provide an important interpretation of the identities as relating to root systems such as for the Askey-Wilson integral and many generalizations \cite{MR783216, MR772878, MR845667, MR1139492, MR1266569}. See \cite{Okazaki:2023hiv} for details of such integrals and identities in the context of 3d $\Ncal = 2$ gauge theories with boundaries.
It is an interesting question as to when integrals related to root systems have an infinite product evaluation, and in which physical systems these can be realized. We present here large classes of such examples.

We now summarize a main result of our work, the conjecture that for the classes of theories described below, the half-indices and the fundamental Wilson line one-point functions can be expressed in terms of an infinite product given by Ramanujan's general theta function. In particular, in the cases with chirals in the fundamental representation, we find that the unflavored half-index and the one-point function of the fundamental Wilson line 
together form a basis for the line defect indices, and this pair has beautiful expressions as Rogers-Ramanujan type functions 
in terms of Ramanujan's general theta function (\ref{r_theta}) with the form
\begin{align}
\label{general_h_Rtheta}
\mathbb{II}^{G_k}_{\mathcal{N}}(q)&=C_1 \frac{f(a_1,b_1)}{f(-q)}, \\
\label{general_Wfund_Rtheta}
\langle W_{\tiny \yng(1)}\rangle^{G_k}(q)&=C_2 \frac{f(a_2,b_2)}{f(-q)}, 
\end{align}
where $a_i$, $b_i$ and $C_i$ are the theory dependent parameters 
for the $G_k$ CS theories with (anti)fundamental chiral multiplets listed as follows:
\begin{align}
\begin{array}{c|c|c|c|c|c|c}
G_k&C_1&a_1&b_1&C_2&a_2&b_2 \\ \hline 
SU(N)_{-N-1}&1&(-1)^N q^{\frac{N}{2}}&(-1)^Nq^{\frac{N}{2}}&-q^{\frac12}&(-1)^N q^{\frac{N}{2}-1}&(-1)^N q^{\frac{N}{2}+1} \\
USp(2n)_{-n-\frac32}&1&-q^{n+1}&-q^{n+2}&-q^{\frac12}&-q^n&-q^{n+3} \\
SO(N)_{-N,\zeta,\chi}&1& (-1)^{\epsilon} \chi q^{\frac{N}{2}} & (-1)^{\epsilon} \chi q^{\frac{N}{2}} &-2q^{\frac12}& (-1)^{\epsilon} \chi q^{\frac{N-2}{2}} & (-1)^{\epsilon} \chi q^{\frac{N+2}{2}} \\
O(N)_{-N,\zeta,+}&1&q^{2N}&q^{2N}&-2q^{\frac12}&q^{2N-2}&q^{2N+2} \\
O(N)_{-N,\zeta,-}&-q^{\frac{N}{2}}&1&q^{4N}&-(-1)^{\epsilon}2q^{\frac{N-1}{2}}&q^2&q^{4N-2} \\
{G_{2_{-5}}}&1&-q^{\frac32}&-q&-q^{\frac12}&-q^{\frac12}&-q^2 \\
\end{array}
\end{align}
where for the orthogonal groups we have written $N = 2n + \epsilon$ with $\epsilon \in \{0, 1\}$ and we have included the fugacities $\zeta$ and $\chi$ for the discrete $\Zb_2$ magnetic and charge conjugation symmetries.
These form a global $\Zb_2 \times \Zb_2$ symmetry of $SO(N)$ gauge theories. Gauging or partial gauging of this global symmetry produces theories with other orthogonal gauge groups, $Spin(N)$, $Pin_{\pm}(N)$ or $O(N)_{\pm}$. Details of this in relation to superconformal indices were worked out in \cite{Aharony:2013kma} and the manifestation for 3d half-indices is detailed in \cite{Okazaki:2021pnc}.
In fact, all results here are independent of $\zeta$
since the matter we include is not charged under the magnetic $\Zb_2$ symmetry and the vector multiplet with Neumann boundary conditions is also not sensitive to this symmetry. The fugacity $\zeta$ would couple to magnetic flux so there is dependence when calculating full indices or half-indices where the vector multiplet has Dirichlet boundary conditions.
While the case with $USp(2)_{-\frac52}$ $\cong$ $SU(2)_{-\frac52}$ 
where the pair of (\ref{general_h_Rtheta}) and (\ref{general_Wfund_Rtheta}) realizes the Rogers-Ramanujan functions 
\begin{align}
\mathbb{II}_{\mathcal{N}}^{SU(2)_{-\frac52}}&=\frac{f(-q^2,-q^3)}{f(-q)}, \qquad 
\langle W_{\tiny \yng(1)}\rangle^{SU(2)_{-\frac52}}=-q^{\frac12}\frac{f(-q,-q^4)}{f(-q)}
\end{align}
was found in \cite{Okazaki:2024paq}, here
we find more general results for other gauge groups.

Also we propose general formulas of the half-indices and the one-point functions 
for the $SU(N)$ CS theories coupled to an adjoint chiral multiplet 
in terms of Ramanujan's general theta function as eta-products of the form 
\begin{align}
\label{eta_prod}
\prod_{i} \eta(d_i \tau)^{m_i}, 
\end{align}
where $d_i \in \Nb$ and $m_i \in \Zb$ \footnote{We use the term eta-product whereas alternatively these may also be referred to as eta-quotients in general with the term eta-product reserved for the case where all $m_i \in \Nb$.} and the Dedekind eta function is defined by
\begin{align}
\label{eta}
\eta(\tau)&:=q^{\frac{1}{24}}\prod_{n=1}^{\infty}(1-q^n) 
\nonumber\\
&=q^{\frac{1}{24}}f(-q) \; .
\end{align}
It satisfies the following conditions 
\begin{align}
\eta(\tau+1)&=e^{\frac{\pi i}{12}}\eta(\tau),\\
\eta(-1/\tau)&=\sqrt{\tau/i} \eta(\tau)
\end{align}
and hence is a modular form of weight $\frac{1}{2}$.
See appendix~\ref{app_modular} for our definition of a modular form and properties of eta-products.

There exist linear identities for eta-products.
In the case that the eta-products are modular functions (modular forms of weight zero) on some $\Gamma_0(N)$ there is an algorithmic way to prove such identities. This has been formulated and automated in Maple packages provided by Garvan \cite{garvan2019tutorialmapleetapackage}. We use this later to prove some interesting identities which allow us in some examples to reexpress a sum of eta-products, derived by analytically evaluating a half-index, as a single eta-product.

It is observed by Ono \cite{MR2020489} that 
every holomorphic modular form for $SL(2,\mathbb{Z})$ is expressible as a linear combination of eta-products of level $4$. 
As we find that the half-indices and line defect correlators can be also generated by eta-products,  
it is an interesting question to classify the spaces of line defect half-indices for the CS theories with the adjoint chiral multiplet. 

One class of examples where we find results as eta-products is for the $SU(N)_{-2N}$ CS theories with an adjoint chiral multiplet. In particular, we conjecture that 
\begin{align}
\label{hsunadj_tcore}
\mathbb{II}^{SU(N)_{-2N}}_{\mathcal{N},D}(q)
&=q^{-\frac{N^2-1}{12}}\frac{\eta(2N\tau)^N}{\eta(2\tau)} \; , \\
\label{wn_sunadj_tcore}
\langle W_{2Nk+N} \rangle^{SU(N)_{-2N}}
&=(-1)^{N+1}q^{(N-1)k(k+1) -\frac{(N-1)(N-2)}{12}}
\frac{\eta(2\tau)^3\eta(N\tau)^2\eta(2N\tau)^{N-4}}{\eta(\tau)^2} \; , \\
\label{w2n_sunadj_tcore}
\langle W_{2Nk+2N}\rangle^{SU(N)_{-2N}}
&=Nq^{(N-1)(k+1)^2-\frac{N^2-1}{12}}
\frac{\eta(2N\tau)^N}{\eta(2\tau)}
\end{align}
with other charged Wilson line one-point functions vanishing.
Remarkably, the half-index (\ref{hsunadj_tcore}) is identified with the generating function for the $N$-core partitions \cite{MR1055707}. 
According to the equivalent matrix integral expression, the eta-products (\ref{hsunadj_tcore})-(\ref{w2n_sunadj_tcore}) can be also viewed as the closed-form expressions for 
the Schur index \cite{Gadde:2011ik,Gadde:2011uv} and the Schur line defect index \cite{Dimofte:2011py,Cordova:2016uwk} of 4d $\mathcal{N}=2$ pure $SU(N)$ SYM theory.

The eta-products appear as the central objects in the study of the phenomena of ``moonshine'' that relates the sporadic simple group $M_{24}$ to the modular form via some McKay-Thompson series \cite{MR554399,MR822240,MR822235,MR759465,MR1376550}. 
Interestingly, we find McKay-Thompson series in terms of eta-products for half-indices of symplectic gauge theories with an adjoint chiral as well as a fundamental chiral.

Numerous results with eta-product form suggest there should be a physical interpretation of the modular transformation properties of these expressions. We also note that in cases where the results can be expressed as an infinite product, this hints that there may be a confining duality, although it is often not straightforward to make this interpretation physically.
Note that while details of modularity is an interesting avenue which has been explored in some cases \cite{Cheng:2018vpl, Cheng:2019uzc, Costello:2020ndc, Cheng:2022rqr, Ferrari:2023fez} but is not well understood in general for 3d half-indices, that is not our main purpose in this paper. Instead, we are more focussed on presenting classes of theories where the unflavored half-index can be written in infinite product form.

To derive some of the exact results we use the Jacobi triple product identity which can be written in the equivalent forms
\begin{align}
\label{JacobiTripleProduct_q12}
(q)_{\infty} (q^{\frac{1}{2}} x^{\pm}; q)_{\infty} & = \sum_{m \in \Zb} (-1)^m q^{\frac{1}{2} m^2} x^m, 
 \\
\label{JacobiTripleProduct_q}
(q)_{\infty} (x^{-1}; q)_{\infty} (qx; q)_{\infty} & = \sum_{m \in \Zb} (-1)^m q^{\frac{1}{2}m(m+1)} x^m \; .
\end{align}
This is equivalent to
\begin{align}
\label{JacobiTripleProduct}
(q)_{\infty} (x^{\pm}; q)_{\infty} & = (1-x) \sum_{m \in \Zb} (-1)^m q^{\frac{1}{2}m(m+1)} x^m
\end{align}
and this form is used in several derivations to replace products of $q$-Pochhammer symbols with sums. In the case of Neumann boundary condition for the vector multiplet, this often enables the half-index to be expressed as an integral of multiple sums. The integrals can then be evaluated simply using the Cauchy residue theorem and in some cases the resulting sum can be interpreted as a half-index with Dirichlet boundary conditions for a dual vector multiplet.

\subsection{Structure}
\label{sec_str}
The paper is organized as follows.
In section \ref{sec_CS_SUN} we study the $SU(N)$ CS theories with chiral multiplets in the (anti)fundamental and adjoint representations. We consider various cases with different numbers of 3d (anti)fundamental chirals with the gauge anomaly cancellation being achieved by choosing an appropriate Chern-Simons level. Most of our results are conjectures based on calculation of the half-indices or Wilson line one-point functions as $q$-series to high order in $q$. However, in some cases we also present derivations of analytic results. The main focus is in demonstrating/conjecturing expressions which are infinite prodcuts related to Rogers-Ramanujan type functions. We also comment on the simple relation to $U(N)$ theories.
In section \ref{sec_CS_Usp} we consider similar theories but with symplectic gauge group where we derive and conjecture results of a similar form in terms of Rogers-Ramanujan type functions or eta-products. In some cases we conjecture results which are not infinite products but instead given in terms of a finite sum of eta-products.
In section \ref{sec_CS_SO} we consider similar theories but with orthogonal gauge groups. Here we consider the discrete $\Zb_2$ magnetic and charge conjugation symmetries which allow us to describe the half-indices for the various orthogonal groups with different global structures. Again we find results in terms of Rogers-Ramanujan type functions or eta-products. However, there are also many cases where we find that the Wilson line one-point functions are $q$-series with a finite number of terms.
Exceptional groups can also be considered and in section~\ref{sec_CS_G2} we present results for the simplest case of $G_2$.
One common feature for all the types of gauge group we consider is that in many cases we find that the Wilson line one-points functions for general representations can be written simply in terms of the half-index or the fundamental Wilson line one-point function.

\subsection{Future works}
\label{sec_future}

\begin{itemize}

\item 
The line defect half-indices examined in this work are associated with Neumann boundary conditions for the gauge fields and the Wilson line operators in the CS theory. 
One expects to have their dual descriptions involving Dirichlet boundary conditions for the dual gauge fields and the vortex line operators, 
for which the line defect half-indices take the form of certain infinite series. 
As we have found that the Neumann half-indices are given as certain infinite product form, 
such dualities of the boundary conditions can provide us with the Rogers-Ramanujan type identities 
as non-trivial equalities between the infinite product and the infinite series. 
We will leave this interesting possibility for future work. 

\item 
While we find the general formula of the line defect half-indices for the CS theory of (special)unitary gauge group with an adjoint chiral, 
we are interested in exploring more general formulas for the cases with symplectic, orthogonal and exceptional gauge groups by potentially taking different field content with various R-charge assignments. 

\item For lower ranks we have demonstrated the formulas of the line defect half-indices 
while for higher ranks we have numerically checked them in this work. 
It would be nice to complete the analytical proofs for arbitrary rank. 

\item It would be interesting to examine the $q$-difference equations 
satisfied by the line defect half-indices by taking into account the flavor fugacities as studied in \cite{Okazaki:2024paq}. 

\end{itemize}

\section{$SU(N)$ CS theories}
\label{sec_CS_SUN}
While the $SU(N)$ CS theories with chiral multiplets in the (anti)fundamental representations were studied in \cite{Okazaki:2024paq}, 
we generalize the expressions in terms of Ramanujan's general theta function for the one-point functions in this section. 
We also investigate the $SU(N)$ CS theories of level $k$ with an adjoint chiral multiplet, $\Phi$. 
Such theories can be labeled by $r_{\Phi}\in \mathbb{Z}$, the R-charge of the adjoint chiral multiplet which we will take to be zero. 
We note that when $r_{\Phi}=2$, the theories are the so-called $T[L(k,1),SU(N)]$ or simply $T[M_3]$ theories
which can be constructed from M5-branes wrapping the 3-manifold $M_3$ 
$=$ $L(k,1)$ $\cong$ $S^3/\mathbb{Z}_k$ \cite{Gukov:2015sna} but in such cases although there is interest in the study of $q$-series associated to 3-manifolds in the context of 3d-3d correspondence, such as homological blocks labelled $\widehat{Z}$ in \cite{Gukov:2016gkn, Gukov:2017kmk, Cheng:2018vpl, Cheng:2019uzc, Costello:2020ndc, Cheng:2022rqr, Ferrari:2023fez}, expected to be related to physical half-indices, they are not the same as the half-indices we construct. In particular, with the boundary conditions we consider in this article, the half-indices for these theories are not $q$-series with zeroth order term simply $1$. Specifically, if we calculate the half index of these theories with Dirichlet boundary conditions for the adjoint chiral, they will vanish in the unflavored limit. This happens because the contribution from the adjoint chiral with $r_{\Phi}=2$ will contain powers of $(t^{-1}; q)_{\infty}$ in the numerator for Dirichlet boundary condition, where $t$ is a flavor fugacity for the adjoint chiral. Therefore such theories will have different properties from the ordinary unitary superconformal gauge theories which arise for other values of R-charge. In particular we will focus on the case with $r_{\Phi}=0$ and will see that we have half-indices with well-defined $q$-series of the expected form.\footnote{Conversely, for $r_{\Phi}=0$ we cannot take Neumann boundary condition for the adjoint chiral in the unflavored case since the half-indices would diverge in that limit.} We also note that the expressions calculated for $T[M_3]$ \cite{Gukov:2015sna} include contributions to cancel gauge anomalies which are not generally identifiable as 2d chiral or Fermi multiplets. The physical interpretation of these theories and boundary conditions is therefore not clear, and it is not obvious how properties such as mock modularity would generalize to half-indices where we restrict to including only 2d multiplets to cancel gauge anomalies.
We can also introduce $N_f$ fundamental and $N_a$ anti-fundamental 3d chiral multiplets with Dirichlet boundary conditions and we choose these chirals to have R-charge $1$.

The field content is summarized as
\begin{align}
\label{SUN_Adj_Nf_Na_charges}
\begin{array}{c|c|c|c|c|c|c|c}
& \textrm{bc} & SU(N) & SU(N_f) & SU(N_a) & U(1)_A & U(1)_B & U(1)_R \\ \hline
\textrm{VM} & \mathcal{N} & {\bf Adj} & {\bf 1} & {\bf 1} & 0 & 0 & 0 \\
\Phi & \textrm{D} & {\bf Adj} & {\bf 1} & {\bf 1} & 0 & 1 & 0 \\
Q_I & \textrm{D} & {\bf N} & {\bf N_f} & {\bf 1} & 1 & 0 & 1 \\
\overline{Q}_i & \textrm{D} & {\bf \overline{N}} & {\bf 1} & {\bf N_a} & 1 & 0 & 1
\end{array}
\end{align}

We can easily calculate the gauge and 't Hooft anomalies as follows \cite{Dimofte:2017tpi},
\begin{align}
\label{bdy_suN_adj_Nf_anom}
\Acal & = \underbrace{{N\Tr(s^2)} + \frac{N^2 - 1}{2}r^2}_{\textrm{VM}, \; \Ncal} + \underbrace{{N\Tr(s^2)} + \frac{N^2 - 1}{2} (b-r)^2}_{\Phi, \; D} +
\nonumber \\
 & + \underbrace{\left( \frac{N_f}{2}\Tr(s^2) + \frac{N}{2}\Tr(x^2) + \frac{N}{2}N_f a^2 \right)}_{Q_I, \; D}
  + \underbrace{\left( \frac{N_a}{2}\Tr(s^2) + \frac{N}{2}\Tr( \tilde{x}^2) + \frac{N}{2}N_a a^2 \right)}_{\overline{Q}_i, \; D}
  \nonumber \\
  = & \left( 2N + \frac{N_f + N_a}{2} \right) \Tr(s^2) + \frac{N}{2} \Tr(x^2) + \frac{N}{2} \Tr( \tilde{x}^2) + \frac{N}{2}(N_f + N_a) a^2 +
  \nonumber \\
  & + \frac{N^2 - 1}{2}b^2 - (N^2 - 1)br + \left( N^2 - 1 \right) r^2
\end{align}
in the case with an adjoint chiral, while without the adjoint chiral the result is easily seen to be
\begin{align}
\label{bdy_suN_Nf_anom}
\Acal & = \left( N + \frac{N_f + N_a}{2} \right) \Tr(s^2) + \frac{N}{2} \Tr(x^2) + \frac{N}{2} \Tr( \tilde{x}^2) + \frac{N}{2}(N_f + N_a) a^2 +
  \nonumber \\
  & + \frac{1}{2} \left( N^2 - 1 \right) r^2
 \; .
\end{align}

Matching anomalies is important is cases of conjectured dualities. However, the most crucial point, especially since we will be focussing on unflavored half-indices, is that we cancel the gauge anomaly in cases where the vector multiplet has Neumann boundary condition. This can be done through the introduction of 2d boundary $\Ncal = (0,2)$ supermultiplets, either chiral multiplets or Fermi multiplets, or by adjusting the Chern-Simons level. Here we choose not to introduce 2d matter so to
cancel the gauge anomaly we need to take CS level 
\begin{align}
k & = -2N - \frac{1}{2}(N_f + N_a)
\end{align}
in the case with an adjoint chiral and
\begin{align}
k & = -N - \frac{1}{2}(N_f + N_a)
\end{align}
without an adjoint chiral.

It is worthwhile noting that fundamental and anti-fundamental chirals contribute in exactly the same way to the half-index, and similarly a 2d fundamental Fermi multiplet contributes the same way as two fundamental 3d chiral multiplets with Dirichlet boundary conditions, other than the global fugacities. So, if we set the global fugacities to one there may be many different theories with the same specialized half-index.

The contributions to the half-index from each multiplet can be deduced from \cite{Gadde:2013sca, Yoshida:2014ssa, Dimofte:2017tpi}. For the vector multiplet we have
\begin{align}
\label{h_suN_VM}
\frac{(q)_{\infty}^{N-1}}{N!} \oint \left( \prod_{i = 1}^{N-1} \frac{ds_i}{2\pi i s_i} \right) \left( \prod_{i \ne j}^N (s_i s_j^{-1};q)_{\infty} \right)
\end{align}
where $\prod_{i = 1}^N s_i = 1$.

If we fix the R-charge of the adjoint chirals to $r_{\Phi} = 0$ and for the (anti)fundamental chirals to $r = 1$ then the contribution to the half-indices from the adjoint chiral will be a factor in the integrand
\begin{align}
\label{h_suN_Adj}
(q)_{\infty}^{N-1} \left( \prod_{i \ne j}^N (q s_i s_j^{-1};q)_{\infty} \right)
\end{align}
and from the fundamental and antifundamental chirals will be a factor in the integrand
\begin{align}
\label{h_suN_NfNa}
\prod_{i = 1}^N \left( \prod_{I = 1}^{N_f} (q^{\frac{1}{2}} s_i x_I; q)_{\infty} \right) \left( \prod_{J = 1}^{N_a} (q^{\frac{1}{2}} s_i^{-1} \tilde{x}_J; q)_{\infty} \right)
\end{align}
We will mostly consider unflavored (half-)indices where the global flavor fugacities for $SU(N_f)$, $SU(N_a)$, $U(1)_A$ and $U(1)_B$ are fixed. In particular we will always fix the $U(1)_A$ and $U(1)_B$ fugacities to one. For the adjoint chiral there is a simple interpretation of specialising the $U(1)_B$ fugacity to $1$. This corresponds to imposing a $D_c$ boundary condition for $\Phi$ \cite{Dimofte:2017tpi}, meaning that the Dirichlet condition sets the boundary value to a non-zero value $c$, thereby breaking the $U(1)_B$ flavor symmetry. Since the adjoint chiral has R-charge $0$, this does not break the R-symmetry.

There are two types of specialization where it may be possible to calculate analytic results using the Jacobi triple product formula. The first is the case where $N_f = N_a$ and we specialize the flavor fugacities by setting $\tilde{x}_I = x_I^{-1}$. This gives contribution
\begin{align}
\prod_{i = 1}^N \prod_{I = 1}^{N_f} (q^{\frac{1}{2}} s_i^{\pm} x_I^{\pm}; q)_{\infty}
\end{align}
from the fundamental and antifundamental chirals. Note that this can also be interpreted as the contribution from $N_f$ fundamental 2d Fermi multiplets rather than 3d chiral multiplets with Dirichlet boundary conditions. Further specializing to $x_I = 1$ will typically give additional simplification.
This can be viewed as a mathematical simplification of the expression for the half-index but it may also have an interpretation in terms of interesting boundary conditions for the 3d fundamental and anti-fundamental chirals or it may be possible to choose a superpotential to break the flavor symmetry. We note that this cannot be $D_c$ boundary conditions for such chirals as we have given them R-charge $1$ and we don't want to break the R-symmetry.

The second case is $SU(2)$ where, noting that the anti-fundamental and fundamental representations are the same, we set $N_a = 0$ and also specialize the flavor fugacities $x_I = 1$. This gives contribution
\begin{align}
   (q^{\frac{1}{2}} s^{\pm}; q)_{\infty}^{N_f} & = \frac{1}{(q)_{\infty}^{N_f}} \sum_{m_I \in \Zb} (-1)^{\sum_I m_I} q^{\frac{1}{2} \sum_I m_I^2} s^{\sum_I m_I}
\end{align}

The half-index can be modified to give line defect correlators by introducing BPS line operators which are perpendicular to the boundary. 
When we choose the Neumann boundary condition for the gauge field 
and introduce the Wilson line operator $W_{\mathcal{R}}$ transforming in the representation $\mathcal{R}$ of gauge group, 
the line defect correlator is evaluated by inserting the associated character $\chi_{\mathcal{R}}$ in the matrix integral. 
In the case of $SU(N)$ gauge group, 
the character of the irreducible representation labeled by the Young diagram $\lambda$ is the Schur polynomial
\begin{align}
\chi^{\mathfrak{su}(N)}_{\lambda}(s)
&=\frac{\det s_{j}^{\lambda_i+N-i}}{\det s_j^{N-j}}. 
\end{align}
One can take another basis for the set of Wilson lines $W_{k}$ labeled by the power symmetric functions of degree $k$
\begin{align}
p_k(s)
&=\sum_{i=1}^{N}s_i^k, 
\end{align}
where $\prod_{i}s_i=1$. 
We call them the Wilson lines of charge $k$. 

\subsection{$SU(N)_{-N-1}$}
\label{sec_SUN_-N-1}
For the $SU(N)$ CS theory with level $k=-N-1$ whose vector multiplet satisfies the Neumann boundary condition, 
the gauge anomaly free boundary condition is realized by coupling $N_f=N_a=1$ chirals obeying the Dirichlet boundary condition. 
The unflavored half-index is given by \cite{Okazaki:2024paq}
\begin{align}
\label{h_SUN_-N-1}
\II^{SU(N)_{-N-1}}_{\mathcal{N}}(q)
&=\frac{\varphi((-1)^N q^{N/2})}{f(-q)}
=\frac{f((-1)^N q^{N/2}, (-1)^N q^{N/2})}{f(-q)}, 
\end{align}
where $f(a,b)$ and $f(-q)$ are Ramanujan's general theta function (\ref{r_theta}) and (\ref{r_theta_f}). 

Here we observe that the unflavored one-point function of the fundamental Wilson line can be written also
in terms of Ramanujan's general theta function. 
We find that 
\begin{align}
\label{W1_SUN_-N-1}
\langle W_{\tiny \yng(1)}\rangle^{SU(N)_{-N-1}}(q)
&=-q^{\frac12}\frac{f((-1)^N q^{N/2-1},(-1)^N q^{N/2+1} )}{f(-q)}. 
\end{align}
The one-point functions of the Wilson lines in other representations can be given in terms of the 
half-index (\ref{h_SUN_-N-1}) and the one-point function (\ref{W1_SUN_-N-1}). 

The half-index (\ref{h_SUN_-N-1}) and the one-point function (\ref{W1_SUN_-N-1}) agree with the following infinite series
\begin{align}
\label{h_U1_N}
\II^{SU(N)_{-N-1}}_{\mathcal{N}}(q)
&=\frac{1}{(q;q)_{\infty}}\sum_{m\in \mathbb{Z}}
(-1)^{Nm}q^{\frac{Nm^2}{2}}, \\
\label{T1_U1_N}
\langle W_{\tiny \yng(1)}\rangle^{SU(N)_{-N-1}}(q)
&=\frac{1}{(q;q)_{\infty}}\sum_{m\in \mathbb{Z}}
(-1)^{Nm+(N-1)}q^{\frac{Nm^2}{2}+(N-1)m}. 
\end{align}
The infinite series expression (\ref{h_U1_N}) (resp. (\ref{T1_U1_N})) simply indicates that 
the theories considered have a dual description as pure $U(1)_N$ CS theories
whose vector multiplets obeys the Dirichlet boundary condition without (resp. with) a vortex line. 

\subsection{$SU(N)_{-2N}$ with adjoint}
\label{sec_SUN_adj_-2N}
Now we examine the theory with an adjoint chiral multiplet but without any fundamental chirals, i.e.\ $N_f = 0$.

\subsubsection{$SU(2)_{-4}$ with adjoint}
\label{sec_CS_SU2_adj_-4}
We begin with 3d $\mathcal{N} = 2$ $SU(2)$ CS theory of level $k=-4$ with boundary, 
where the vector multiplet has Neumann boundary condition and the adjoint chiral multiplet has Dirichlet boundary condition. 
The half-index is evaluated by the matrix integral
\begin{align}
\label{h_su2_adj_k-4}
\mathbb{II}_{\mathcal{N},D}^{SU(2)_{-4}}(q)
&= \frac{(q)_{\infty}^2}{2} \oint \frac{ds}{2\pi is}
(s^{\pm 2};q)_{\infty} (qs^{\pm 2};q)_{\infty}. 
\end{align}
One finds that it is described by the eta-product 
\begin{align}
\label{h_su2_k4_product}
\mathbb{II}_{\mathcal{N},D}^{SU(2)_{-4}}&=
q^{-\frac14}\frac{\eta(4\tau)^2}{\eta(2\tau)}
\nonumber\\
&=\prod_{n=1}^{\infty}\frac{(1-q^{4n})^2}{1-q^{2n}}. 
\end{align}
It can be expanded as
\begin{align}
\label{h_su2_k4_2}
\mathbb{II}_{\mathcal{N},D}^{SU(2)_{-4}}(q)
&=\sum_{n\ge0}q^{2t_n}
\nonumber\\
&=1+q^2+q^6+q^{12}+q^{20}+q^{30}+q^{42}+q^{56}+\cdots
\end{align}
where $t_n$ $=$ $n(n+1)/2$ are the triangular numbers. 

Since the eta-product in (\ref{h_su2_k4_product}) can be also expressed as a Jacobi theta function (see e.g. \cite{MR1032476}), 
we obtain the infinite series expression 
\begin{align}
\mathbb{II}_{\mathcal{N},D}^{SU(2)_{-4}}(q)
&=\frac12 q^{-\frac14} 
\sum_{m\in \mathbb{Z}}q^{(m+\frac12)^2}
\nonumber\\
&=\frac12 q^{-\frac14} \vartheta_2(2\tau). 
\end{align}

Indeed, these results are easily derived.
Using the Jacobi triple product identity \eqref{JacobiTripleProduct_q} twice we have
\begin{align}
    \II^{SU(2)_{-4}}_{\mathcal{N},D} (q)& = \frac{1}{2} (q)_{\infty}^2 \oint \frac{ds}{2 \pi i s} (s^{\pm 2}; q)_{\infty} (q s^{\pm 2}; q)_{\infty}
    \nonumber \\
    & = \frac{1}{2} \oint \frac{ds}{2 \pi i s} \sum_{m, n \in \Zb} (-1)^{m + n} q^{\frac{1}{2}m(m + 1) + \frac{1}{2}n(n + 1)} s^{2(m - n)}
    \nonumber \\
    \label{h_su2_adj_sum1}
    & = \frac{1}{2} \sum_{m \in \Zb} q^{m(m + 1)} \; .
\end{align}
Now, setting $x = -1$ and replacing $q$ with $q^2$ in \eqref{JacobiTripleProduct_q} we have
\begin{align}
    \sum_{m \in \Zb} q^{m(m + 1)} & = (q^2; q^2)_{\infty} (-1; q^2)_{\infty} (-q^2; q^2)_{\infty}
    \nonumber \\
    & = \prod_{n = 1}^{\infty} (1 - q^{2n}) 2(1 + q^{2n}) (1 + q^{2n})
    \nonumber \\
    & = 2 \prod_{n = 1}^{\infty} \frac{(1 - q^{4n})^2}{1 - q^{2n}} \; .
\end{align}
Hence
\begin{align}
\label{h_su2_adj}
\II^{SU(2)_{-4}}_{\mathcal{N},D}(q)& = q^{-\frac14}\frac{\eta(4\tau)^2}{\eta(2\tau)}
\end{align}
as anticipated. 

Alternatively, we can write 
\begin{align}
\II^{SU(2)_{-4}}_{\mathcal{N},D}(q)& =
\sum_{m\in \mathbb{Z}}
q^{4m^2+2m}
\end{align}
which can be derived starting from \eqref{h_su2_adj_sum1} and splitting the sum into sum over even and odd integers with the replacements $m \to 2m$ and $m \to -(2m+1)$.

The one-point functions of the charged Wilson lines $W_n$ of 3d $\mathcal{N} = 2$ $SU(2)$ CS theory with level $k=-4$ 
obeying the previous boundary conditions are given by
\begin{align}
\label{Wn_su2_adj_k-4}
\langle W_{n}\rangle^{SU(2)_{-4}}
&= \frac{(q)_{\infty}^2}{2} \oint \frac{ds}{2\pi is}
(s^{\pm 2};q)_{\infty} (qs^{\pm 2};q)_{\infty} (s^n+s^{-n}). 
\end{align}

We find that 
\begin{align}
\label{wn_su2_k4a}
\langle W_{4k+2}\rangle^{SU(2)_{-4}}
&=-q^{k(k+1)}\frac{\eta(2\tau)^5}{\eta(\tau)^2\eta(4\tau)^2}
\nonumber\\
&=-q^{k(k+1)}\prod_{n=1}^{\infty}\frac{(1-q^{2n})^5}{(1-q^n)^2(1-q^{4n})^2}, \\
\label{wn_su2_k4b}
\langle W_{4k+4}\rangle^{SU(2)_{-4}}
&=2q^{(k+1)^2-\frac14}\frac{\eta(4\tau)^2}{\eta(2\tau)}
\nonumber\\
&=2q^{(k+1)^2}\prod_{n=1}^{\infty}\frac{(1-q^{4n})^2}{1-q^{2n}}, \\
\label{wn_su2_k4b}
\langle W_{2k+1}\rangle^{SU(2)_{-4}}&=0
\end{align}
for $k\ge0$. 
For even $n$, they are non-trivial whereas they vanish when $n$ is odd. 
Note that the expression (\ref{wn_su2_k4b}) is obtained by multiplying the half-index (\ref{h_su2_k4_product}) by $2q^{(k+1)^2}$. 

According to Jacobi's identity (see e.g. \cite{MR3053701})
\begin{align}
\prod_{n=1}^{\infty}
\frac{(1-q^{2n})^5}{(1-q^n)^2(1-q^{4n})^2}
&=\sum_{m\in \mathbb{Z}}q^{m^2}
=\vartheta_3(2\tau) 
\end{align}
and the eta-product identity, 
the one-point functions (\ref{wn_su2_k4a}) and (\ref{wn_su2_k4b}) can be also expressed as the infinite series
\begin{align}
\langle W_{4k+2}\rangle^{SU(2)_{-4}}
&=-q^{k(k+1)}\sum_{m\in \mathbb{Z}}q^{m^2}
\nonumber\\
&=-q^{k(k+1)}\vartheta_3(2\tau), \\
\langle W_{4k+4}\rangle^{SU(2)_{-4}}
&= q^{(k+1)^2-\frac14} 
\sum_{m\in \mathbb{Z}}q^{(m+\frac12)^2}
\nonumber\\
&= q^{(k+1)^2-\frac14} \vartheta_2(2\tau). 
\end{align}

Similarly, we can analytically show the results as follows: 
\begin{align}
    \langle s^{2 \alpha} \rangle^{SU(2)_{-4}} & = \frac{1}{2} (q)_{\infty}^2 \oint \frac{ds}{2 \pi i s} (s^{\pm 2}; q)_{\infty} (q s^{\pm 2}; q)_{\infty} s^{2 \alpha}
    \nonumber \\
    & = \frac{1}{2} \oint \frac{ds}{2 \pi i s} \sum_{m, n \in \Zb} (-1)^{m + n} q^{\frac{1}{2}m(m + 1) + \frac{1}{2}n(n + 1)} s^{2(m - n + \alpha)}
    \nonumber \\
    & = \frac{1}{2} (-1)^{\alpha} \sum_{n \in \Zb} q^{\frac{1}{2}(n - \alpha)(n - \alpha + 1) + \frac{1}{2}n(n + 1)}
    \; .
\end{align}
Note that in the second line we can see that the expression vanishes if $\alpha$ is not an integer, demonstrating that the expression vanishes for odd powers of $s$.

Now in the case where $\alpha = 2k$ we can define $m = n - k$ to find
\begin{align}
    \sum_{n \in \Zb} q^{\frac{1}{2}(n - \alpha)(n - \alpha + 1) + \frac{1}{2}n(n + 1)} & 
    = \sum_{m \in \Zb} q^{\frac{1}{2}(m - k)(m - k + 1) + \frac{1}{2}(m + k)(m + k + 1)}
    \nonumber \\
    & = q^{k^2} \sum_{m \in \Zb} q^{m(m + 1)}
    = 2 q^{k^2} \prod_{n = 1}^{\infty} \frac{(1 - q^{4n})^2}{1 - q^{2n}}
\end{align}
while if instead $\alpha = 2k + 1$ we have
\begin{align}
    \sum_{n \in \Zb} q^{\frac{1}{2}(n - \alpha)(n - \alpha + 1) + \frac{1}{2}n(n + 1)} & = \sum_{m \in \Zb} q^{\frac{1}{2}(m - k - 1)(m - k) + \frac{1}{2}(m + k)(m + k + 1)}
    \nonumber \\
    & = q^{k(k + 1)} \sum_{m \in \Zb} q^{m^2}
    = q^{k(k + 1)} \prod_{n=1}^{\infty} \frac{(1-q^{2n})^5}{(1-q^n)^2(1-q^{4n})^2} \; .
\end{align}
So, we see that
\begin{align}
\label{h_su2_adj_s4k}
    \langle s^{4k}\rangle^{SU(2)_{-4}} & = q^{k^2} \prod_{n = 1}^{\infty} \frac{(1 - q^{4n})^2}{1 - q^{2n}}
\end{align}
and
\begin{align}
\label{h_su2_adj_s4kp2}
    \langle s^{4k + 2} \rangle^{SU(2)_{-4}} & = - \frac{1}{2} q^{k(k + 1)} \prod_{n=1}^{\infty} \frac{(1-q^{2n})^5}{(1-q^n)^2(1-q^{4n})^2} \; .
\end{align}

We note that all these results involve eta-products of weight $\frac{1}{2}$.

\subsubsection{$SU(3)_{-6}$ with adjoint}
\label{sec_CS_su3k6}
Next consider the $SU(3)$ CS theory with an adjoint chiral multiplet. 
In the absence of the fundamental flavors, the theory coupled to an adjoint chiral has level $k=-6$. 
The half-index reads
\begin{align}
\label{h_su3_m6_int}
\mathbb{II}_{\mathcal{N},D}^{SU(3)_{-6}}
&=\frac{(q)_{\infty}^4}{6} \oint \prod_{i=1}^{2}\frac{ds_i}{2\pi is_i}
\prod_{i\neq j}^{3} \left( \frac{s_i}{s_j};q \right)_{\infty} \left(q\frac{s_i}{s_j};q \right)_{\infty}, 
\end{align}
where $s_3=s_1^{-1}s_2^{-1}$.

Using the Jacobi triple product expression \eqref{JacobiTripleProduct_q} we can replace the integrand with an infinite sum over $6$ integer variables $m_{ij}$ where $i, j, \in \{1, 2, 3\}$ and we fix $m_{ii} = 0$. The two integrals can then be explicitly performed to remove two of the sums, leaving a sum over 4 variables. This gives the result
\begin{align}
\label{h_su3_m6_4sum}
    \mathbb{II}_{\mathcal{N},D}^{SU(3)_{-6}}
     & = \frac{1}{6(q)_{\infty}^2} \sum_{m_{12}, m_{13}, m_{21}, m_{23} \in \Zb} (-1)^{m_{12} + m_{21}} q^{X} \; ,
     \nonumber \\
    X & = \frac{1}{2} m_{12} (m_{12} + 1) + \frac{1}{2} m_{21} (m_{21} + 1) + (m_{12} - m_{21})^2
    \nonumber \\
    & + m_{13}(m_{13} + m_{12} - m_{21} + 1) + m_{23}(m_{23} - m_{12} + m_{21} + 1)
\end{align}

Using Mathematica, it can be expanded as
\begin{align}
1+q^2+2q^4+2q^8+q^{10}+2q^{12}+q^{16}+2q^{18}+2q^{20} + 2q^{24} + 2q^{28} + 3q^{32} +\cdots
\end{align}
We therefore find that to high order in $q$ the half-index agrees with the following eta-product: 
\begin{align}
\mathbb{II}_{\mathcal{N},D}^{SU(3)_{-6}}
&=q^{-\frac23}\frac{\eta(6\tau)^3}{\eta(2\tau)}
\nonumber\\
&=\prod_{n=1}^{\infty}
\frac{(1-q^{6n})^{3}}{1-q^{2n}}. 
\end{align}

It follows that the half-index can be rewritten as
\begin{align}
\label{h_su3_m6_2sum}
\mathbb{II}_{\mathcal{N},D}^{SU(3)_{-6}}
&=\sum_{m_1,m_2\in \mathbb{Z}}
q^{3(m_1^2+m_2^2)+3(m_1+m_2)^2+2m_1+4m_2}
\end{align}
although it is not obvious how to prove the result by deriving this expression from the sum over 4 variables \eqref{h_su3_m6_4sum}.

The one-point functions of the charged Wilson lines $W_n$ are evaluated from the matrix integral 
\begin{align}
\langle W_{n}\rangle^{SU(3)_{-6}}
&=\frac{(q)_{\infty}^4}{6} \oint \prod_{i=1}^{2}\frac{ds_i}{2\pi is_i}
\prod_{i\neq j}^{3} \left( \frac{s_i}{s_j};q \right)_{\infty} \left(q\frac{s_i}{s_j};q \right)_{\infty}(s_1^n+s_2^n+s_1^{-n}s_2^{-n}), 
\end{align}
where $s_3=s_1^{-1}s_2^{-1}$. 

We find that they are non-trivial only when the charge $n$ is divisible by $3$, and in particular are given by
\begin{align}
\langle W_{6k+3}\rangle^{SU(3)_{-6}}
&=q^{2k(k+1)-\frac16}\frac{\eta(2\tau)^3\eta(3\tau)^2}{\eta(\tau)^2\eta(6\tau)}
\nonumber\\
&=
q^{2k(k+1)}
\prod_{n=1}^{\infty}
\frac{(1-q^{2n})^3(1-q^{3n})^2}{(1-q^n)^2(1-q^{6n})}, \\
\langle W_{6k+6}\rangle^{SU(3)_{-6}}
&=3q^{2(k+1)^2-\frac23}\frac{\eta(6\tau)^3}{\eta(2\tau)}
\nonumber\\
&=3q^{2(k+1)^2}\prod_{n=1}^{\infty}
\frac{(1-q^{6n})^3}{(1-q^{2n})}, \\
\langle W_{n}\rangle^{SU(3)_{-6}}&=0, \qquad \textrm{for $n\equiv 1,2 \mod3$}. 
\end{align}
We note that all these results involve eta-products of weight $1$, so for $SU(2)$ and $SU(3)$ we find eta-products of weight half the rank of the gauge group.

\subsubsection{General case}
\label{sec_CSsuN_-2Nadj}
Let us discuss the general 3d $\mathcal{N} = 2$ $SU(N)$ CS theory coupled to an adjoint chiral with boundary, 
where the vector multiplet has Neumann boundary condition and the adjoint chiral multiplet has Dirichlet boundary conditions. 
The half-index is evaluated from the integral 
\begin{align}
\label{h_sun_k-2n_integral}
\II^{SU(N)_{-2N}}_{\mathcal{N}, D} &
=\frac{1}{N!} (q;q)_{\infty}^{2(N-1)}
\oint \prod_{i=1}^{N - 1}\frac{ds_i}{2\pi is_i}
\prod_{i\neq j}^{N} \left( \frac{s_i}{s_j};q \right)_{\infty} \left(q\frac{s_i}{s_j};q \right)_{\infty}, 
\end{align}
where $\prod_{i=1}^Ns_i=1$. Again, we can use the Jacobi triple product formula to express this half-index as a multiple sum. However, after integrating this leaves a sum over $(N-1)^2$ variables but we conjecture an expression in terms of a sum over $(N-1)$ variables.

We conjecture that the half-index (\ref{h_sun_k-2n_integral})  is given by
\begin{align}
\label{h_sun_k2n}
\II^{SU(N)_{-2N}}_{\mathcal{N}, D}
&=q^{-\frac{N^2-1}{12}}\frac{\eta(2N\tau)^N}{\eta(2\tau)}
\nonumber\\
&=\prod_{n = 1}^{\infty} \frac{(1 - q^{2Nn})^N}{1 - q^{2n}}
 = \frac{(q^{2N}; q^{2N})_{\infty}^N}{(q^2; q^2)_{\infty}} \; . 
\end{align}
We note that this is an eta-product of weight $\frac{1}{2}(N-1)$, generalizing the pattern noted for the previous examples.

Interestingly, the function (\ref{h_sun_k2n}) is identified with the generating function for the number $c_N(n)$ of $N$-core partitions, 
the partitions to which the associated Young diagrams have no hook length divisible by $N$ \cite{MR1055707}. 
Hence 
\begin{align}
\label{tcore_partition}
\II^{SU(N)_{-2N}}_{\mathcal{N}, D}&=\sum_{n\ge 0} c_N(n)q^{2n}. 
\end{align}
This implies that 
3d $\mathcal{N} = 2$ $SU(N)$ CS theory of level $k=-2N$ with an adjoint chiral multiplet  admits 
the gauge invariant BPS local operators living at the boundary 
which are one-to-one with the $N$-core partitions! 

For prime $N>3$ it follows that \cite{MR4400889}
\begin{align}
\frac{\eta(N\tau)^N}{\eta(\tau)}
\in \mathcal{M}_{\frac{N-1}{2}}(\Gamma_{0}(N),\chi_N) \; ,
\end{align}
where $\mathcal{M}_{k}(\Gamma_0(N),\chi_N)$ is the vector space of holomorphic forms on $\Gamma_0(N)$ with the Legendre symbol $\chi_N$. 
The half-index is simply obtained by replacing $\tau$ with $2\tau$ and multiplying it by a monomial in $q$. Therefore, the half-index multiplied by a suitable power of $q$ belongs to $\mathcal{M}_{k}(\Gamma_0(N),\chi_N)$.

There is a bijection on $t$-core partitions from which the generating function can be rewritten as the infinite series \cite{MR1055707}. 
Therefore we have 
\begin{align}
\II^{SU(N)_{-2N}}_{\mathcal{N}, D}
&=\sum_{m_1,\cdots,m_{N-1}\in \mathbb{Z}}
q^{N\sum_{I=1}^{N-1}m_I^2+N(\sum_{I=1}^{N-1}m_I)^2+2\sum_{I=1}^{N-1} (I-N) m_I}
\nonumber \\
&=\sum_{m_1,\cdots,m_{N-1}\in \mathbb{Z}}
q^{N\sum_{I=1}^{N}m_I^2 + 2\sum_{I=1}^{N} I m_I} 
\end{align}
where $m_N = - \sum_{I=1}^{N-1}m_I$.
It is tempting to try to interpret this result as the half-index of a dual theory with gauge group $SU(N)_{2N}$ with Dirichlet boundary conditions for the vector multiplet. The absence of the vector multiplet contribution, leaving only the monopole contribution, would correspond to having an adjoint chiral of R-charge zero with $D_c$ boundary condition. However, the origin of the contribution to the exponent of $q$ which is linear in $m_I$ is not clear.

In the large $N$ limit, it becomes the generating function for ordinary partitions 
\begin{align}
\lim_{N\rightarrow \infty}
\II^{SU(N)_{-2N}}_{\mathcal{N}, D} & = \prod_{n=1}^{\infty}\frac{1}{1-q^{2n}}. 
\end{align}

Also we find that the one-point functions of the charged Wilson line operators are non-trivial only when the charges are multiples of $N$. 
We conjecture that they also have expressions as eta-products with the structure depending on whether the charge is $0$ or $N$ modulo $2N$
\begin{align}
\label{wn_suN_k2Na}
\langle W_{2Nk+N} \rangle^{SU(N)_{-2N}}
&=(-1)^{N+1}q^{(N-1)k(k+1) -\frac{(N-1)(N-2)}{12}}
\frac{\eta(2\tau)^3\eta(N\tau)^2\eta(2N\tau)^{N-4}}{\eta(\tau)^2}
\nonumber\\
&=(-1)^{N+1}q^{(N-1)k(k+1)}
\prod_{n=1}^{\infty}
\frac{(1-q^{2n})^3 (1-q^{Nn})^{2} (1-q^{2Nn})^{N-4}}
{(1-q^n)^2}, \\
\label{wn_suN_k2Nb}
\langle W_{2Nk+2N}\rangle^{SU(N)_{-2N}}
&=Nq^{(N-1)(k+1)^2-\frac{N^2-1}{12}}
\frac{\eta(2N\tau)^N}{\eta(2\tau)}
\nonumber\\
&=Nq^{(N-1)(k+1)^2}\prod_{n=1}^{\infty}
\frac{(1-q^{2Nn})^N}{1-q^{2n}}, \\
\label{wn_suN_k2Nc}
\langle W_{n}\rangle^{SU(N)_{-2N}}
&=0\qquad \textrm{for $n\not\equiv 0,N \mod 2N$}. 
\end{align}
Note that all these results are of the form of an eta-product of weight $\frac{1}{2}(N-1)$ multiplied by a monomial in $q$.

We remark that the matrix integrals (\ref{h_sun_k2n}), (\ref{wn_suN_k2Na}) and (\ref{wn_suN_k2Nb}) 
for the half-index and the line defect correlators
are equivalent to those of the Schur index \cite{Gadde:2011ik,Gadde:2011uv} and of the Schur line defect indices \cite{Dimofte:2011py,Cordova:2016uwk} of 4d $\mathcal{N}=2$ pure $SU(N)$ SYM theory. 

\subsection{$SU(N)_{-2N-N_f/2}$ with adjoint}
\label{sec_suN_-2N-Nf/2+adj}
In this section we consider the $SU(N)$ CS theory coupled to $N_f$ fundamental chirals as well as an adjoint chiral. As previously noted, since we have set the global fugacities to one, there are other interpretations of the fundamental chiral contributions as anti-fundamental chirals or with two replaced by a 2d fundamental Fermi. We present some results for $N_f = 2$, including a general conjecture for $SU(N)$, then consider an example with $N_f = 1$.

\subsubsection{$SU(2)_{-5}$ with adjoint}
\label{sec_CS_su2_-5}
In the case with $N_f = 2$, we have the $SU(2)$ CS theory with an adjoint chiral multiplet and level $k=-5$. 

We find that the unflavored half-index exactly agrees with the following eta-product: 
\begin{align}
\mathbb{II}_{\mathcal{N},D}^{SU(2)_{-5}}
&=q^{-\frac16}\frac{\eta(2\tau)^5 \eta(5\tau)}{\eta(\tau)^3\eta(4\tau)^2}
\nonumber\\
&=\prod_{n=1}^{\infty}
\frac{(1-q^{2n})^5 (1-q^{5n})}{(1-q^{n})^3(1-q^{4n})^2}. 
\end{align}
As for the case with $N_f = 0$ we see that the weight of the eta-product is half the rank.

In fact, we can calculate this exactly as follows
\begin{align}
\mathbb{II}_{\mathcal{N},D}^{SU(2)_{-5}}
&= \frac{(q)_{\infty}^2}{2} \oint \frac{ds}{2\pi is}
(s^{\pm 2};q)_{\infty} (qs^{\pm 2};q)_{\infty}
(q^{\frac12}s^{\pm};q)_{\infty}^2
\nonumber \\
 & = \frac{1}{(q)_{\infty}^2} \left( \prod_{n = 1}^{\infty} \frac{(1 - q^{2n})^4(1 - q^{10n})^5}{(1 - q^{n})^2(1 - q^{5n})^2(1 - q^{20n})^2} - \right.
\nonumber \\
 & \left. - q \prod_{n = 1}^{\infty} \frac{(1 - q^{2n})^{10}(1 - q^{20n})^2}{(1 - q^n)^4(1 - q^{4n})^4(1 - q^{10n})} \right). 
\end{align}

The matching of these two expressions is equivalent to a linear eta-product identity
\begin{align}
      \frac{\eta(4\tau)^2 \eta(10\tau)^5}{\eta(\tau) \eta(2\tau) \eta(5\tau)^3 \eta(20\tau)^2} - \frac{\eta(2\tau)^5 \eta(20\tau)^2}{\eta(\tau)^3 \eta(4\tau)^2 \eta(5\tau) \eta(10\tau)}
      & = 1, 
\end{align}
where we note that the two eta-products on the left are modular functions. We have proved this identity\footnote{We thank one of the referees for pointing out that this identity was found by Michael Somos and can be found as t20\_12\_78 in the file eta07.gp at \cite{Somos}.} using the qseries and ETA packages for Maple created by Garvan \cite{garvan2019tutorialmapleetapackage}.

Unlike the case without fundamental chirals, 
the one-point functions of the Wilson line operators do not seem to have simple expressions in terms of eta-products. 

\subsubsection{$SU(3)_{-7}$ with adjoint}
\label{sec_CS_su3k7}
To gain more insight, we proceed by examining the $SU(3)$ CS theory of level $-7$ with $N_f=2$ fundamental chirals and an adjoint chiral. 
We find that the unflavored half-index is given by the eta-product, again of weight half the rank of the gauge group and where $\prod_{i=1}^3 s_i=1$,
\begin{align}
\mathbb{II}_{\mathcal{N},D}^{SU(3)_{-7}}
\label{h_su3_-7}
 & = \frac{1}{3!} (q;q)_{\infty}^{4}
\oint \prod_{i=1}^{2}\frac{ds_i}{2\pi is_i}
\left(
\prod_{i\neq j}^{3} \left( \frac{s_i}{s_j};q \right)_{\infty} \left(q\frac{s_i}{s_j};q \right)_{\infty}
\right)
\prod_{i = 1}^3
(q^{\frac12}s_i^{\pm};q)_{\infty}^2
\nonumber \\
&=q^{-\frac{13}{24}}
\frac{\eta(3\tau/2)^2 \eta(7\tau)^2}{\eta(\tau)\eta(3\tau)}
\nonumber\\
&=\prod_{n=1}^{\infty}\frac{(1-q^{\frac{3n}{2}})^2 (1-q^{7n})^2}{(1-q^n) (1-q^{3n})}. 
\end{align}

\subsubsection{General case}
\label{sec_CSsuN_-2N1adj}
Now we propose the general formula for the half-index for 3d $\mathcal{N} = 2$ $SU(N)$ CS theory of level $-2N-1$ (with an adjoint chiral and $N_f = 2$)
where the vector multiplet has Neumann boundary condition and the adjoint and (anti)fundamental chiral multiplets have Dirichlet boundary conditions. 
We conjecture that 
\begin{align}
\label{h_suN_-2N-1}
\II^{SU(N)_{-2N-1}}_{\mathcal{N}, D}
 & = \frac{1}{N!} (q;q)_{\infty}^{2(N-1)}
\oint \prod_{i=1}^{N - 1}\frac{ds_i}{2\pi is_i}
\left(
\prod_{i\neq j}^{N} \left( \frac{s_i}{s_j};q \right)_{\infty} \left(q\frac{s_i}{s_j};q \right)_{\infty}
\right)
\prod_{i = 1}^N
(q^{\frac12}s_i^{\pm};q)_{\infty}^2
\nonumber \\
& = \prod_{n = 1}^{\infty} \frac{(1 - (-1)^N q^{Nn})^2 (1 - q^{Nn})^3 (1 - q^{(2N+1)n})^{N-1}}{(1 - q^{n}) (1 - (-1)^N q^{\frac{N}{2}n})^2 (1 - q^{2Nn})^2}
\nonumber\\
& = \begin{cases}
q^{-\frac{2N^2-N-2}{24}}\frac{\eta(N\tau)^5\eta((2N+1)\tau)^{N-1}}{\eta(\tau)\eta(N\tau/2)^2\eta(2N\tau)^2} & \textrm{for $N$ even}\cr
q^{-\frac{2N^2-N-2}{24}}\frac{\eta(N\tau/2)^2 \eta((2N+1)\tau)^{N-1}}{\eta(\tau)\eta(N\tau)} & \textrm{for $N$ odd}\cr
\end{cases}
\nonumber \\
 & = \begin{cases}
 \prod_{n = 1}^{\infty} \frac{(1 - q^{Nn})^5 (1 - q^{(2N+1)n})^{N-1}}{(1 - q^{n}) (1 - q^{\frac{N}{2}n})^2 (1 - q^{2Nn})^2} & \textrm{for $N$ even}\cr
 \prod_{n = 1}^{\infty} \frac{(1 - q^{\frac{N}{2}n})^2 (1 - q^{(2N+1)n})^{N-1}}{(1 - q^{n}) (1 - q^{Nn})} & \textrm{for $N$ odd}\cr
\end{cases}
\end{align}
where as before the weight of the eta-products is half the rank of the gauge group.

\subsubsection{$SU(2)_{-9/2}$ with adjoint}
\label{sec_CS_su_-9/2}
When instead we take $N_f=1$, the $SU(2)$ CS theory with a single adjoint chiral multiplet has level $k=-9/2$. 
In this case, the unflavored half-index is given by
\begin{align}
\mathbb{II}_{\mathcal{N},D}^{SU(2)_{-9/2}}(x;q)
&= \frac{(q)_{\infty}^2}{2} \oint \frac{ds}{2\pi is}
(s^{\pm 2};q)_{\infty} (qs^{\pm 2};q)_{\infty}
(q^{\frac12}s^{\pm}x;q)_{\infty}. 
\end{align}

We find that the unflavored half-index is expressible as the weight-$\frac{1}{2}$ eta-product
\begin{align}
\mathbb{II}_{\mathcal{N},D}^{SU(2)_{-9/2}}(x;q)
&=q^{-\frac{5}{24}}\frac{\eta(3\tau)^2}{\eta(\tau)}
\nonumber\\
&=\prod_{n=1}^{\infty}
\frac{(1-q^{3n})^2}{(1-q^{n})}. 
\end{align}

Indeed we can calculate this exactly as follows
\begin{align}
\mathbb{II}_{\mathcal{N},D}^{SU(2)_{-9/2}}(x=1;q)
&= \frac{(q)_{\infty}^2}{2} \oint \frac{ds}{2\pi is}
(s^{\pm 2};q)_{\infty} (qs^{\pm 2};q)_{\infty}
(q^{\frac12}s^{\pm};q)_{\infty}
\nonumber \\
 & = \langle (q^{\frac12}s^{\pm};q)_{\infty} \rangle^{SU(2)_{-4}} = \frac{1}{(q)_{\infty}} \sum_{m \in \Zb} (-1)^m q^{\frac{1}{2}m^2} \langle s^m \rangle^{SU(2)_{-4}}
\nonumber \\
 & = \frac{1}{(q)_{\infty}} \sum_{k \in \Zb} \left( q^{8k^2} \langle s^{4k} \rangle^{SU(2)_{-4}} + q^{2(2k+1)^2} \langle s^{4k+2} \rangle^{SU(2)_{-4}} \right)
\nonumber \\
 & = \frac{1}{(q)_{\infty}} \sum_{k \in \Zb} \left( q^{9k^2} \prod_{n = 1}^{\infty} \frac{(1 - q^{4n})^2}{1 - q^{2n}} - \frac{1}{2} q^2 q^{9k(k+1)} \prod_{n = 1}^{\infty} \frac{(1 - q^{2n})^5}{(1 - q^n)^2 (1 - q^{4n})^2} \right)
\nonumber \\
 & = \frac{1}{(q)_{\infty}} \left( \prod_{n = 1}^{\infty} \frac{(1 - q^{4n})^2 (1 - q^{18n})^5}{(1 - q^{2n}) (1 - q^{9n})^2 (1 - q^{36n})^2} - \right.
\nonumber \\
 & \left. - q^2 \prod_{n = 1}^{\infty} \frac{(1 - q^{2n})^5 (1 - q^{36n})^2}{(1 - q^{n})^2 (1 - q^{4n})^2 (1 - q^{18n})} \right). 
\end{align}

The matching of these two expressions is equivalent to a linear eta-product identity
\begin{align}
     \frac{\eta(4\tau)^2 \eta(18\tau)^5}{\eta(2\tau) \eta(3\tau)^2 \eta(9\tau)^2 \eta(36\tau)^2} - \frac{\eta(2\tau)^5 \eta(36\tau)^2}{\eta(\tau)^2 \eta(3\tau)^2 \eta(4\tau)^2 \eta(18\tau)}
    & = 1
\end{align}
where we note that the two eta-products on the left are modular functions. We have proved this identity\footnote{We thank one of the referees for pointing out that this identity was found by Michael Somos and can be found as t36\_12\_126 in the file eta07.gp at \cite{Somos}.} using the qseries and ETA packages for Maple created by Garvan \cite{garvan2019tutorialmapleetapackage}.

\begin{comment}
\section{$U(N)$ CS theories}
\label{sec_CS_uN}
The half-indices for $U(N)$ CS theory are simply those for the $SU(N)$ theory multiplied by the factor $(q)_{\infty}^2$
\begin{align}
    \II^{U(N)} & = (q)_{\infty}^2 \prod_{n = 1}^{\infty} \frac{(1 - q^{2Nn})^N}{1 - q^{2n}}. 
\end{align}
\end{comment}

\subsection{$U(N)$ vs $SU(N)$}
\label{sec_uN_suN}
The results for $U(N)$ CS theories are essentially the same as for the $SU(N)$ theories we have considered above. The main difference in the half-indices is that for $U(N)$ we have $N$ independent gauge fugacities, say $t_i$, whereas for $SU(N)$ there are $N$ gauge fugacities, say $s_i$, with a constraint $s_N = \prod_{i=1}^{N-1}s_i^{-1}$. However, we can write $t_i = t^{1/N} s_i$ and it is easy to see that
\begin{align}
    \prod_{i = 1}^N \oint \frac{dt_i}{t_i} & = \oint \frac{dt}{t} \prod_{i = 1}^{N-1} \oint \frac{ds_i}{s_i} \; .
\end{align}
Also note that the contribution to the half-index from a vector multiplet 
or an adjoint chiral is the same for $U(N)$ as for $SU(N)$ up to an overall factor of $(q)_{\infty}$ for each. 

Hence we see that with Neumann boundary conditions for the vector multiplet and Dirichlet for the adjoint chiral,
\begin{align}
    \langle \prod_{i = 1}^N t_i^{\alpha_i} \rangle & = \left\{ \begin{array}{lcl} 0 & , & \sum_{i = 1}^N \alpha_i \ne 0 \\ (q)_{\infty}^2 \langle \prod_{i = 1}^{N-1} s_i^{\alpha_i} \rangle  & , & \sum_{i = 1}^N \alpha_i = 0 \end{array} \right.
\end{align}

For example if we consider $U(2)$ with an adjoint chiral we have

\begin{comment}
\subsection{$U(2)$}
\label{sec_CS_u2}
Using the Jacobi triple product identity \eqref{JacobiTripleProduct_q} twice we have
\begin{align}
    \II^{U(2)} & = \frac{1}{2} (q)_{\infty}^4 \oint \frac{ds_1}{2 \pi i s_1} \frac{ds_2}{2 \pi i s_2} (s_1^{\pm} s_2^{\mp}; q)_{\infty} (q s_1^{\pm} s_2^{\mp}; q)_{\infty}
    \nonumber \\
    & = \frac{1}{2} (q)_{\infty}^2 \oint \frac{ds_1}{2 \pi i s_1} \frac{ds_2}{2 \pi i s_2} \sum_{m, n \in \Zb} (-1)^{m + n} q^{\frac{1}{2}m(m + 1) + \frac{1}{2}n(n + 1)} (s_1 s_2^{-1})^{m - n}
    \nonumber \\
    & = \frac{1}{2} (q)_{\infty}^2 \oint \frac{ds_2}{2 \pi i s_2} \sum_{m \in \Zb} q^{m(m + 1)}
    \nonumber \\
    & = \frac{1}{2} (q)_{\infty}^2 \sum_{m \in \Zb} q^{m(m + 1)} \; .
\end{align}
Now, setting $x = -1$ and replacing $q$ with $q^2$ in \eqref{JacobiTripleProduct_q} we have
\begin{align}
    \sum_{m \in \Zb} q^{m(m + 1)} & = (q^2; q^2)_{\infty} (-1; q^2)_{\infty} (-q^2; q^2)_{\infty}
    \nonumber \\
    & = \prod_{n = 1}^{\infty} (1 - q^{2n}) 2(1 + q^{2n}) (1 + q^{2n})
    \nonumber \\
    & = 2 \prod_{n = 1}^{\infty} \frac{(1 - q^{4n})^2}{1 - q^{2n}} \; .
\end{align}
\end{comment}

\begin{align}
    \II^{U(2)}_{\mathcal{N}} & = (q)_{\infty}^2 \prod_{n = 1}^{\infty} \frac{(1 - q^{4n})^2}{1 - q^{2n}}
\end{align}
using \eqref{h_su2_adj}.

Similarly we can calculate
\begin{align}
    \langle s_1^{\alpha} s_2^{\beta} \rangle^{U(2)} & = \frac{1}{2} (q)_{\infty}^4 \oint \frac{ds_1}{2 \pi i s_1} \frac{ds_2}{2 \pi i s_2} (s_1^{\pm} s_2^{\mp}; q)_{\infty} (q s_1^{\pm} s_2^{\mp}; q)_{\infty} s_1^{\alpha} s_2^{\beta}
\end{align}

\begin{comment}
    \nonumber \\
    & = \frac{1}{2} (q)_{\infty}^2 \oint \frac{ds_1}{2 \pi i s_1} \frac{ds_2}{2 \pi i s_2} \sum_{m, n \in \Zb} (-1)^{m + n} q^{\frac{1}{2}m(m + 1) + \frac{1}{2}n(n + 1)} s_1^{m - n + \alpha} s_2^{-m + n + \beta}
    \nonumber \\
    & = \frac{1}{2} (q)_{\infty}^2 (-1)^{\alpha} \oint \frac{ds_2}{2 \pi i s_2} \sum_{n \in \Zb} q^{\frac{1}{2}(n - \alpha)(n - \alpha + 1) + \frac{1}{2}n(n + 1)} s_2^{\alpha + \beta}
    \nonumber \\
    & = \left \{ \begin{array}{lcl} 0 & , & \alpha + \beta \ne 0 \\ \frac{1}{2} (q)_{\infty}^2 (-1)^{\alpha} \sum_{n \in \Zb} q^{\frac{1}{2}(n - \alpha)(n - \alpha + 1) + \frac{1}{2}n(n + 1)} & , & \alpha + \beta = 0 \; . \end{array} \right.
\end{align}

Now in the case where $\alpha = 2k$ we can define $m = n - k$ to find
\begin{align}
    \sum_{n \in \Zb} q^{\frac{1}{2}(n - \alpha)(n - \alpha + 1) + \frac{1}{2}n(n + 1)} & = \sum_{m \in \Zb} q^{\frac{1}{2}(m - k)(m - k + 1) + \frac{1}{2}(m + k)(m + k + 1)}
    \nonumber \\
    & = q^{k^2} \sum_{m \in \Zb} q^{m(m + 1)}
    = 2 q^{k^2} \prod_{n = 1}^{\infty} \frac{(1 - q^{4n})^2}{1 - q^{2n}}
\end{align}
while if instead $\alpha = 2k + 1$ we have
\begin{align}
    \sum_{n \in \Zb} q^{\frac{1}{2}(n - \alpha)(n - \alpha + 1) + \frac{1}{2}n(n + 1)} & = \sum_{m \in \Zb} q^{\frac{1}{2}(m - k - 1)(m - k) + \frac{1}{2}(m + k)(m + k + 1)}
    \nonumber \\
    & = q^{k(k + 1)} \sum_{m \in \Zb} q^{m^2}
    = q^{k(k + 1)} \prod_{n=1}^{\infty} \frac{(1-q^{2n})^5}{(1-q^n)^2(1-q^{4n})^2} \; .
\end{align}
So, we see that

\end{comment}

with results using \eqref{h_su2_adj_s4k}
\begin{align}
    \langle s_1^{2k} s_2^{-2k} \rangle^{U(2)} & = q^{k^2} (q)_{\infty}^2 \prod_{n = 1}^{\infty} \frac{(1 - q^{4n})^2}{1 - q^{2n}}
\end{align}
and \eqref{h_su2_adj_s4kp2}
\begin{align}
    \langle s_1^{2k + 1} s_2^{-(2k + 1)} \rangle^{U(2)} & = - \frac{1}{2} q^{k(k + 1)} (q)_{\infty}^2 \prod_{n=1}^{\infty} \frac{(1-q^{2n})^5}{(1-q^n)^2(1-q^{4n})^2} \; .
\end{align}

\section{$USp(2n)$ CS theories}
\label{sec_CS_Usp}
Here we consider $USp(2n)_k$ gauge theories with $N_f$ fundamental chirals 
where the vector multiplet has Neumann boundary condition and the fundamental chirals, $Q_I$, have R-charge $1$ and Dirichlet boundary conditions. We may also consider an adjoint chiral $\Phi$ with R-charge $0$ and Dirichlet boundary condition.

We summarize the field content and charges in the following table
\begin{align}
\label{USp_Adj_Nf_charges}
\begin{array}{c|c|c|c|c|c|c}
& \textrm{bc} & USp(2n) & SU(N_f) & U(1)_A & U(1)_B & U(1)_R \\ \hline
\textrm{VM} & \mathcal{N} & {\bf Adj} & {\bf 1} & 0 & 0 & 0 \\
\Phi & \textrm{D} & {\bf Adj} & {\bf 1} & 0 & 1 & 0 \\
Q_I & \textrm{D} & {\bf 2n} & {\bf N_f} & 1 & 0 & 1
\end{array}
\end{align}

We can easily calculate the gauge and 't Hooft anomalies \cite{Okazaki:2021pnc} as follows,
\begin{align}
\label{bdy_USp_adj_Nf_anom}
\Acal & = \underbrace{{(n+1)\Tr(s^2)} + \frac{n(2n+1)}{2}r^2}_{\textrm{VM}, \; \Ncal} + \underbrace{{(n+1)\Tr(s^2)} + \frac{n(2n+1)}{2} (b-r)^2}_{\Phi, \; D}
\nonumber \\
 & + \underbrace{\left( \frac{N_f}{2} \Tr(s^2) + n\Tr(x^2) + nN_fa^2 \right)}_{Q_I, \; D}
  \nonumber \\
  = & \left( 2n + 2 + \frac{N_f}{2} \right) \Tr(s^2) + n \Tr(x^2) + nN_f a^2 +
  \nonumber \\
  & + \frac{n(2n+1)}{2}b^2 - n(2n+1) br + n(2n+1) r^2
\end{align}
in the case with an adjoint chiral, while without the adjoint chiral the result is easily seen to be
\begin{align}
\label{bdy_USp_Nf_anom}
\Acal & = \left( n + 1 + \frac{N_f}{2} \right) \Tr(s^2) + n \Tr(x^2) + nN_f a^2 + \frac{n(2n+1)}{2} r^2
 \; .
\end{align}

To cancel the gauge anomaly we need to take CS level 
\begin{align}
k & = -2n - 2 - \frac{N_f}{2}
\end{align}
in the case with an adjoint chiral and
\begin{align}
k & = -n - 1 - \frac{N_f}{2}
\end{align}
without an adjoint chiral.

Similar to the case with unitary gauge group, the only distinction between two fundamental chirals with Dirichlet boundary conditions and a 2d fundamental Fermi multiplet is in the details of the global symmetries. Therefore we cannot distinguish between these cases if we set the flavor fugacities to $1$.

The half-index takes the form \cite{Gadde:2013sca, Yoshida:2014ssa, Dimofte:2017tpi, Okazaki:2021pnc}
\begin{align}
\label{h_USp2n_k}
\mathbb{II}_{\mathcal{N}}^{USp(2n)_k}
&=\frac{(q)_{\infty}^{n}}{n! 2^n} \prod_{i=1}^n \oint \frac{ds_i}{2\pi i s_i}
\left( \prod_{i \ne j}^n (s_i s_j^{-1}; q)_{\infty} \right) \left( \prod_{i \le j}^n (s_i^{\pm} s_j^{\pm}; q)_{\infty} \right)
\nonumber\\
&\times 
\prod_{\alpha = 1}^{N_f} \prod_{i = 1}^n (q^{1/2} s_i^{\pm} x_{\alpha}; q)_{\infty} \; .
\end{align}
without an adjoint chiral and
\begin{align}
\label{h_USp2n_adj_k}
&
\mathbb{II}_{\mathcal{N}, D}^{USp(2n)_k}
\nonumber\\
&=\frac{(q)_{\infty}^{2n}}{n! 2^n} \prod_{i=1}^n \oint \frac{ds_i}{2\pi i s_i}
\left( \prod_{i \ne j}^n (s_i s_j^{-1}; q)_{\infty} (q s_i s_j^{-1}; q)_{\infty} \right) \left( \prod_{i \le j}^n (s_i^{\pm} s_j^{\pm}; q)_{\infty} (q s_i^{\pm} s_j^{\pm}; q)_{\infty} \right)
\nonumber\\
&\times 
\prod_{\alpha = 1}^{N_f} \prod_{i = 1}^n (q^{1/2} s_i^{\pm} x_{\alpha}; q)_{\infty}
\end{align}
with an adjoint chiral.

Again the correlators of the BPS Wilson line operators in the representation $\mathcal{R}$ can be evaluated by introducing the characters. 
In the case of $USp(2n)$ gauge group, 
the character of the irreducible representation with highest weight 
labeled by the Young diagram $\lambda$ is given by \cite{MR1153249}
\begin{align}
\label{usp2n_character}
\chi^{\mathfrak{usp}(2n)}_{\lambda}(s)
&=\frac{\det (s_j^{\lambda_i+n-i+1}-s_j^{-(\lambda_i+n-i+1)})}
{\det(s_j^{n-i+1}-s_j^{-(n-i+1)})}. 
\end{align}
For example, the character of the fundamental representation is 
\begin{align}
\chi^{\mathfrak{usp}(2n)}_{\tiny \yng(1)}(s)
&=\sum_{i=1}^{n}(s_i+s_i^{-1})
\end{align}
We consider another basis for the set of Wilson lines $W_{k}$ labeled by the power symmetric functions
\begin{align}
p_k(s)+p_k(s^{-1})
&=\sum_{i=1}^{n}s_i^k+s_i^{-k}. 
\end{align}
Similarly, we call them the Wilson lines of charge $k$. 

\subsection{$USp(2n)_{-n-1}$}
\label{sec_USp2n_-n-1}
Let us begin with the pure CS theory of symplectic gauge group, i.e.\ $N_f = 0$ and no adjoint chiral.

\subsubsection{$USp(4)_{-3}$}
\label{sec_USp4_k3}
With $n = 2$, one finds the $USp(4)$ pure CS theory of level $k=-3$ 
whose vector multiplet obeys the Neumann boundary condition. 
As the theory has no matter fields, there is no non-trivial gauge invariant BPS local operators at the boundary 
so that the half-index is trivial 
\begin{align}
\label{h_USp4_k3}
\mathbb{II}_{\mathcal{N}}^{USp(4)_{-3}}
&=\frac{(q)_{\infty}^{2}}{8} \prod_{i=1}^2 \oint \frac{ds_i}{2\pi i s_i}
\left( \prod_{i \ne j}^2 (s_i s_j^{-1}; q)_{\infty} \right) \left( \prod_{i \le j}^2 (s_i^{\pm} s_j^{\pm}; q)_{\infty} \right)=1.
\end{align}

Non-trivial one-point functions appear for the Wilson lines with even charges
\begin{align}
\langle W_2\rangle^{USp(4)_{-3}}&=-1-q,\\
\langle W_4\rangle^{USp(4)_{-3}}&=-1-q^2,\\
\langle W_6\rangle^{USp(4)_{-3}}&=q+q^2+q^4+q^5,\\
\langle W_8\rangle^{USp(4)_{-3}}&=-q^4-q^8,\\
\langle W_{10}\rangle^{USp(4)_{-3}}&=-q^5-q^{10},\\
\langle W_{12}\rangle^{USp(4)_{-3}}&=q^8+q^{10}+q^{14}+q^{16}. 
\end{align}

We conjecture that the non-trivial one-point function of the Wilson line in the symmetric representation is given by
\begin{align}
\label{wsym_USp4_k3}
\langle W_{(6k)}\rangle^{USp(4)_{-3}}&=q^{k(3k+2)}, \nonumber\\ 
\langle W_{(6k-4)}\rangle^{USp(4)_{-3}}&=-q^{k(3k-2)}
\end{align}
for $k=1,2,\cdots$. 
We have numerically checked the formula (\ref{wsym_USp4_k3}) up to rank-$15$ symmetric Wilson line.  
We see that  the local operators counted by the one-point functions (\ref{wsym_USp4_k3}) 
carry the charges which are given by the generalized octagonal numbers. 
It is instructive to compare the expressions (\ref{wsym_USp4_k3}) with the one-point function for the $USp(2)$ $\cong$ $SU(2)$ pure CS theory of level $k=-2$ in \cite{Okazaki:2024paq}
\begin{align}
\label{wsym_USp2_k2}
\langle W_{(4k)}\rangle^{USp(2)_{-2}}&=q^{k(2k+1)}, \nonumber\\
\langle W_{(4k-2)}\rangle^{USp(2)_{-2}}&=-q^{k(2k-1)}, 
\end{align}
for which the local operator carries the triangular number charge, 
or equivalently, generalized hexagonal numbers. 

\subsubsection{$USp(6)_{-4}$}
\label{sec_USp6_k4}
For $n = 3$, we have the $USp(6)$ pure CS theory of level $k=-4$ 
whose vector multiplet obeys the Neumann boundary condition. 
While the half-index is trivial, 
we obtain the non-trivial one-point functions of the Wilson lines with even charges. 
For example, 
\begin{align}
\langle W_2\rangle^{USp(6)_{-4}}&=-1-q,\\
\langle W_4\rangle^{USp(6)_{-4}}&=-1-q^2,\\
\langle W_6\rangle^{USp(6)_{-4}}&=-1-q^3,\\
\langle W_8\rangle^{USp(6)_{-4}}&=q+q^2+q^3+q^5+q^6+q^7,\\
\langle W_{10}\rangle^{USp(6)_{-4}}&=-q^5-q^{10}. 
\end{align}

For the symmetric Wilson lines, we conjecture that 
\begin{align}
\label{wsym_USp6_k4}
\langle W_{(8k)}\rangle^{USp(6)_{-4}}&=q^{k(4k+3)}, \nonumber\\  
\langle W_{(8k-6)}\rangle^{USp(6)_{-4}}&=-q^{k(4k-3)}. 
\end{align}
We have numerically confirmed the conjectural results (\ref{wsym_USp6_k4}) 
up to the rank-$11$ symmetric Wilson line. 
We see that the charges carried by the local operators are identified with the generalized decagonal numbers. 

\subsubsection{General case}
\label{sec_USpCS2n_-n-1}
We conjecture that the non-trivial one-point functions of the Wilson lines in the symmetric representation are given by 
\begin{align}
\label{wsym_USp2n_-n-1}
\langle W_{(2(n+1)k)}\rangle^{USp(2n)_{-n-1}}&=q^{k((n+1)k-n)}, \nonumber\\  
\langle W_{(2(n+1)k-2n)}\rangle^{USp(2n)_{-n-1}}&=-q^{k((n+1)k+n)}. 
\end{align}
This indicates that 
the boundary local operators at the end of the symmetric Wilson line carry the the generalized $(2n+4)$-gonal number charges. 

We can define a grand canonical ensemble as
\begin{align}
    & \sum_{k \in \Zb} \langle W_{(2(n+1)k)}\rangle^{USp(2n)_{-n-1}} \Lambda^{(n+1)k}
    + \sum_{k \in \Zb} \langle W_{(2(n+1)k - 2n)}\rangle^{USp(2n)_{-n-1}} \Lambda^{(n+1)k - n}
    \nonumber \\
    & = q^{-\frac{1}{4}(n-1)} \left( \Lambda^{\frac{1}{2}(n+1)} \vartheta_2\left( (n+1)z + \tau; 2(n+1)\tau \right) - \Lambda^{-\frac{3}{2}n - \frac{1}{2}} \vartheta_2\left( (n+1)z - \tau; 2(n+1)\tau \right) \right)
\end{align}
where $q = e^{2\pi i \tau}$ and $\Lambda = e^{2 \pi i z}$. The appearance of Jacobi theta functions indicates interesting modular properties and it would be interesting to investigate the physical interpretation of this.

\subsection{$USp(2n)_{-n-3/2}$}
\label{sec_USp2n_-n-3/2}
We now consider the case with a single chiral multiplet, i.e.\ $N_f = 1$ but still no adjoint chiral.

\subsubsection{$USp(4)_{-7/2}$}
\label{sec_USp4_k-72}
For $n=2$, the CS theory has gauge group $USp(4)$ and level $k=-7/2$. 
It is coupled to a fundamental chiral with Dirichlet boundary condition so that the half-index is non-trivial, unlike for $N_f = 0$. 
The half-index is evaluated from the matrix integral
\begin{align}
\label{h_USp4_k72_int}
\mathbb{II}_{\mathcal{N}}^{USp(4)_{-7/2}}
&=\frac{(q)_{\infty}^{2}}{8} \prod_{i=1}^2 \oint \frac{ds_i}{2\pi i s_i}
\left( \prod_{i \ne j}^2 (s_i s_j^{-1}; q)_{\infty} \right) \left( \prod_{i \le j}^2 (s_i^{\pm} s_j^{\pm}; q)_{\infty} \right)
\prod_{i=1}^2
(q^{1/2} s_i^{\pm}x;q)_{\infty}. 
\end{align}
It can be expanded as
\begin{align}
1+x^2q+(x^2+x^4)q^2+(x^2+x^4)q^3+(x^2+2x^4)q^4+(x^2+2x^4+x^6)q^5+\cdots
\end{align}
which in the unflavored limit, beomes
\begin{align}
&
1+q+2q^2+2q^3+3q^4+4q^5+6q^6+7q^7+10q^8+12q^9+16q^{10}
\nonumber\\
&+19q^{11}+25q^{12}+30q^{13}+\cdots. 
\end{align}
We find that it agrees with 
\begin{align}
\label{h_USp4_k-72}
\mathbb{II}_{\mathcal{N}}^{USp(4)_{-7/2}}
&=\frac{f(-q^3,-q^4)}{f(-q)}
\nonumber\\
&=\prod_{n=1}^{\infty}\frac{1}{(1-q^{7n-1})(1-q^{7n-2})(1-q^{7n-5})(1-q^{7n-6})}. 
\end{align}

The one-point function of the Wilson line in the fundamental representation is given by
\begin{align}
\label{Wfund_USp4_k-72_int}
\langle W_{\tiny \yng(1)}\rangle^{USp(4)_{-7/2}}
&=\frac{(q)_{\infty}^{2}}{8} \prod_{i=1}^2 \oint \frac{ds_i}{2\pi i s_i}
\left( \prod_{i \ne j}^2 (s_i s_j^{-1}; q)_{\infty} \right) \left( \prod_{i \le j}^2 (s_i^{\pm} s_j^{\pm}; q)_{\infty} \right)
\prod_{i=1}^2
(q^{1/2} s_i^{\pm}x;q)_{\infty}
\nonumber\\
&\times (s_1+s_2+s_1^{-1}+s_2^{-1}). 
\end{align}
We find that in the unflavored limit it can be expanded as
\begin{align}
&-q^{\frac12}
(1+q+q^2+2q^3+3q^4+3q^5+5q^6+6q^7+8q^8+13q^{10}
\nonumber\\
&+16q^{11}+21q^{12}+25q^{13}+\cdots). 
\end{align}
We conjecture that 
\begin{align}
\label{Wfund_USp4_k-72}
\langle W_{\tiny \yng(1)}\rangle^{USp(4)_{-7/2}}
&=-q^{\frac12}\frac{f(-q^2,-q^5)}{f(-q)}
\nonumber\\
&=-q^{\frac12}\prod_{n=1}^{\infty}\frac{1}{(1-q^{7n-1})(1-q^{7n-3})(1-q^{7n-4})(1-q^{7n-6})}. 
\end{align}

Also we find that the unflavored one-point functions of symmetric Wilson lines are simply given in terms of the half-index (\ref{h_USp4_k-72}) 
and the one-point function (\ref{Wfund_USp4_k-72}) of the fundamental Wilson line 
\begin{align}
\langle W_{\tiny \yng(2)}\rangle^{USp(4)_{-7/2}}(q)&=q^{\frac12}\mathbb{II}_{\mathcal{N}}^{USp(4)_{-7/2}}(q), \\
\langle W_{\tiny \yng(3)}\rangle^{USp(4)_{-7/2}}(q)&=q^{\frac32}\langle W_{\tiny \yng(1)}\rangle^{USp(4)_{-7/2}}(q), \\
\langle W_{\tiny \yng(4)}\rangle^{USp(4)_{-7/2}}(q)&=0, \\
\langle W_{\tiny \yng(5)}\rangle^{USp(4)_{-7/2}}(q)&=0, \\
\langle W_{\tiny \yng(6)}\rangle^{USp(4)_{-7/2}}(q)&=0, \\
\langle W_{\tiny \yng(7)}\rangle^{USp(4)_{-7/2}}(q)&=-q^{\frac{11}{2}}\mathbb{II}_{\mathcal{N}}^{USp(4)_{-7/2}}(q), \\
\langle W_{\tiny \yng(8)}\rangle^{USp(4)_{-7/2}}(q)&=-q^{\frac{13}{2}}\langle W_{\tiny \yng(1)}\rangle^{USp(4)_{-7/2}}(q), \\
\langle W_{\tiny \yng(9)}\rangle^{USp(4)_{-7/2}}(q)&=-q^{8}\langle W_{\tiny \yng(1)}\rangle^{USp(4)_{-7/2}}(q), \\
\langle W_{\tiny \yng(10)}\rangle^{USp(4)_{-7/2}}(q)&=-q^{10}\mathbb{II}_{\mathcal{N}}^{USp(4)_{-7/2}}(q). 
\end{align}

We remark that 
the unflavored half-index (\ref{h_USp4_k-72}) and the one-point function (\ref{Wfund_USp4_k-72}) of the fundamental Wilson line 
resemble the Rogers-Ramanujan functions which appear in the $SU(2)\cong USp(2)$ CS theory with level $k=-5/2$ \cite{Okazaki:2024paq} . 
In fact, they are identified with the functions associated with the Rogers-Selberg identities \cite{MR1577136} which generalize the Rogers-Ramanujan identities! 
We have
\begin{align}
\mathbb{II}_{\mathcal{N}}^{USp(4)_{-7/2}}(q)
&=\frac{(q^2;q^2)_{\infty}}{(q;q)_{\infty}}
\sum_{n=0}^{\infty}\frac{q^{2n^2}}{(q^2;q^2)_n (-q;q)_{2n}}, \\
\langle W_{\tiny \yng(1)}\rangle^{USp(4)_{-7/2}}
&=\frac{(q^2;q^2)_{\infty}}{(q;q)_{\infty}}
\sum_{n=0}^{\infty}\frac{q^{2n^2+2n}}{(q^2;q^2)_n (-q;q)_{2n}}. 
\end{align}

\subsubsection{$USp(6)_{-9/2}$}
\label{sec_USp6_k92}
For $n=3$, 
we have $USp(6)$ CS theory with level $k=-9/2$ 
and a fundamental chiral with Dirichlet boundary condition. 

We find that the unflavored half-index agrees with the following $q$-series: 
\begin{align}
\label{h_USp6_k92}
\mathbb{II}_{\mathcal{N}}^{USp(6)_{-9/2}}
&=\frac{f(-q^4,-q^5)}{f(-q)}
\nonumber\\
&=\prod_{n=1}^{\infty}
\frac{1}{(1-q^{9n-1})(1-q^{9n-2})(1-q^{9n-3})(1-q^{9n-6})(1-q^{9n-7})(1-q^{9n-8})}
\nonumber\\
&=1+q+2q^2+3q^3+4q^4+5q^5+8q^6+10q^7+\cdots. 
\end{align}

We find that the unflavored one-point function coincides with 
\begin{align}
\label{Wfund_USp6_k92}
\langle W_{\tiny \yng(1)}\rangle^{USp(6)_{-9/2}}
&=-q^{1/2} \frac{f(-q^3,-q^6)}{f(-q)}
\nonumber\\
&=
-q^{\frac12}\prod_{n=1}^{\infty}
\frac{1}{(1-q^{9n-1})(1-q^{9n-2})(1-q^{9n-4})(1-q^{9n-5})(1-q^{9n-7})(1-q^{9n-8})}
\nonumber\\
&=-q^{\frac12}
(1+q+2q^2+2q^3+4q^4+5q^5+7q^6+9q^7+\cdots). 
\end{align}

The unflavored one-point functions of symmetric Wilson lines are expressible 
in terms of the half-index (\ref{h_USp6_k92}) and the one-point function (\ref{Wfund_USp6_k92}) of the fundamental Wilson line. 
For example, 
\begin{align}
\langle W_{\tiny \yng(2)}\rangle^{USp(6)_{-9/2}}(q)&=q^{\frac12}\langle W_{\tiny \yng(1)}\rangle^{USp(6)_{-9/2}}(q), \\
\langle W_{\tiny \yng(3)}\rangle^{USp(6)_{-9/2}}(q)&=q^{\frac32}\mathbb{II}_{\mathcal{N}}^{USp(6)_{-9/2}}(q), \\
\langle W_{\tiny \yng(4)}\rangle^{USp(6)_{-9/2}}(q)&=0, \\
\langle W_{\tiny \yng(5)}\rangle^{USp(6)_{-9/2}}(q)&=0, \\
\langle W_{\tiny \yng(6)}\rangle^{USp(6)_{-9/2}}(q)&=0, \\
\langle W_{\tiny \yng(7)}\rangle^{USp(6)_{-9/2}}(q)&=0, \\
\langle W_{\tiny \yng(8)}\rangle^{USp(6)_{-9/2}}(q)&=0. 
\end{align}

The one-point function $\langle W_{\tiny \yng(9)}\rangle^{USp(6)_{-9/2}}(q)$ is non-zero and we conjecture that it and higher symmetric representation Wilson lines can be written as a product of $\mathbb{II}_{\mathcal{N}}^{USp(6)_{-9/2}}(q)$ or $\langle W_{\tiny \yng(1)}\rangle^{USp(6)_{-9/2}}(q)$ with a monomial in $q^{\frac{1}{2}}$.

We note that the half-index (\ref{h_USp6_k92}) can be rewritten as the infinite series 
according to Bailey's modulus $9$ analog of the Rogers-Ramanujan identities \cite{MR22816,MR49225}
\begin{align}
\mathbb{II}_{\mathcal{N}}^{USp(6)_{-9/2}}
&=\frac{(q^3;q^3)_{\infty}}{(q;q)_{\infty}}
\sum_{n=0}^{\infty}
\frac{(q;q)_{3n}}{(q^3;q^3)_{n}(q^3;q^3)_{2n}}q^{3n^2}. 
\end{align}

\subsubsection{General case}
\label{sec_CSUSp2n_-n-3/2}
We observe that 
the unflavored Neumann half-index and the one-point functions of the Wilson line operators 
for $USp(2n)_{-n-3/2}$ CS theory with a fundamental chiral multiplet 
are expressible in terms of Ramanujan's general theta function. 
We conjecture that
\begin{align}
\label{h_USp2n_-n-3/2}
\mathbb{II}_{\mathcal{N}}^{USp(2n)_{-n-3/2}}
&=\frac{f(-q^{n+1},-q^{n+2})}{f(-q)}
\end{align}
and 
\begin{align}
\label{Wfund_USp2n_-n-3/2}
\langle W_{\tiny \yng(1)}\rangle^{USp(2n)_{-n-3/2}}
&=-q^{\frac12} \frac{f(-q^{n},-q^{n+3})}{f(-q)}. 
\end{align}
For $n=1$ the half-index (\ref{h_USp2n_-n-3/2}) and the one-point function (\ref{Wfund_USp2n_-n-3/2}) 
reproduces the Rogers-Ramanujan functions  as proved in \cite{Okazaki:2024paq}. 
It would be interesting to investigate further the modular properties of these half-indices and one-point functions for general $n$.

\subsection{$USp(2n)_{-2n-2}$ with adjoint}
\label{sec_USp2n_adj_-2n-2}
Let us now consider symplectic gauge theories with an adjoint chiral and in this section we set $N_f = 0$. The case of $n=1$ corresponds to $SU(2)$ with an adjoint chiral discussed in section~\ref{sec_CS_SU2_adj_-4}.

\subsubsection{$USp(4)_{-6}$ with adjoint}
\label{sec_USp4_adj_k-6}
With $n = 2$ we have the $USp(4)$ CS theory with level $k=-6$. 
The half-index reads 
\begin{align}
\label{h_USp4adj_k6}
\mathbb{II}_{\mathcal{N}, D}^{USp(4)_{-6}}
&=\frac{(q)_{\infty}^{4}}{8} \prod_{i=1}^2 \oint \frac{ds_i}{2\pi i s_i}
\left( \prod_{i \ne j}^2 (s_i s_j^{-1}; q)_{\infty} (q s_i s_j^{-1}; q)_{\infty} \right) \left( \prod_{i \le j}^2 (s_i^{\pm} s_j^{\pm}; q)_{\infty} (q s_i^{\pm} s_j^{\pm}; q)_{\infty} \right)
\end{align}
which, as for gauge group $SU(N)$ with an adjoint chiral, has weight given by half the rank of the gauge group.

The unflavored half-index agrees with the following eta-product: 
\begin{align}
\mathbb{II}_{\mathcal{N}, D}^{USp(4)_{-6}}
&=q^{-\frac{5}{6}}\frac{\eta(4\tau)\eta(6\tau)\eta(12\tau)}{\eta(2\tau)}
\nonumber\\
&=\prod_{n=1}^{\infty}
\frac{(1-q^{4n}) (1-q^{6n}) (1-q^{12n})}{(1-q^{2n})}. 
\end{align}

The one-point functions of the Wilson lines with even charges are non-trivial 
while those of the Wilson lines with odd charges vanish. 
We find that the unflavored one-point functions of the Wilson lines with charge $2$, $4$ and $6$ 
can be expressed as the eta-products
\begin{align}
\label{Wch2_USp4adj_k6}
\langle W_{2}\rangle^{USp(4)_{-6}}
&=-q^{-\frac23}
\frac{\eta(2\tau)^4\eta(3\tau)^2\eta(12\tau)^2}{\eta(\tau)^2\eta(4\tau)^2\eta(6\tau)^2}
\nonumber\\
&=-\prod_{n=1}^{\infty}
\frac{(1-q^{2n})^4 (1-q^{3n})^2 (1-q^{12n})^2}
{(1-q^n)^2 (1-q^{4n})^2 (1-q^{6n})^2}, 
\end{align}
\begin{align}
\label{Wch4_USp4adj_k6}
\langle W_{4}\rangle^{USp(4)_{-6}}
&=-q^{-\frac{1}{6}}
\frac{\eta(4\tau)^2\eta(6\tau)^4}{\eta(2\tau)^2\eta(12\tau)^2}
\nonumber\\
&=-\prod_{n=1}^{\infty}
\frac{(1-q^{4n})^2 (1-q^{6n})^4}
{(1-q^{2n})^2(1-q^{12n})^2}, 
\end{align}
\begin{align}
\label{Wch6_USp4adj_k6}
\langle W_{6}\rangle^{USp(4)_{-6}}
&=2q^{\frac{2}{3}}
\frac{\eta(2\tau)^2\eta(6\tau)^4}{\eta(\tau)\eta(3\tau)\eta(4\tau)\eta(12\tau)}
\nonumber\\
&=2q\prod_{n=1}^{\infty}
\frac{(1-q^{2n})^2 (1-q^{6n})^4}
{(1-q^{n}) (1-q^{3n}) (1-q^{4n}) (1-q^{12n})}. 
\end{align}

For the Wilson lines with larger charges 
they are expressible in terms of the 
half-index (\ref{h_USp4adj_k6}) and one-point functions (\ref{Wch2_USp4adj_k6}), (\ref{Wch4_USp4adj_k6}) (\ref{Wch6_USp4adj_k6}) for lower charges
\begin{align}
\langle W_{8}\rangle^{USp(4)_{-6}}&=q^2 \langle W_{4}\rangle^{USp(2n)_{-6}}, \\
\langle W_{10}\rangle^{USp(4)_{-6}}&=q^4 \langle W_{2}\rangle^{USp(2n)_{-6}}, \\
\langle W_{12}\rangle^{USp(4)_{-6}}&=4q^6 \mathbb{II}_{\mathcal{N}, D}^{USp(4)_{-6}}, \\
\langle W_{14}\rangle^{USp(4)_{-6}}&=q^8 \langle W_{2}\rangle^{USp(2n)_{-6}}, \\
\langle W_{16}\rangle^{USp(4)_{-6}}&=q^{10} \langle W_{4}\rangle^{USp(2n)_{-6}}, \\
\langle W_{18}\rangle^{USp(4)_{-6}}&=q^{12} \langle W_{6}\rangle^{USp(2n)_{-6}}. 
\end{align}

\subsection{$USp(2n)_{-2n-2-N_f/2}$ with adjoint}
\label{sec_USp2n_adj_-2n-2-Nf/2-Na/2}
Next consider the $USp(2n)$ CS theory with both an adjoint and $N_f$ fundamental chirals. 

\subsubsection{$USp(4)_{-13/2}$ with adjoint}
\label{sec_USp4_adj_k-13/2}
For $n=2$ and $N_f=1$, $USp(4)$ CS theory has level $k=-13/2$. 
We find that 
the unflavored half-index can be expanded as
\begin{align}
\mathbb{II}_{\mathcal{N}, D}^{USp(4)_{-13/2}}
&=1+q+3q^2+3q^3+4q^4+6q^5+10q^6+10q^7
\nonumber\\
&+16q^8+19q^9+26q^{10}+31q^{11}+41q^{12}+\cdots
\end{align}
This agrees with 
\begin{align}
q^{\frac13} \left(
5+\frac{\eta(\tau)^2}{\eta(13\tau)^2}
+13\frac{\eta(13\tau)^2}{\eta(\tau)^2}
\right)^{\frac13}, 
\end{align} 
which is the Mckay-Thompson series 
of class $39B$ for Monster group \cite{MR1400423,MR1291027} up to the overall $q$ factor. 

\subsubsection{$USp(6)_{-17/2}$ with adjoint}
\label{sec_USp6_adj_k-17/2}
With $n=3$ and $N_f=1$ we have the $USp(6)$ CS theory with level $k=-17/2$. 
We find that the unflavored half-index has an expansion
\begin{align}
1+q+3q^2+4q^3+6q^4+7q^5+13q^6+16q^7+\cdots. 
\end{align}
This coincides with 
\begin{align}
\frac{\psi(q^2)\varphi(q^{17})-q^4\varphi(q)\psi(q^{34})}{f(-q)f(-q^{17})}, 
\end{align}
which is the Mckay-Thompson series of class $34a$ for Monster group \cite{MR1400423,MR1291027} up to the overall $q$ factor.

\section{$SO(N)$ CS theories}
\label{sec_CS_SO}
In this section we investigate $SO(N)_k$ gauge theories with $N_f$ fundamental chirals, $Q_I$, 
where the vector multiplet has Neumann boundary condition 
and the fundamental chirals have Dirichlet boundary conditions.
We may also consider an adjoint chiral $\Phi$ with R-charge $0$ and Dirichlet boundary condition.
We include discrete fugacities $\zeta$ for the $\Zb_2^{\Mcal}$ magnetic symmetry 
and $\chi$ for $\Zb_2^{\Ccal}$ charge conjugation symmetry, although for the theories we consider there is no dependence on the magnetic fugacity $\zeta$.

We summarize the field content and charges in the following table
\begin{align}
\label{SO_Adj_Nf_charges}
\begin{array}{c|c|c|c|c|c|c}
& \textrm{bc} & SO(N) & SU(N_f) & U(1)_A & U(1)_B & U(1)_R \\ \hline
\textrm{VM} & \mathcal{N} & {\bf Adj} & {\bf 1} & 0 & 0 & 0 \\
\Phi & \textrm{D} & {\bf Adj} & {\bf 1} & 0 & 1 & 0 \\
Q_I & \textrm{D} & {\bf N} & {\bf N_f} & 1 & 0 & 1
\end{array}
\end{align}

We can easily calculate the gauge and 't Hooft anomalies \cite{Okazaki:2021pnc} as follows,
\begin{align}
\label{bdy_SO_adj_Nf_anom}
\Acal & = \underbrace{{(N-2)\Tr(s^2)} + \frac{N(N-1)}{4}r^2}_{\textrm{VM}, \; \Ncal} + \underbrace{{(N-2)\Tr(s^2)} + \frac{N(N-1)}{4} (b-r)^2}_{\Phi, \; D}
\nonumber \\
 & + \underbrace{\left( N_f \Tr(s^2) + \frac{N}{2} \Tr(x^2) + \frac{N N_f}{2} a^2 \right)}_{Q_I, \; D}
  \nonumber \\
  = & \left( 2N - 4 + N_f \right) \Tr(s^2) + \frac{N}{2} \Tr(x^2) + \frac{NN_f}{2} a^2 +
  \nonumber \\
  & + \frac{N(N-1)}{4}b^2 - \frac{N(N-1)}{2} br + \frac{N(N-1)}{2} r^2
\end{align}
in the case with an adjoint chiral, while without the adjoint chiral the result is easily seen to be
\begin{align}
\label{bdy_SO_Nf_anom}
\Acal & = \left( N - 2 + N_f \right) \Tr(s^2) + \frac{N}{2} \Tr(x^2) + \frac{N N_f}{2} a^2  + \frac{N(N-1)}{4} r^2
 \; .
\end{align}

To cancel the gauge anomaly we need to take CS level 
\begin{align}
k & = -2N + 4 - N_f
\end{align}
in the case with an adjoint chiral and
\begin{align}
k & = -N + 2 - N_f
\end{align}
without an adjoint chiral.

Similar to the previous case with unitary or symplectic gauge groups, the only distinction between two fundamental chirals with Dirichlet boundary conditions and a 2d fundamental Fermi multiplet is in the details of the global symmetries. Therefore we cannot distinguish between these cases if we set the flavor fugacities to $1$.

We can write the half-index without an adjoint chiral as \cite{Gadde:2013sca, Yoshida:2014ssa, Dimofte:2017tpi, Okazaki:2021pnc}
\begin{align}
\label{h_SON_k_zeta_chi}
\mathbb{II}_{\mathcal{N}}^{SO(N)_{k, \zeta \chi}}
&=
\frac{(q)_{\infty}^{n}}{n! 2^{n - 1 + \epsilon}} \prod_{i=1}^n \oint \frac{ds_i}{2\pi i s_i}
(\chi s_i^{\pm}; q)_{\infty}^{\epsilon}
\prod_{i < j}^n (s_i^{\pm} s_j^{\mp}; q)_{\infty} (s_i^{\pm} s_j^{\pm}; q)_{\infty}
 \nonumber \\
& \times
\prod_{\alpha = 1}^{N_f} \left( (\chi q^{\frac12} x_\alpha; q)_{\infty}^{\epsilon} \prod_{i = 1}^n (q^{\frac12} s_i^{\pm} x_\alpha; q)_{\infty} \right)
\end{align}
where $N = 2n + \epsilon$ with $n \in \Zb$ and $\epsilon \in \{0, 1\}$, except in the case where $\chi = -1$ and $\epsilon = 0$ where we have
\begin{align}
\label{h_SO2n_k_zeta_m}
\mathbb{II}_{\mathcal{N}}^{SO(2n)_{k, \zeta -}}
&=
\frac{(q)_{\infty}^{n-1} (-q; q)_{\infty}}{(n-1)! 2^{n - 1}} \prod_{i=1}^{n-1} \oint \frac{ds_i}{2\pi i s_i}
(s_i^{\pm}; q)_{\infty} (-s_i^{\pm}; q)_{\infty}
\prod_{i < j}^{n-1} (s_i^{\pm} s_j^{\mp}; q)_{\infty} (s_i^{\pm} s_j^{\pm}; q)_{\infty}
 \nonumber \\
& \times
\prod_{\alpha = 1}^{N_f} \left( (\pm q^{\frac12} x_\alpha; q)_{\infty} \prod_{i = 1}^{n-1} (q^{\frac12} s_i^{\pm} x_\alpha; q)_{\infty} \right) \; .
\end{align}

Similarly, with an adjoint chiral we have
\begin{align}
\label{h_SON_k_zeta_chi_adj}
\mathbb{II}_{\mathcal{N},D}^{SO(N)_{k, \zeta \chi}}
&=
\frac{(q)_{\infty}^{2n}}{n! 2^{n - 1 + \epsilon}} \prod_{i=1}^n \oint \frac{ds_i}{2\pi i s_i}
(\chi s_i^{\pm}; q)_{\infty}^{\epsilon} (\chi q s_i^{\pm}; q)_{\infty}^{\epsilon}
\nonumber\\
&\times 
\prod_{i < j}^n (s_i^{\pm} s_j^{\mp}; q)_{\infty} (s_i^{\pm} s_j^{\pm}; q)_{\infty} (q s_i^{\pm} s_j^{\mp}; q)_{\infty} (q s_i^{\pm} s_j^{\pm}; q)_{\infty}
 \nonumber \\
& \times
\prod_{\alpha = 1}^{N_f} \left( (\chi q^{\frac12} x_\alpha; q)_{\infty}^{\epsilon} \prod_{i = 1}^n (q^{\frac12} s_i^{\pm} x_\alpha; q)_{\infty} \right), 
\end{align}
except in the case where $\chi = -1$ and $\epsilon = 0$ where we have
\begin{align}
\label{h_SO2n_k_zeta_m_adj}
\mathbb{II}_{\mathcal{N},D}^{SO(2n)_{k, \zeta -}}
&=
\frac{(q)_{\infty}^{2n-2} (-q; q)_{\infty}^2}{(n-1)! 2^{n - 1}} \prod_{i=1}^{n-1} \oint \frac{ds_i}{2\pi i s_i}
(s_i^{\pm}; q)_{\infty} (-s_i^{\pm}; q)_{\infty}
(q s_i^{\pm}; q)_{\infty} (-q s_i^{\pm}; q)_{\infty}
 \nonumber \\
& \times
\prod_{i < j}^{n-1} (s_i^{\pm} s_j^{\mp}; q)_{\infty} (s_i^{\pm} s_j^{\pm}; q)_{\infty}
(q s_i^{\pm} s_j^{\mp}; q)_{\infty} (q s_i^{\pm} s_j^{\pm}; q)_{\infty}
 \nonumber \\
& \times
\prod_{\alpha = 1}^{N_f} \left( (\pm q^{\frac12} x_\alpha; q)_{\infty} \prod_{i = 1}^{n-1} (q^{\frac12} s_i^{\pm} x_\alpha; q)_{\infty} \right) \; .
\end{align}

We can compute the correlators of the BPS Wilson line operators in the representation $\mathcal{R}$ 
by inserting the characters in the matrix integral. 
In the case of $SO(2n+1)$ and $SO(2n)$ gauge groups with discrete fugacity $\chi = +$, 
the characters of the irreducible representation with highest weight 
labeled by the Young diagram $\lambda$ are given by \cite{MR1153249}
\begin{align}
\label{so2n+1_character}
\chi^{\mathfrak{so}(2n+1)}_{\lambda}(s)
&=\frac{\det (s_j^{\lambda_i+n-i+1/2}-s_j^{-(\lambda_i+n-i+1/2)})}
{\det(s_j^{n-i+1/2}-s_j^{-(n-i+1/2)})}, \\
\label{so2n_character}
 \chi^{\mathfrak{so}(2n)}_{\lambda}(s)
&=\frac{\det (s_j^{\lambda_i+n-i}+s_j^{-(\lambda_i+n-i)}) + \det (s_j^{\lambda_i+n-i} - s_j^{-(\lambda_i+n-i)})}
{\det(s_j^{n-i}+s_j^{-(n-i)})},
\end{align}
For example, the characters of the fundamental representation are 
\begin{align}
\chi^{\mathfrak{so}(2n+1)}_{\tiny \yng(1)}(s)
&=1+\sum_{i=1}^{n}(s_i+s_i^{-1}), \\
\chi^{\mathfrak{so}(2n)}_{\tiny \yng(1)}(s)
&=\sum_{i=1}^{n}(s_i+s_i^{-1}). 
\end{align}
Similarly, we also introduce another basis for the Wilson lines $W_{k}$ labeled by 
$1+p_k(s)+p_k(s^{-1})$ and $p_k(s)+p_k(s^{-1})$ for $SO(2n+1)$ and $SO(2n)$ respectively, 
which we refer to as the charged Wilson lines.

For $SO(2n)$ with $\chi=-$ we instead employ the character (\ref{usp2n_character}) of the $USp(2n-2)$.
In the case of gauge group $SO(2n+1)$ with $\chi = -$ we can use the character for the case with $\chi = +$ up to a change of sign for some terms. In particular, for an even (odd) rank representation each term in the character will be a product of an even (odd) number of fugacities $\{ s_i^{\pm}, \chi \}$. Therefore, terms in the character with a product of an odd (even) number of fugacities $\{ s_i^{\pm} \}$ must have a factor of $\chi$, and hence will change sign in the case of $\chi = -$ compared to he case of $\chi = +$.

\subsection{$SO(2n)_{-2n+2}$}
\label{sec_SO2n_-2n+2}
Let us begin with the theories of even rank orthogonal gauge groups, i.e.\ with $\epsilon=0$, and with no adjoint or fundamental chirals. 

\subsubsection{$SO(4)_{-2}$}
\label{sec_SO4_k-2}
For $n=2$ we have the $SO(4)$ pure CS theory with level $k=-2$ obeying the Neumann boundary condition. 
The half-index is trivial
\begin{align}
\label{h_SO4_k2}
\mathbb{II}_{\mathcal{N}}^{SO(4)_{-2,\zeta +}}
&=\frac{(q)_{\infty}^2}{4}
\prod_{i=1}^2 \oint \frac{ds_i}{2\pi is_i}
\prod_{i<j}^2 
(s_i^{\pm}s_{j}^{\mp};q)_{\infty}
(s_i^{\pm}s_{j}^{\pm};q)_{\infty}
=1. 
\end{align}

The half-index is decorated by the insertion of the Wilson lines. 
The one-point function of the charge-$n$ Wilson line is evaluated as
\begin{align}
\label{W_SO4_k2}
\langle W_n\rangle^{SO(4)_{-2,\zeta +}}
&=\frac{(q)_{\infty}^2}{4}
\prod_{i=1}^2 \oint \frac{ds_i}{2\pi is_i}
\prod_{i<j}^2 
(s_i^{\pm}s_{j}^{\mp};q)_{\infty}
(s_i^{\pm}s_{j}^{\pm};q)_{\infty} (s_1^n+s_2^n+s_1^{-n}+s_2^{-n})
\end{align}
It is again straightforward to calculate this using the Jacobi triple product formula.
The non-trivial one-point function appears for even $n$. 
We find 
\begin{align}
\langle W_{2k}\rangle^{SO(4)_{-2,\zeta +}}&=
q^{k(k-1)}+2q^{k^2}+q^{k(k+1)}. 
\end{align}

Indeed we can derive the general result that if $\alpha + \beta$ is odd then
$\langle s_1^{\alpha} s_2^{\beta} \rangle^{SO(4)_{-2}} = 0$ while if $\alpha + \beta$ is even then
\begin{align}
    \langle s_1^{\alpha} s_2^{\beta} \rangle^{SO(4)_{-2}} & =
    \frac{(-1)^{\beta}}{4} q^{\frac{1}{4}(\alpha^2 + \beta^2)} \left( q^{\frac{1}{2}\alpha} + q^{-\frac{1}{2}\alpha} + q^{\frac{1}{2}\beta} + q^{-\frac{1}{2}\beta} \right). 
\end{align}

We can also derive the one-point functions of Wilson lines in the (anti)symmetric representations. 
We have 
\begin{align}
\langle W_{(2k)} \rangle^{SO(4)_{-2,\zeta +}}&=q^{k(k+1)}, \\
\langle W_{\tiny \yng(1,1)} \rangle^{SO(4)_{-2, \zeta +}}&=-q. 
\end{align}
while those for the odd rank symmetric representations vanish. 

For $\chi=-$ there is no non-trivial antisymmetric Wilson line. 
The one-point functions of the symmetric Wilson lines involves the sign factor with a single term
\begin{align}
\langle W_{(2k)} \rangle^{SO(4)_{-2,\zeta -}}&=(-1)^kq^{k(k+1)}. 
\end{align} 

\subsubsection{$SO(6)_{-4}$}
\label{sec_SO6_k-4}
When $n=3$ the theory is the $SO(6)$ pure CS theory with level $k=-4$. 
Again the half-index is trivial as there is no boundary BPS local operator. 
One finds non-trivial boundary BPS local operators in the presence of the Wilson line. 
We find that the one-point functions of the Wilson lines with even charges are non-trivial. 
For example, 
\begin{align}
\langle W_{2}\rangle^{SO(6)_{-4,\zeta +}}&=1+q,\\
\langle W_{4}\rangle^{SO(6)_{-4,\zeta +}}&=1+q+2q^2+q^3+q^4,\\
\langle W_{6}\rangle^{SO(6)_{-4,\zeta +}}&=q^3+q^6,\\
\langle W_{8}\rangle^{SO(6)_{-4,\zeta +}}&=q^4+q^6+2q^8+q^{10}+q^{12}, \\
\langle W_{10}\rangle^{SO(6)_{-4,\zeta +}}&=q^{10}+q^{15}
\end{align}

We find that the one-point functions of the Wilson lines in the rank-$4k$ symmetric representations are non-trivial 
while those in the other symmetric representations vanish. 
We find that 
\begin{align}
\langle W_{(4k)}\rangle^{SO(6)_{-4,\zeta +}}&=q^{2k(k+1)}, \\
\langle W_{(4k)}\rangle^{SO(6)_{-4,\zeta -}}&=(-1)^kq^{2k(k+1)}. 
\end{align}
Also we have the non-vanishing one-point function of the rank-2 antisymmetric Wilson line
\begin{align}
\langle W_{\tiny \yng(1,1)}\rangle^{SO(6)_{-4, \zeta \chi}}&=-q. 
\end{align}
which is independent of $\chi$. 

\subsubsection{General case}
\label{sec_SO2n_-2n+2}
For general the $SO(2n)$ pure CS theory with level $k=-2n+2$ obeying the Neumann boundary condition, the half-index is trivial. 
On the other hand, the one-point functions of the Wilson lines in the rank-$(2(n-1)k)$ symmetric representations where $k=1,2,\cdots$ are non-trivial. 
We conjecture that 
\begin{align}
\langle W_{(2(n-1)k)}\rangle^{SO(2n)_{-2n+2, \zeta +}}&=q^{(n-1)k(k+1)}. 
\end{align}

We can define a grand canonical ensemble as
\begin{align}
    \sum_{k \in \Zb} \langle W_{(2(n-1)k)}\rangle^{SO(2n)_{-2n+2, \zeta +}} \Lambda^{k}
    & = q^{-\frac{1}{4}(n-1)} \Lambda^{-\frac{1}{2}} \vartheta_2\left( z; 2(n-1)\tau \right)
\end{align}
where $q = e^{2\pi i \tau}$ and $\Lambda = e^{2 \pi i z}$, where agin we note the appearance of a Jacobi theta function indicating interesting modular transformation properties.

With $\chi=-$ we conjecture that 
\begin{align}
\langle W_{(2(n-1)k)}\rangle^{SO(2n)_{-2n+2, \zeta +}}&=(-1)^kq^{(n-1)k(k+1)}. 
\end{align}

\subsection{$SO(2n)_{-2n+1}$}
\label{sec_SO2n_-2n+1}
We now include a single fundamental chiral multiplet, i.e.\ $N_f = 1$.

\subsubsection{$SO(2)_{-1}$}
\label{sec_SO2_k-1}
When the CS theory has a chiral multiplet, the half-index is non-trivial. 
With $n=1$, we have the $SO(2)$ CS theory with level $k=-1$ and a single chiral multiplet. 
The half-index is evaluated as
\begin{align}
\mathbb{II}_{\mathcal{N}}^{SO(2)_{-1,\zeta +}}(x;q)
&=(q)_{\infty} \oint \frac{ds}{2\pi is} 
(q^{1/2}s^{\pm}x;q)_{\infty}. 
\end{align}
It can be expanded as
\begin{align}
1+(x^2-1)q+(x^2-1)q^2+(x^2-1)q^4+(x^2-1)^2q^5+2x^2(x^2-1)q^6+\cdots. 
\end{align}
Turning off the flavor fugacity by setting $x$ to $1$, it becomes trivial
\begin{align}
\mathbb{II}_{\mathcal{N}}^{SO(2)_{-1,\zeta +}}(q)=1. 
\end{align}

We find that the unflavored one-point functions are given by
\begin{align}
\langle W_k\rangle^{SO(2)_{-1,\zeta +}}(q)&=(-1)^k q^{\frac{k^2}{2}}. 
\end{align}

These results are easily derived using the Jacobi triple product formula \eqref{JacobiTripleProduct}.

\subsubsection{$SO(4)_{-3}$}
\label{sec_SO4_k-3}
For $n=2$ one finds the $SO(4)$ CS theory with level $k=-3$. 
The half-index is given by
\begin{align}
\label{h_SO4_k-3}
\mathbb{II}_{\mathcal{N}}^{SO(4)_{-3,\zeta +}}(x;q)
&=\frac{(q)_{\infty}^2}{4}
\left( \prod_{i=1}^2 \oint \frac{ds_i}{2\pi is_i} \right)
\left(~ \prod_{i<j}^2 
(s_i^{\pm}s_{j}^{\mp};q)_{\infty}
(s_i^{\pm}s_{j}^{\pm};q)_{\infty} \right)
\prod_{i=1}^2 (q^{\frac12}s_i^{\pm}x;q)_{\infty} \; .
\end{align}
It can be expanded as
\begin{align}
&
1+(x^4-x^2)q^2+(x^4-x^2)q^3+(x^4-x^2)q^4+(x^4-x^2)q^5
\nonumber\\
&+(x^4-x^2)q^6+(-x^2+2x^4-x^6)q^7
+(-x^2+2x^4-2x^6+x^8)q^8+\cdots
\end{align}
In the unflavored limit, it becomes trivial 
\begin{align}
\mathbb{II}_{\mathcal{N}}^{SO(4)_{-3}}(q)=1. 
\end{align}
Hence there are exact cancellations between boundary bosonic and fermionic BPS local operators. 

The one-point function is given by
\begin{align}
\label{W_SO4_k-3}
&
\langle W_n\rangle^{SO(4)_{-3,\zeta +}}(x;q)
\nonumber\\
&=\frac{(q)_{\infty}^2}{4}
\prod_{i=1}^2 \oint \frac{ds_i}{2\pi is_i}
\prod_{i<j}^2 
(s_i^{\pm}s_{j}^{\mp};q)_{\infty}
(s_i^{\pm}s_{j}^{\pm};q)_{\infty}
(q^{\frac12}s_i^{\pm}x;q)_{\infty}
(s_1^n+s_2^n+s_1^{-n}+s_2^{-n}). 
\end{align}

Turning off the flavor fugacities, the one-point function gets simplified. 
We obtain
\begin{align}
\langle W_1\rangle^{SO(4)_{-3,\zeta +}}(q)&=-q^{1/2}, \\
\langle W_2\rangle^{SO(4)_{-3,\zeta +}}(q)&=1, \\
\langle W_3\rangle^{SO(4)_{-3,\zeta +}}(q)&=-q^{\frac12}-2q^{3/2}-q^{5/2}, \\
\langle W_4\rangle^{SO(4)_{-3,\zeta +}}(q)&= q^4, \\
\langle W_5\rangle^{SO(4)_{-3,\zeta +}}(q)&=-q^{5/2}, \\
\langle W_6\rangle^{SO(4)_{-3,\zeta +}}(q)&=q^{4}+2q^{6}+q^8, \\
\langle W_7\rangle^{SO(4)_{-3,\zeta +}}(q)&=-q^{21/2}, \\
\langle W_8\rangle^{SO(4)_{-3,\zeta +}}(q)&=q^{8}, \\
\langle W_9\rangle^{SO(4)_{-3,\zeta +}}(q)&=-q^{21/2}-2q^{27/2}-q^{33/2}. 
\end{align}

Again, we can calculate these results exactly using the Jacobi triple product formula to find
\begin{align}
    \langle s_1^{\alpha} s_2^{\beta} \rangle^{SO(4)_{-3,\zeta +}} & =
    \frac{(-1)^{\beta}}{4(q)_{\infty}^2} q^{\alpha^2 + \alpha\beta + \frac{1}{2}\beta^2} \left( q^{\frac{1}{2}\alpha}h(4\alpha + 2\beta + 1) + q^{-\frac{1}{2}\alpha}h(4\alpha + 2\beta - 1) \right)
    \nonumber \\
    & \times \left( h(-2\alpha - 2\beta + 1) + h(-2\alpha - 2\beta -1) \right)
\end{align}
where we have defined
\begin{align}
    h(\lambda) & = (q^{\frac{3 \pm \lambda}{2}}; q^3)_{\infty} (q^3; q^3)_{\infty} = h(-\lambda)
\end{align}
and in particular
\begin{align}
    h(2l + 1) & = (q^{1 - l}; q^3)_{\infty} (q^{2 + l}; q^3)_{\infty} (q^3; q^3)_{\infty} \; .
\end{align}

It is straightforward to check that
\begin{align}
    h(2l + 1) & = \hat{h}(l) q^{-\frac{l(l+1)}{6}} (q)_{\infty}
\end{align}
where
\begin{align}
    \hat{h}(l) & = \left\{ \begin{array}{lcr}
    (-1)^{\frac{l+1}{3}} & , & l \equiv -1 \mod 3 \\
    (-1)^{\frac{l}{3}} & , & l \equiv 0 \mod 3 \\
    0 & , & l \equiv 1 \mod 3 \\
    \end{array} \right.
\end{align}
so
\begin{align}
    \langle s_1^{\alpha} s_2^{\beta} \rangle^{SO(4)_{-3,\zeta +}} & =
    \frac{1}{4} q^{\frac{1}{6}\alpha^2 + \frac{1}{6}\beta^2} \left( (-1)^{\beta} \hat{h}(2\alpha + \beta) q^{\frac{1}{6}(\alpha - \beta)} + (-1)^{\alpha} \hat{h}(\alpha + 2\beta) q^{\frac{1}{6}(-\alpha + \beta)} \right)
    \nonumber \\
    & \times \left( \hat{h}(\alpha + \beta) q^{\frac{1}{6}(-\alpha - \beta)} + \hat{h}(-\alpha - \beta) q^{\frac{1}{6}(\alpha + \beta)} \right)
\end{align}
which gives
\begin{align}
    \langle W_{\alpha} \rangle^{SO(4)_{-3,\zeta +}}(q) & = q^{\frac{\alpha^2}{6}} \left( \hat{h}(2\alpha) q^{\frac{\alpha}{6}} + (-1)^{\alpha} \hat{h}(\alpha) q^{-\frac{\alpha}{6}} \right) \left( \hat{h}(-\alpha) q^{\frac{\alpha}{6}} + \hat{h}(\alpha) q^{-\frac{\alpha}{6}} \right)
    \nonumber \\
    & = 
    (-1)^{\alpha} q^{\frac{\alpha^2}{6}} \left( \hat{h}(-\alpha) q^{\frac{\alpha}{6}} + \hat{h}(\alpha) q^{-\frac{\alpha}{6}} \right)^2
    \nonumber \\
    & = \left\{ \begin{array}{lcr}
    (-1)^{\alpha} q^{\frac{\alpha(\alpha - 2)}{6}} & , & \alpha \equiv -1 \mod 3 \\
    (-1)^{\alpha} \left( q^{\frac{\alpha(\alpha + 2)}{6}} + 2q^{\frac{\alpha^2}{6}} + q^{\frac{\alpha(\alpha - 2)}{6}} \right) & , & \alpha \equiv 0 \mod 3 \\
    (-1)^{\alpha} q^{\frac{\alpha(\alpha + 2)}{6}} & , & \alpha \equiv 1 \mod 3 \\
    \end{array} \right.
\end{align}

The unflavored one-point functions of symmetric Wilson lines are given by
\begin{align}
\langle W_{\tiny \yng(2)}\rangle^{SO(4)_{-3,\zeta +}}(q)&=0, \\
\langle W_{\tiny \yng(3)}\rangle^{SO(4)_{-3,\zeta +}}(q)&=-q^{\frac52}, \\
\langle W_{\tiny \yng(4)}\rangle^{SO(4)_{-3,\zeta +}}(q)&=q^{4}, \\
\langle W_{\tiny \yng(5)}\rangle^{SO(4)_{-3,\zeta +}}(q)&=0, \\
\langle W_{\tiny \yng(6)}\rangle^{SO(4)_{-3,\zeta +}}(q)&=q^8, \\
\langle W_{\tiny \yng(7)}\rangle^{SO(4)_{-3,\zeta +}}(q)&=-q^{\frac{21}{2}}, \\
\langle W_{\tiny \yng(8)}\rangle^{SO(4)_{-3,\zeta +}}(q)&=0, \\
\langle W_{\tiny \yng(9)}\rangle^{SO(4)_{-3,\zeta +}}(q)&=-q^{\frac{33}{2}}, \\
\langle W_{\tiny \yng(10)}\rangle^{SO(4)_{-3,\zeta +}}(q)&=q^{20}. 
\end{align}

For $\chi=-$ the half-index and the one-point function of the fundamental Wilson line are identical to those for $\chi=+$. 
For higher rank symmetric Wilson line one-point functions we find results which are either the same or differ by an overall sign
\begin{align}
\langle W_{\tiny \yng(2)}\rangle^{SO(4)_{-3,\zeta -}}(q)&=0, \\
\langle W_{\tiny \yng(3)}\rangle^{SO(4)_{-3,\zeta -}}(q)&=q^{\frac52}, \\
\langle W_{\tiny \yng(4)}\rangle^{SO(4)_{-3,\zeta -}}(q)&=-q^{4}, \\
\langle W_{\tiny \yng(5)}\rangle^{SO(4)_{-3,\zeta -}}(q)&=0, \\
\langle W_{\tiny \yng(6)}\rangle^{SO(4)_{-3,\zeta -}}(q)&=q^8, \\
\langle W_{\tiny \yng(7)}\rangle^{SO(4)_{-3,\zeta -}}(q)&=-q^{\frac{21}{2}}, \\
\langle W_{\tiny \yng(8)}\rangle^{SO(4)_{-3,\zeta -}}(q)&=0, \\
\langle W_{\tiny \yng(9)}\rangle^{SO(4)_{-3,\zeta -}}(q)&=q^{\frac{33}{2}}, \\
\langle W_{\tiny \yng(10)}\rangle^{SO(4)_{-3,\zeta -}}(q)&=-q^{20}. 
\end{align}

\subsubsection{$SO(6)_{-5}$}
\label{sec_SO6_k-5}
With $n=3$ and $N_f=1$ we have the $SO(6)$ CS theory with level $k=-5$ and a single fundamental chiral. 
While the unflavored half-index becomes unity, the unflavored line defect correlators are non-trivial. 
The unflavored charged Wilson line one-point functions are given by
\begin{align}
\langle W_1\rangle^{SO(6)_{-5,\zeta +}}(q)&=-q^{\frac{1}{2}}, \\
\langle W_2\rangle^{SO(6)_{-5,\zeta +}}(q)&=1, \\
\langle W_3\rangle^{SO(6)_{-5,\zeta +}}(q)&=-q^{\frac32}, \\
\langle W_4\rangle^{SO(6)_{-5,\zeta +}}(q)&=1, \\
\langle W_5\rangle^{SO(6)_{-5,\zeta +}}(q)&=-q^{\frac12}-q^{\frac32}-2q^{\frac52}-q^{\frac72}-q^{\frac92}, \\
\langle W_6\rangle^{SO(6)_{-5,\zeta +}}(q)&=q^6, \\
\langle W_7\rangle^{SO(6)_{-5,\zeta +}}(q)&=-q^{\frac72}, \\
\langle W_8\rangle^{SO(6)_{-5,\zeta +}}(q)&=0, \\
\langle W_9\rangle^{SO(6)_{-5,\zeta +}}(q)&=-q^{\frac92}, \\
\langle W_{10}\rangle^{SO(6)_{-5,\zeta +}}(q)&=-q^{6}. 
\end{align}

The non-trivial unflavored one-point functions of symmetric Wilson lines are
\begin{align}
\langle W_{\tiny \yng(5)}\rangle^{SO(6)_{-5,\zeta +}}(q)&=-q^{\frac92}, \\
\langle W_{\tiny \yng(6)}\rangle^{SO(6)_{-5,\zeta +}}(q)&=q^6. 
\end{align}

Again with $\chi=-$ the one-point functions are the same up to extra sign factors. 
In particular, we find 
\begin{align}
\langle W_{\tiny \yng(5)}\rangle^{SO(6)_{-5,\zeta -}}(q)&=q^{\frac92}, \\
\langle W_{\tiny \yng(6)}\rangle^{SO(6)_{-5,\zeta -}}(q)&=-q^6. 
\end{align}

\subsection{$SO(2n)_{-2n}, O(2n)_{-2n}$}
\label{sec_SO2n_-2n}
We now consider the case with $N_f=2$, and also examine both choices of $\chi = \pm 1$ which then allows us to present results for $O(2n)$ gauge groups.

\subsubsection{$SO(2)_{-2}, O(2)_{-2}$}
\label{sec_SO2_k2}
Let us consider the case with $n=1$
where the $SO(2)$ CS theory has level $k=-2$. 
The half-index for $\chi = +$ is given by 
\begin{align}
\mathbb{II}_{\mathcal{N}}^{SO(2)_{-2,\zeta +}}(x;q)&=
(q)_{\infty}\oint \frac{ds}{2\pi is}
(q^{1/2}s^{\pm}x)_{\infty}(q^{1/2}s^{\pm}x^{-1})_{\infty}. 
\end{align}
Here we have included a $U(1)$ flavor fugacity as would be expected if we replace the two fundamental chirals with a 2d fundamental Fermi multiplet.
We can calculate analytically using the Jacobi triple product formula to find
\begin{align}
\label{h_SO2_k2_exact}
\mathbb{II}_{\mathcal{N}}^{SO(2)_{-2,\zeta +}}(x;q)
& = \frac{1}{(q)_{\infty}} \sum_{m \in \Zb} q^{m^2} x^{2m}
\nonumber\\
&= \frac{f(qx^2, qx^{-2})}{(q)_{\infty}} = \frac{(-q x^{\pm 2}; q^2)_{\infty} (q^2; q^2)_{\infty}}{(q)_{\infty}}. 
\end{align}
The infinite sum here has an obvious interpretation as a dual $U(1)_2$ CS theory with Dirichlet boundary condition for the vector multiplet and no chirals. This dual theory and it's half-index were briefly discussed in \cite{Dimofte:2017tpi} where it was noted that the final product expression gives the vacuum character of the $U(1)_k$ WZW model. The fugacity $x$ corresponds to the dual $U(1)$ gauge fugacity (broken to a global symmetry by the dirichlet boundary condition) although we note that here we don't have the coupling to a topological $U(1)$.

If we also set the flavor fugacity $x = 1$, the unflavored half-index is given by
\begin{align}
\label{h_SO2_k-2}
\mathbb{II}_{\mathcal{N}}^{SO(2)_{-2,\zeta +}}
&=\frac{\varphi(q)}{f(-q)}
=\frac{f(q,q)}{f(-q)}
\nonumber\\
&=\prod_{n=1}^{\infty}
\frac{(1-q^{2n})^5}{(1-q^n)^3(1-q^{4n})^2}. 
\end{align}

When the discrete fugacity $\chi$ is set to $-1$, the half-index reads
\begin{align}
\label{h_SO2_k-2_chi-_exact}
\mathbb{II}_{\mathcal{N}}^{SO(2)_{-2,\zeta -}}(x;q)&=
(-q;q)_{\infty}(\pm q^{\frac12}x;q)_{\infty} (\pm q^{\frac12}x^{-1};q)_{\infty}. 
\end{align}
In the unflavored limit, the half-index (\ref{h_SO2_k-2_chi-_exact}) becomes 
\begin{align}
\mathbb{II}_{\mathcal{N}}^{SO(2)_{-2,\zeta -}}(q)
&=\frac{f(-q,-q)}{f(-q)}
\end{align}
and we can clearly see the general result
\begin{align}
\mathbb{II}_{\mathcal{N}}^{SO(2)_{-2,\zeta \chi}}(q)
&=\frac{f(\chi q, \chi q)}{f(-q)} \; .
\end{align}

The half-index of the $O(2)$ CS theory is obtained by gauging the $\mathbb{Z}_2^{\mathcal{C}}$ charge conjugation symmetry
\begin{align}
\mathbb{II}_{\mathcal{N}}^{O(2)_{-2,\zeta \chi'}}(x;q)&=
\frac12 \left(
\mathbb{II}_{\mathcal{N}}^{SO(2)_{-2,\zeta +}}+\chi' \mathbb{II}_{\mathcal{N}}^{SO(2)_{-2,\zeta -}}
\right). 
\end{align}
Turning off the flavor fugacity $x$, we get
\begin{align}
\mathbb{II}_{\mathcal{N}}^{O(2)_{-2,\zeta +}}(q)&=\frac{f(q^4,q^4)}{f(-q)}, \\
\mathbb{II}_{\mathcal{N}}^{O(2)_{-2,\zeta -}}(q)&=q\frac{f(1,q^8)}{f(-q)}. 
\end{align}

We can again calculate the exact result of the one-point functions of the charged Wilson lines using the Jacobi triple product formula. 
For $\chi=+$ we find
\begin{align}
\langle W_{2k} \rangle^{SO(2)_{-2,\zeta +}} & = 2 \frac{q^{k^2} x^{-2k}}{(q)_{\infty}} f(qx^2, qx^{-2}) \; ,
\nonumber \\
\langle W_{2k + 1} \rangle^{SO(2)_{-2,\zeta +}} & = -4 \frac{q^{k^2 + k + \frac{1}{2}} x^{-2k}}{(q)_{\infty}} f(q^2x^2, x^{-2}) \; .
\end{align}

In the unflavored limit, we find 
\begin{align}
\label{W1_SO2_k-2}
\langle W_{1}\rangle^{SO(2)_{-2,\zeta +}}(q)
&=-4q^{1/2}\frac{f(q^2,1)}{f(-q)}
\nonumber\\
&=-8q^{1/2}\prod_{n=1}^{\infty}
\frac{(1-q^{4n})^2}
{(1-q^n)(1-q^{2n})}. 
\end{align}

More generally, the unflavored one-point functions of the Wilson lines with even charges (resp. odd charges) 
can be expressed in terms of the half-index (\ref{h_SO2_k-2}) (resp. the one-point function (\ref{W1_SO2_k-2})). 
We find that 
\begin{align}
\langle W_{2k}\rangle^{SO(2)_{-2,\zeta +}}&= 2q^{k^2}\mathbb{II}_{\mathcal{N}}^{SO(2)_{-2,\zeta +}}, \\
\langle W_{2k+1}\rangle^{SO(2)_{-2,\zeta +}}&=q^{k(k+1)}\langle W_{1}\rangle^{SO(2)_{-2,\zeta +}}. 
\end{align}

\subsubsection{$SO(4)_{-4}, O(4)_{-4}$}
\label{sec_SO4_k-4}
For $n = 2$ the half-index takes the form
\begin{align}
\label{h_SO4_k-4}
\mathbb{II}_{\mathcal{N}}^{SO(4)_{-4,\zeta +}}(x;q)
&=\frac{(q)_{\infty}^2}{4}
\prod_{i=1}^2 \oint \frac{ds_i}{2\pi is_i}
\prod_{i<j}^2 
(s_i^{\pm}s_{j}^{\mp};q)_{\infty}
(s_i^{\pm}s_{j}^{\pm};q)_{\infty}
\prod_{\alpha=1}^2
(q^{\frac12}a s_i^{\pm}x_{\alpha};q)_{\infty}.
\end{align}
We find that in the unflavored limit it is given by
\begin{align}
\mathbb{II}_{\mathcal{N}}^{SO(4)_{-4,\zeta +}}(q)
&=\frac{\varphi(q^2)}{f(-q)}
=\frac{f(q^2,q^2)}{f(-q)}
\nonumber\\
&=\prod_{n=1}^{\infty}
\frac{(1-q^{4n})^5}{(1-q^n)(1-q^{2n})^2(1-q^{8n})^2}. 
\end{align}
For $\chi=-1$ the half-index is given by
\begin{align}
\label{h_SO4_k-4_chi-}
\mathbb{II}_{\mathcal{N}}^{SO(4)_{-4,\zeta -}}(x;q)
&=
\frac{(q)_{\infty} (-q;q)_{\infty}}{2}
\oint \frac{ds}{2\pi is}
(s^{\pm};q)_{\infty}(-s^{\pm};q)_{\infty}
\nonumber\\
&\times 
(\pm q^{\frac12}x;q)_{\infty}
(\pm q^{\frac12}x^{-1};q)_{\infty}
(q^{\frac12}s^{\pm}x;q)_{\infty}
(q^{\frac12}s^{\pm}x^{-1};q)_{\infty}. 
\end{align}
In the unflavored limit, we find that 
\begin{align}
\mathbb{II}_{\mathcal{N}}^{SO(4)_{-4,\zeta -}}(q)
&=\frac{f(-q^2,-q^2)}{f(-q)}. 
\end{align}
Gauging the $\mathbb{Z}_2^{\mathcal{C}}$ symmetry yields the half-indices of the $O(4)$ CS theory. 
The unflavored half-indices are given by
\begin{align}
\mathbb{II}_{\mathcal{N}}^{O(4)_{-4,\zeta +}}(q)&=\frac{f(q^8,q^8)}{f(-q)}, \\
\mathbb{II}_{\mathcal{N}}^{O(4)_{-4,\zeta -}}(q)&=q^2 \frac{f(1,q^{16})}{f(-q)}. 
\end{align}

For $\chi=+$ the unflavored one-point function of the fundamental Wilson line is given by
\begin{align}
\label{W1_SO4_k-4}
\langle W_{\tiny \yng(1)}\rangle^{SO(4)_{-4,\zeta +}}(q)
&=-2q^{\frac12}\frac{f(q,q^3)}{f(-q)}
\nonumber\\
&=-2 q^{\frac12} \prod_{n=1}^{\infty}(1+q^n)^2
\nonumber\\
&=-2q^{\frac12}\prod_{n=1}^{\infty}\frac{(1-q^{2n})^2}{(1-q^n)^2}. 
\end{align}

Also the one-point functions of the symmetric Wilson lines can be expressed in terms of 
the half-index (\ref{h_SO4_k-4}) and the one-point function (\ref{W1_SO4_k-4})
\begin{align}
\langle W_{\tiny \yng(2)}\rangle^{SO(4)_{-4,\zeta +}}(q)&=q \mathbb{II}_{\mathcal{N}}^{SO(4)_{-4,\zeta +}}(q), \\
\langle W_{\tiny \yng(3)}\rangle^{SO(4)_{-4,\zeta +}}(q)&=0, \\
\langle W_{\tiny \yng(4)}\rangle^{SO(4)_{-4,\zeta +}}(q)&=q^{3} \mathbb{II}_{\mathcal{N}}^{SO(4)_{-4,\zeta +}}(q), \\
\langle W_{\tiny \yng(5)}\rangle^{SO(4)_{-4,\zeta +}}(q)&=q^4 \langle W_{\tiny \yng(1)}\rangle^{SO(4)_{-4,\zeta +}}(q), \\
\langle W_{\tiny \yng(6)}\rangle^{SO(4)_{-4,\zeta +}}(q)&=q^{6} \mathbb{II}_{\mathcal{N}}^{SO(4)_{-4,\zeta +}}(q), \\
\langle W_{\tiny \yng(7)}\rangle^{SO(4)_{-4,\zeta +}}(q)&=0, \\
\langle W_{\tiny \yng(8)}\rangle^{SO(4)_{-4,\zeta +}}(q)&=q^{10} \mathbb{II}_{\mathcal{N}}^{SO(4)_{-4,\zeta +}}(q), \\
\langle W_{\tiny \yng(9)}\rangle^{SO(4)_{-4,\zeta +}}(q)&=q^{12} \langle W_{\tiny \yng(1)}\rangle^{SO(4)_{-4,\zeta +}}(q), \\
\langle W_{\tiny \yng(10)}\rangle^{SO(4)_{-4,\zeta +}}(q)&=q^{15} \mathbb{II}_{\mathcal{N}}^{SO(4)_{-4,\zeta +}}(q), \\
\langle W_{\tiny \yng(11)}\rangle^{SO(4)_{-4,\zeta +}}(q)&=0. 
\end{align}

For $\chi=-$ we get
\begin{align}
\label{W1_SO4_k-4_chi-}
\langle W_{\tiny \yng(1)}\rangle^{SO(4)_{-4,\zeta -}}(q)
&=-2q^{\frac12}\frac{f(-q,-q^3)}{f(-q)}. 
\end{align}
Again the one-point function of the symmetric Wilson lines can be described by 
the half-index (\ref{h_SO4_k-4_chi-}) and (\ref{W1_SO4_k-4_chi-})
\begin{align}
\langle W_{\tiny \yng(2)}\rangle^{SO(4)_{-4,\zeta -}}(q)&=q \mathbb{II}_{\mathcal{N}}^{SO(4)_{-4,\zeta -}}(q), \\
\langle W_{\tiny \yng(3)}\rangle^{SO(4)_{-4,\zeta -}}(q)&=0, \\
\langle W_{\tiny \yng(4)}\rangle^{SO(4)_{-4,\zeta -}}(q)&=-q^{3} \mathbb{II}_{\mathcal{N}}^{SO(4)_{-4,\zeta -}}(q), \\
\langle W_{\tiny \yng(5)}\rangle^{SO(4)_{-4,\zeta -}}(q)&=-q^4 \langle W_{\tiny \yng(1)}\rangle^{SO(4)_{-4,\zeta -}}(q), \\
\langle W_{\tiny \yng(6)}\rangle^{SO(4)_{-4,\zeta -}}(q)&=-q^{6} \mathbb{II}_{\mathcal{N}}^{SO(4)_{-4,\zeta -}}(q), \\
\langle W_{\tiny \yng(7)}\rangle^{SO(4)_{-4,\zeta -}}(q)&=0, \\
\langle W_{\tiny \yng(8)}\rangle^{SO(4)_{-4,\zeta -}}(q)&=q^{10} \mathbb{II}_{\mathcal{N}}^{SO(4)_{-4,\zeta -}}(q), \\
\langle W_{\tiny \yng(9)}\rangle^{SO(4)_{-4,\zeta -}}(q)&=q^{12} \langle W_{\tiny \yng(1)}\rangle^{SO(4)_{-4,\zeta -}}(q), \\
\langle W_{\tiny \yng(10)}\rangle^{SO(4)_{-4,\zeta -}}(q)&=q^{15} \mathbb{II}_{\mathcal{N}}^{SO(4)_{-4,\zeta -}}(q), \\
\langle W_{\tiny \yng(11)}\rangle^{SO(4)_{-4,\zeta -}}(q)&=0. 
\end{align}

The one-point function for the $O(4)$ CS theory is obtained by gauging the $\mathbb{Z}_2^{\mathcal{C}}$ symmetry. 
For the fundamental Wilson line we get
\begin{align}
\langle W_{\tiny \yng(1)}\rangle^{O(4)_{-4,\zeta +}}(q)&=-2q^{\frac12}\frac{f(q^{6},q^{10})}{f(-q)}, \\
\langle W_{\tiny \yng(1)}\rangle^{O(4)_{-4,\zeta -}}(q)&=-2q^{\frac32}\frac{f(q^{2},q^{14})}{f(-q)}. 
\end{align}

\subsubsection{$SO(6)_{-6}, O(6)_{-6}$}
\label{sec_SO6_k-6}
We proceed further by investigating the case with $n=3$.  
We find that the unflavored half-index is given by
\begin{align}
\label{h_SO6_k-6}
\mathbb{II}_{\mathcal{N}}^{SO(6)_{-6,\zeta +}}(q)
&=\frac{\varphi(q^3)}{f(-q)}
=\frac{f(q^3,q^3)}{f(-q)}
\nonumber\\
&=\prod_{n=1}^{\infty}
\frac{(1-q^{6n})^5}{(1-q^n) (1-q^{3n})^2}. 
\end{align}
For $\chi=-$ we find 
\begin{align}
\mathbb{II}_{\mathcal{N}}^{SO(6)_{-6,\zeta -}}(q)
&=\frac{f(-q^3,-q^3)}{f(-q)}. 
\end{align}
Also the half-index of the $O(6)$ CS theory is evaluated as
\begin{align}
\mathbb{II}_{\mathcal{N}}^{O(6)_{-6,\zeta +}}(q)&=\frac{f(q^{12},q^{12})}{f(-q)}, \\
\mathbb{II}_{\mathcal{N}}^{O(6)_{-6,\zeta -}}(q)&=q^3\frac{f(1,q^{24})}{f(-q)}. 
\end{align}

Also we find that the unflavored one-point function of the Wilson line in the fundamental representation is given by Ramanujan's general theta function. 
For $\chi=+$ we get 
\begin{align}
\label{Wfund_SO6_k-6}
\langle W_{\tiny \yng(1)}\rangle^{SO(6)_{-6,\zeta +}}(q)
&=-2q^{\frac12}\frac{f(q^4,q^2)}{f(-q)}. 
\end{align}

The unflavored one-point functions of the symmetric Wilson lines can be expressed 
in terms of the half-index (\ref{h_SO6_k-6}) and the one-point function (\ref{Wfund_SO6_k-6}). 
For example, 
\begin{align}
\langle W_{\tiny \yng(2)}\rangle^{SO(6)_{-6,\zeta +}}(q)&=q \mathbb{II}_{\mathcal{N}}^{SO(6)_{-6,\zeta +}}(q), \\
\langle W_{\tiny \yng(3)}\rangle^{SO(6)_{-6,\zeta +}}(q)&=0, \\
\langle W_{\tiny \yng(4)}\rangle^{SO(6)_{-6,\zeta +}}(q)&=0, \\
\langle W_{\tiny \yng(5)}\rangle^{SO(6)_{-6,\zeta +}}(q)&=0, \\
\langle W_{\tiny \yng(6)}\rangle^{SO(6)_{-6,\zeta +}}(q)&=q^5 \mathbb{II}_{\mathcal{N}}^{SO(6)_{-6,\zeta +}}(q), \\
\langle W_{\tiny \yng(7)}\rangle^{SO(6)_{-6,\zeta +}}(q)&=q^6 \langle W_{\tiny \yng(1)}\rangle^{SO(6)_{-6,\zeta +}}(q). 
\end{align}

With $\chi=-$ we obtain
\begin{align}
\label{Wfund_SO6_k-6_chi-}
\langle W_{\tiny \yng(1)}\rangle^{SO(6)_{-6,\zeta -}}(q)
&=-2q^{1/2}\frac{f(-q^4,-q^2)}{f(-q)}
\end{align}
and 
\begin{align}
\langle W_{\tiny \yng(2)}\rangle^{SO(6)_{-6,\zeta -}}(q)&=q \mathbb{II}_{\mathcal{N}}^{SO(6)_{-6,\zeta -}}(q), \\
\langle W_{\tiny \yng(3)}\rangle^{SO(6)_{-6,\zeta -}}(q)&=0, \\
\langle W_{\tiny \yng(4)}\rangle^{SO(6)_{-6,\zeta -}}(q)&=0, \\
\langle W_{\tiny \yng(5)}\rangle^{SO(6)_{-6,\zeta -}}(q)&=0, \\
\langle W_{\tiny \yng(6)}\rangle^{SO(6)_{-6,\zeta -}}(q)&=-q^5 \mathbb{II}_{\mathcal{N}}^{SO(6)_{-6,\zeta -}}(q), \\
\langle W_{\tiny \yng(7)}\rangle^{SO(6)_{-6,\zeta -}}(q)&=-q^6 \langle W_{\tiny \yng(1)}\rangle^{SO(6)_{-6,\zeta -}}(q). 
\end{align}

For the $O(6)$ CS theory we get
\begin{align}
\langle W_{\tiny \yng(1)}\rangle^{O(6)_{-6,\zeta +}}(q)&=-2q^{\frac12}\frac{f(q^{10},q^{14})}{f(-q)}, \\
\langle W_{\tiny \yng(1)}\rangle^{O(6)_{-6,\zeta -}}(q)&=-2q^{\frac52}\frac{f(q^{2},q^{22})}{f(-q)}. 
\end{align}

\subsubsection{General case}
\label{sec_SO2n_-2n}
We conjecture that the unflavored half-indices of orthogonal gauge groups $SO(2n)$ and $O(2n)$ of level $-2n$ with $N_f=2$ are given by 
\begin{align}
\label{h_SO2n_-2n}
\mathbb{II}_{\mathcal{N}}^{SO(2n)_{-2n,\zeta +}}(q)
&=\frac{\varphi(q^n)}{f(-q)}
=\frac{f(q^n,q^n)}{f(-q)}, \\
\label{h_SO2n_-2n_chi-}
\mathbb{II}_{\mathcal{N}}^{SO(2n)_{-2n,\zeta -}}(q)
&=\frac{f(-q^n,-q^n)}{f(-q)}, \\
\label{h_O2n_-2n_chi+}
\mathbb{II}_{\mathcal{N}}^{O(2n)_{-2n,\zeta +}}(q)
&=\frac{f(q^{4n},q^{4n})}{f(-q)}, \\
\label{h_O2n_-2n_chi-}
\mathbb{II}_{\mathcal{N}}^{O(2n)_{-2n,\zeta -}}(q)
&=q^n \frac{f(1,q^{8n})}{f(-q)}. 
\end{align}
and that the unflavored one-point functions of the fundamental Wilson lines are 
\begin{align}
\label{Wfund_SO2n_-2n}
\langle W_{\tiny \yng(1)}\rangle^{SO(2n)_{-2n,\zeta +}}(q)
&=-2q^{\frac12}\frac{f(q^{n-1},q^{n+1})}{f(-q)}, \\
\label{Wfund_SO2n_-2n_chi-}
\langle W_{\tiny \yng(1)}\rangle^{SO(2n)_{-2n,\zeta -}}(q)
&=-2q^{\frac12}\frac{f(-q^{n-1},-q^{n+1})}{f(-q)}, \\
\label{Wfund_O2n_-2n}
\langle W_{\tiny \yng(1)}\rangle^{O(2n)_{-2n,\zeta +}}(q)
&=-2q^{\frac12}\frac{f(q^{4n-2},q^{4n+2})}{f(-q)}, \\
\label{Wfund_O2n_-2n_chi-}
\langle W_{\tiny \yng(1)}\rangle^{O(2n)_{-2n,\zeta -}}(q)
&=-2q^{n-\frac12}\frac{f(q^{2},q^{8n-2})}{f(-q)}. 
\end{align}

Of course, as was obvious in the examples for $SO(2)$, $SO(4)$ and $SO(6)$, the results for $\chi = +$ and $\chi = -$ are easily combined so give
\begin{align}
\label{h_SO2n_-2n_chi}
\mathbb{II}_{\mathcal{N}}^{SO(2n)_{-2n,\zeta \chi}}(q)
&=\frac{f(\chi q^n, \chi q^n)}{f(-q)}, \\
\label{Wfund_SO2n_-2n_chi}
\langle W_{\tiny \yng(1)}\rangle^{SO(2n)_{-2n,\zeta \chi}}(q)
&=-2q^{\frac12}\frac{f(\chi q^{n-1}, \chi q^{n+1})}{f(-q)} \; .
\end{align}

Furthermore, we conjecture that the one-point functions of the Wilson lines in the symmetric representations will be simply given in terms of 
the half-index (\ref{h_SO2n_-2n}) and (\ref{Wfund_SO2n_-2n}). 
In other words, the unflavored Neumann half-index and the one-point functions of the Wilson line operators 
for $SO(2n)_{-2n}$ CS theory with a fundamental chiral multiplet 
are expressible in terms of Ramanujan's general theta function.

\subsection{$SO(2n)_{-4n+4}$ with adjoint}
\label{sec_so2n_adj_-4n+4}
We now consider the $SO(2n)$ CS theories with an adjoint chiral. 

\subsubsection{$SO(4)_{-4}$ with adjoint}
\label{sec_so4_adj4}
For $n=2$ and $N_f=0$ the theory is the $SO(4)$ CS theory with level $k=-4$. 

We find that in the unflavored limit, the half-index with $\chi=+$ is given by the weight-$1$ (again, half the rank of the gauge group) eta-product
\begin{align}
\label{h_SO4_adj_k-4}
\mathbb{II}_{\mathcal{N}, D}^{SO(4)_{-4,\zeta +}}
&=q^{-\frac12}\frac{\eta(4\tau)^4}{\eta(2\tau)^2}
\nonumber\\
&=\prod_{n=1}^{\infty}
\frac{(1-q^{4n})^4}{(1-q^{2n})^2}. 
\end{align}
This is not surprising if we note the result \ref{h_su2_k4_product} for $SU(2)$ and recall that $SO(4) \simeq SU(2) \times SU(2)$.

For $\chi=-$ we find 
\begin{align}
\label{h_SO4_adj_k-4_chi-}
\mathbb{II}_{\mathcal{N}, D}^{SO(4)_{-4,\zeta -}}
&=1+q^4+q^{12}+q^{24}+\cdots
\end{align}

We conjecture that this has the closed form expression
\begin{align}
\label{h_SO4_adj_k-4_chi-}
\mathbb{II}_{\mathcal{N}, D}^{SO(4)_{-4,\zeta -}}
&=q^{-\frac12}\frac{\eta(8\tau)^2}{\eta(4\tau)}
\nonumber\\
&= \prod_{n=1}^{\infty}\frac{(1-q^{8n})^2}{(1-q^{4n})}. 
\end{align}
We note that this eta-product has weight $\frac{1}{2}$ which is half of the rank minus one. This is not surprising given the similarity of $SO(2n)$ with $\chi = -$ to $SO(2n - 2)$ with $\chi = +$.

The results for $\chi = \pm$ are consistent with a combined expression
\begin{align}
\label{h_SO4_adj_k-4_chi}
\mathbb{II}_{\mathcal{N}, D}^{SO(4)_{-4,\zeta \chi}}
 & = \prod_{n=1}^{\infty} \frac{(1-q^{4n})^2 (1 - \chi q^{4n})^2}{(1-q^{2n}) (1 - \chi q^{2n})}
\end{align}
which is of the form of the half-index for $SU(2) \times SU(2)$ but where the signs of the contributions have been changed in one of the $SU(2)$ factors in the case where $\chi = -$.

The one-point function of the Wilson line in the fundamental representation vanishes. 
The one-point function of the Wilson line of charge $+2$ is given by the eta-product 
\begin{align}
\label{W2_SO4_adj_k-4}
\langle W_{2}\rangle^{SO(4)_{-4,\zeta +}}
&=\frac{\eta(2\tau)^{10}}{\eta(\tau)^4\eta(4\tau)^4}
\nonumber\\
&=\prod_{n=1}^{\infty}\frac{(1-q^{2n})^{10}}{(1-q^{n})^4(1-q^{4n})^4}. 
\end{align}

More generally, the one-point functions of the Wilson line operators with even charges can be expressed in terms of two kinds of eta-products 
(\ref{h_SO4_adj_k-4}) and (\ref{W2_SO4_adj_k-4}) while those with odd charges vanish. 
Therefore we have
\begin{align}
\langle W_{4k+2}\rangle^{SO(4)_{-4,\zeta +}}&=q^{2k(k+1)}\langle W_{2}\rangle^{SO(4)_{-4,\zeta +}}, \\
\langle W_{4k+4}\rangle^{SO(4)_{-4,\zeta +}}&=4q^{2(k+1)^2}\mathbb{II}_{\mathcal{N}, D}^{SO(4)_{-4,\zeta +}}, \\
\langle W_{2k+1}\rangle^{SO(4)_{-4,\zeta +}}&=0
\end{align}
for $k\ge0$.  We note that all these results are given as the squares of the $SU(2)$ with adjoint chiral results from section~\ref{sec_CS_SU2_adj_-4}.

\begin{align}
\langle W_{2}\rangle^{SO(4)_{-4,\zeta -}}
&=-1-2q^2-2q^8-2q^{18}+\cdots
\end{align}

This is consistent with the general result
\begin{align}
\label{W2_SO4_adj_k-4_chi}
\langle W_{2}\rangle^{SO(4)_{-4,\zeta \chi}}
&= \chi \prod_{n=1}^{\infty}\frac{(1-q^{2n})^5 (1- \chi q^{2n})^5}{(1-q^{n})^2 (1-q^{4n})^2 (1- \chi q^{n})^2 (1- \chi q^{4n})^2}. 
\end{align}

\subsubsection{$SO(6)_{-8}$ with adjoint}
\label{sec_so6_adjk-8}
Let us consider the case with $n=3$ and $N_f=0$.  
We have the $SO(6)$ CS theory with level $k=-8$ and an adjoint chiral. 
We find that the unflavored half-index with $\chi=+$ is given by the weight-$\frac{3}{2}$ eta-product
\begin{align}
\label{h_SO6_adj_k-8}
\mathbb{II}_{\mathcal{N}, D}^{SO(6)_{-8,\zeta +}}
&=q^{-\frac{5}{12}}\frac{\eta(8\tau)^4}{\eta(2\tau)}
\nonumber\\
&=\prod_{n=1}^{\infty}
\frac{(1-q^{8n})^4}{(1-q^{2n})}, 
\end{align}
which of course agrees with the $SU(4)$ result \eqref{h_sun_k2n}.

For $\chi=-$ we conjecture that 
\begin{align}
\label{h_SO6_adj_k-8chi-}
\mathbb{II}_{\mathcal{N}, D}^{SO(6)_{-8,\zeta -}}
&=q^{-\frac{5}{12}}
\frac{\eta(4\tau)^2\eta(16\tau)^2}{\eta(2\tau)\eta(8\tau)}, 
\end{align}
where, as for the $SO(4)$ case, the weight of the eta-product is reduced by one half compared to the case of $\chi = +$.

We can write a unified expression as
\begin{align}
\label{h_SO6_adj_k-8_chi}
\mathbb{II}_{\mathcal{N}, D}^{SO(6)_{-8,\zeta \chi}}
&=\prod_{n=1}^{\infty}
\frac{(1-q^{4n}) (1-q^{8n})^2 (1 - \chi q^{8n})^2}{(1-q^{2n}) (1 - \chi q^{4n})}. 
\end{align}

While the one-point function of the Wilson line of odd charges vanish, 
the one-point functions of the Wilson lines with even charges are non-trivial. 
For the charged Wilson lines $W_2$ and $W_4$ we have
\begin{align}
\label{W2_SO6_adj_k-8}
\langle W_{2}\rangle^{SO(6)_{-8,\zeta +}}
&=q^{-\frac{25}{24}}
\frac{\eta(2\tau)^5\eta(8\tau)^4}{\eta(\tau)^2\eta(4\tau)^4}
\nonumber\\
&=\prod_{n=1}^{\infty}
\frac{(1-q^{2n})^5 (1-q^{8n})^4}{(1-q^{n})^2(1-q^{4n})^4}, \\
\langle W_{4}\rangle^{SO(6)_{-8,\zeta +}}
&=1+2q+5q^2+2q^3+8q^4+4q^5+5q^6+6q^7+8q^8+2q^9+16q^{10}+\cdots. 
\end{align}
The unflavored one point function of $\langle W_2\rangle$ is given by the eta-product, however, 
the expression for 
$\langle W_4\rangle$ does not seem to be so simply expressed in terms of the eta-products. Note that these Wilson lines are not the same as those considered for $SU(4)$ in section~\ref{sec_CSsuN_-2Nadj}

The one-point functions with larger charges can be expressed 
in terms of the half-index (\ref{h_SO6_adj_k-8}) and the one-point functions of the charged Wilson lines $W_2$ and $W_4$. 
For example, 
\begin{align}
\langle W_{6}\rangle^{SO(6)_{-8,\zeta +}}&=q^2 \langle W_{2}\rangle^{SO(6)_{-8,\zeta +}}, \\
\langle W_{8}\rangle^{SO(6)_{-8,\zeta +}}&=6q^4 \mathbb{II}_{\mathcal{N}, D}^{SO(6)_{-8,\zeta +}}, \\
\langle W_{10}\rangle^{SO(6)_{-8,\zeta +}}&=q^6 \langle W_{2}\rangle^{SO(6)_{-8,\zeta +}}, \\
\langle W_{12}\rangle^{SO(6)_{-8,\zeta +}}&=q^8 \langle W_{4}\rangle^{SO(6)_{-8,\zeta +}}. 
\end{align}

\subsection{$SO(2n)_{-4n+4-N_f}$ with adjoint}
\label{sec_so2n_adj_-4n+4-Nf}
We also propose several formulas of the half-indices for the $SO(2n)$ CS theories coupled to both adjoint and fundamental chirals in terms of the eta-products. 

\subsubsection{$SO(2)_{-1}$ with adjoint}
\label{sec_so2_adj1}

In the case with $n=1$ and $N_f=1$ 
we get the $SO(2)$ CS theory with level $k=-1$. 
The unflavored index is evaluated as
\begin{align}
\label{h_SO2_adj_k-1}
\mathbb{II}_{\mathcal{N}, D}^{SO(2)_{-1,\zeta +}}
&=q^{-\frac{1}{24}}\eta(\tau)
\nonumber\\
&=\prod_{n=1}^{\infty}(1-q^n), \\
\label{h_SO2_adj_k-1chi-}
\mathbb{II}_{\mathcal{N}, D}^{SO(2)_{-1,\zeta -}}
&=q^{-\frac{1}{24}}\frac{\eta(2\tau)}{\eta(\tau)}
\nonumber\\
&=\prod_{n=1}\frac{1-q^{2n}}{1-q^n}. 
\end{align}

Subsequently, 
we obtain the half-indices for the $O(2)$ CS theory 
\begin{align}
\label{h_O2_adj_k-1}
\mathbb{II}_{\mathcal{N}, D}^{O(2)_{-1,\zeta +}}&=\sum_{m=0}^{\infty}
\frac{q^{m(2m+1)}}{(q;q)_{2m}}, \\
\label{h_O2_adj_k-1chi-}
\mathbb{II}_{\mathcal{N}, D}^{O(2)_{-1,\zeta +}}&=\sum_{m=0}^{\infty}
\frac{q^{m(2m-1)}}{(q;q)_{2m-1}}. 
\end{align}
The half-index (\ref{h_O2_adj_k-1}) (resp.  (\ref{h_O2_adj_k-1chi-})) is identical to the generating function for the partitions into distinct parts in such a way that the number of parts is even (resp. odd). 

The unflavored one-point functions of the symmetric Wilson lines for the $SO(2)$ CS theory with $\chi=+$ are obtained from the half-index
\begin{align}
\langle W_{n}\rangle^{SO(2)_{-1,\zeta +}}
&=(-1)^n q^{\frac{n^2}{2}}\mathbb{II}_{\mathcal{N}, D}^{SO(2)_{-1,\zeta +}}. 
\end{align}

All these results are easily derived analytically.

\subsubsection{$SO(4)_{-5}$ with adjoint}
\label{sec_so4_adj5}
With $n=2$ and $N_f=1$, we conjecture that the $SO(4)$ CS theory with level $k=-5$. The unflavored half-indices are expressible as 
\begin{align}
\label{h_SO4_adj_k-5}
\mathbb{II}_{\mathcal{N}, D}^{SO(4)_{-5,\zeta +}}
&=q^{-\frac{5}{12}}\eta(5\tau)^2
\nonumber\\
&=\prod_{n=1}^{\infty}(1-q^{5n})^2, \\
\mathbb{II}_{\mathcal{N}, D}^{SO(4)_{-5,\zeta -}}
&=q^{-\frac{5}{12}}\eta(10\tau)
\nonumber\\
&=\prod_{n=1}^{\infty}(1-q^{10n}). 
\end{align}

\subsubsection{$SO(4)_{-6}$ with adjoint}
\label{sec_so4_adj6}
When $n=2$ and $N_f=2$, we have the $SO(4)$ CS theory with level $k=-6$. 
We find that the unflavored half-indices agree with the following eta-products: 
\begin{align}
\label{h_SO4_adj_k-6}
\mathbb{II}_{\mathcal{N}, D}^{SO(4)_{-6,\zeta +}}
&=q^{-\frac13}
\frac{\eta(2\tau)^2\eta(6\tau)^4}{\eta(\tau)\eta(3\tau)\eta(4\tau)\eta(12\tau)}
\nonumber\\
&=\prod_{n=1}^{\infty}
\frac{(1-q^{2n})^2 (1-q^{6n})^4}
{(1-q^n)(1-q^{3n})(1-q^{4n})(1-q^{12n})}, \\
\label{h_SO4_adj_k-6chi-}
\mathbb{II}_{\mathcal{N}, D}^{SO(4)_{-6,\zeta -}}
&=q^{-\frac13}
\frac{\eta(2\tau)\eta(3\tau)\eta(12\tau)}{\eta(\tau)\eta(4\tau)\eta(6\tau)}
\nonumber\\
&=\prod_{n=1}^{\infty}
\frac{(1-q^{2n})(1-q^{3n})(1-q^{12n})}{(1-q^n)(1-q^{4n})(1-q^{6n})}. 
\end{align}

\subsection{$SO(2n+1)_{-2n+1}$}
\label{sec_SO2n+1_-2n+1}
Next we consider the orthogonal gauge theories with odd ranks, i.e.\ with $\epsilon=1$. 

\subsubsection{$SO(3)_{-1}$}
\label{sec_SO3_k-1}
For $n=1$ and $N_f=0$, and without an adjoint chiral, the theory is pure $SO(3)$ CS theory with level $k=-1$. 
When the gauge field satisfies the Neumann boundary condition, 
as there are no non-trivial BPS local operators at the boundary, the half-index is trivial. 

The one-point functions of the charged Wilson lines are given by
We have the half-index
\begin{align}
\langle W_{n}\rangle^{SO(3)_{-1,\zeta +}}(q)&=\frac12 (q)_{\infty}\oint \frac{ds}{2\pi is}(\chi s^{\pm};q)_{\infty}
(1+s^n+s^{-n}). 
\end{align}
One finds
\begin{align}
\langle W_{n}\rangle^{SO(3)_{-1,\zeta +}}&=
1+(-1)^n q^{\frac{n(n-1)}{2}}+(-1)^n q^{\frac{n(n+1)}{2}}. 
\end{align}

Also we find that 
the one-point functions of the Wilson lines in the symmetric representations are given by
\begin{align}
\langle W_{(k)}\rangle^{SO(3)_{-1,\zeta +}}&=(-1)^k q^{\frac{k(k+1)}{2}}. 
\end{align}

For the $SO(3)$ with $\chi=-$, we consider the Wilson line $W_{(k)}$ 
associated with the modified character of the symmetric representation 
in such a way that powers of the fugacities of the form $s^{n}$ are replaced by $(-1)^{n + k} s^{n}$ thereby switching the sign of all odd or even powers depending on the parity of $k$. 
For example, 
\begin{align}
\chi_{\tiny \yng(1)}^{\mathfrak{so}(3)_{-}}(s)&=-1+s+s^{-1}, \\
\chi_{\tiny \yng(2)}^{\mathfrak{so}(3)_{-}}(s)&= 1 + s^2 - s - s^{-1} + s^{-2}. 
\end{align}
We get 
\begin{align}
\langle W_{(k)}\rangle^{SO(3)_{-1,\zeta-}}&= q^{\frac{k(k+1)}{2}}. 
\end{align}

All these results are easily checked analytically using the Jacobi triple product to evaluate
\begin{align}
\langle s^n \rangle^{SO(3)_{-1,\zeta \chi}}&=
\frac{1}{2} (-\chi)^n \left( q^{\frac{n(n-1)}{2}} + q^{\frac{n(n+1)}{2}} \right) \; . 
\end{align}

\subsubsection{$SO(5)_{-3}$}
\label{sec_SO5_k-3}
For $n=2$ and $N_f=0$, the theory is pure $SO(5)$ CS theory with level $k=-3$. 
Again the half-index is trivial. 

The one-point function of the charged Wilson line is given by
\begin{align}
\label{h_SO5_k3}
\langle W_n\rangle^{SO(5)_{-3,\zeta +}}
&=\frac{(q)_{\infty}^2}{8}
\prod_{i=1}^2 \oint \frac{ds_i}{2\pi is_i}
(s_i^{\pm};q)_{\infty}
\prod_{i<j}^2 
(s_i^{\pm}s_{j}^{\mp};q)_{\infty}
(s_i^{\pm}s_{j}^{\pm};q)_{\infty}
\nonumber\\
&\times 
(1 + s_1^n+s_2^n+s_1^{-n}+s_2^{-n}).
\end{align}
For example, 
\begin{align}
\langle W_1\rangle^{SO(5)_{-3,\zeta +}}&=0,\\
\langle W_2\rangle^{SO(5)_{-3,\zeta +}}&=1+q,\\
\langle W_3\rangle^{SO(5)_{-3,\zeta +}}&=-q-q^2-q^3,\\
\langle W_4\rangle^{SO(5)_{-3,\zeta +}}&=1+q^2,\\
\langle W_5\rangle^{SO(5)_{-3,\zeta +}}&=1+q^5,\\
\langle W_6\rangle^{SO(5)_{-3,\zeta +}}&=1+q^3+q^5+q^7+q^9,\\
\langle W_7\rangle^{SO(5)_{-3,\zeta +}}&=1-q^7,\\
\langle W_8\rangle^{SO(5)_{-3,\zeta +}}&=1+q^{12}. 
\end{align}

We also find that the one-point function of the symmetric Wilson line is given by 
\begin{align}
\langle W_{(3k)}\rangle^{SO(5)_{-3,\zeta +}}&=(-1)^{k} q^{3k(k+1)}. 
\end{align}
For $\chi=-$ we obtain
\begin{align}
\langle W_{(3k)}\rangle^{SO(5)_{-3,\zeta -}}&=q^{3k(k+1)}. 
\end{align}

\subsubsection{General case}
\label{sec_CS_SO2n+1_-2n+1}
We conjecture that the symmetric Wilson line one-point function is given by
\begin{align}
\langle W_{((2n-1)k)}\rangle^{SO(2n+1)_{-2n+1,\zeta +}}&=(-1)^k q^{(2n-1)k(k+1)}, \\
\langle W_{((2n-1)k)}\rangle^{SO(2n+1)_{-2n+1,\zeta -}}&=q^{(2n-1)k(k+1)}. 
\end{align}

We can then define a grand canonical ensemble as
\begin{align}
    \sum_{k \in \Zb} \langle W_{((2n-1)k)}\rangle^{SO(2n+1)_{-2n+1, \zeta +}} \Lambda^{k}
    & = i q^{-\frac{1}{4}(2n-1)} \Lambda^{-\frac{1}{2}} \vartheta_1\left( z; 2(2n-1)\tau \right) \; ,
    \\
    \sum_{k \in \Zb} \langle W_{((2n-1)k)}\rangle^{SO(2n+1)_{-2n+1, \zeta -}} \Lambda^{k}
    & = q^{-\frac{1}{4}(2n-1)} \Lambda^{-\frac{1}{2}} \vartheta_2\left( z; 2(2n-1)\tau \right)
\end{align}
where $q = e^{2\pi i \tau}$ and $\Lambda = e^{2 \pi i z}$.

\subsection{$SO(2n+1)_{-2n}$}
\label{sec_SO2n+1_-2n}
Let us consider the cases with matter fields for which the half-indices are non-trivial where we have one fundamental chiral, without an adjoint chiral. 

\subsubsection{$SO(3)_{-2}$}
\label{sec_SO3_k-2}
When $n=1$ we have the $SO(3)$ CS theory with level $k=-2$. 
We have the half-index
\begin{align}
\mathbb{II}_{\mathcal{N}}^{SO(3)_{-3,\zeta \chi}}(q)&=\frac12 (q)_{\infty}\oint \frac{ds}{2\pi is}(\chi s^{\pm};q)_{\infty}
(\chi q^{\frac12}x;q)_{\infty}
(q^{\frac12}s^{\pm}x;q)_{\infty}. 
\end{align}

In the unflavored limit the half-index becomes trivial
\begin{align}
\mathbb{II}_{\mathcal{N}}^{SO(3)_{-3,\zeta +}}(q)&=1. 
\end{align}

The unflavored one-point function of the charged Wilson line is given by 
\begin{align}
\langle W_{k}\rangle^{SO(3)_{-2,\zeta +}}(q)&=1+(-1)^k q^{\frac{k(k-1)}{4}}+(-1)^k q^{\frac{k(k+1)}{4}}. 
\end{align}

The unflavored one-point function of the symmetric Wilson line is 
\begin{align}
\langle W_{(k)}\rangle^{SO(3)_{-2,\zeta +}}(q)&=(-1)^k q^{\frac{k(k+1)}{4}}. 
\end{align}

For $\chi=-$ we find
\begin{align}
\langle W_{(k)}\rangle^{SO(3)_{-2,\zeta -}}(q)&=(-1)^{\frac{k(k+1)}{2}}q^{\frac{k(k+1)}{4}}. 
\end{align}

Again, we can use the Jacobi triple product to derive these results. In particular we find
\begin{align}
    \langle s^n \rangle & = (-1)^n \frac{(\chi q^{1/2}; q)_{\infty}}{2 (q)_{\infty}} q^{\frac{1}{2}n^2} \sum_{m \in \Zb} \chi^m \left( q^{m^2 + \frac{1}{2}(2n-1)m} + q^{m^2 + \frac{1}{2}(2n+1)m} \right)
    \nonumber \\
    & = (-1)^n \frac{(\chi q^{1/2}; q)_{\infty} (q^2; q^2)_{\infty}}{2 (q)_{\infty}} q^{\frac{1}{2}n^2} \Big( (-\chi q^{n + 1/2}; q^2)_{\infty} (-\chi q^{-n + 3/2}; q^2)_{\infty} 
    \nonumber \\
     & + (-\chi q^{-n + 1/2}; q^2)_{\infty} (-\chi q^{n + 3/2}; q^2)_{\infty} \Big) \; .
\end{align}
After some manipulation this simplifies to
\begin{align}
     \langle s^{n = 2m} \rangle & = \frac{1}{2} \chi^m \left( q^{\frac{n(n-1)}{4}} + q^{\frac{n(n+1)}{4}} \right), \\
     \langle s^{n = 2m+1} \rangle & = -\frac{1}{2} \chi^m \left( \chi q^{\frac{n(n-1)}{4}} + q^{\frac{n(n+1)}{4}} \right), 
\end{align}
and the above Wilson line results follow.

\subsubsection{$SO(5)_{-4}$}
\label{sec_SO5_k-4}
For $n=2$ we have the $SO(5)$ CS theory with level $k=-4$. 
The half-index is given by
\begin{align}
\label{h_SO5_k4}
\mathbb{II}_{\mathcal{N}}^{SO(5)_{-4,\zeta +}}
&=\frac{(q)_{\infty}^2}{8}
\prod_{i=1}^2 \oint \frac{ds_i}{2\pi is_i}
(s_i^{\pm};q)_{\infty}
\prod_{i<j}^2 
(s_i^{\pm}s_{j}^{\mp};q)_{\infty}
(s_i^{\pm}s_{j}^{\pm};q)_{\infty}
(q^{\frac12}x;q)_{\infty}
(q^{\frac12}s_{i}^{\pm}x;q)_{\infty}.
\end{align}
It can be expanded as
\begin{align}
&1+(x^3-x^5)q^{5/2}+(x^3-x^5)q^{7/2}+(x^3-x^5)q^{9/2}
+(x^3-x^5)q^{11/2}+(x^3-x^5)q^{13/2}
\nonumber\\
&
+(x^3-x^5)q^{15/2}
+(x^3-x^5)q^{17/2}
+(x^6-x^8)q^{9}
+(x^3-x^5)q^{19/2}
+(x^6-2x^8+x^{10})q^{10}+\cdots.
\end{align}
In the unflavored limit, it becomes trivial
\begin{align}
\mathbb{II}_{\mathcal{N}}^{SO(5)_{-4,\zeta +}}(q)&=1. 
\end{align}

The one-point function of the charged Wilson line is evaluated as
\begin{align}
\label{W_SO5_k4}
\langle W_n \rangle^{SO(5)_{-4,\zeta +}}
&=\frac{(q)_{\infty}^2}{8}
\prod_{i=1}^2 \oint \frac{ds_i}{2\pi is_i}
(s_i^{\pm};q)_{\infty}
\prod_{i<j}^2 
(s_i^{\pm}s_{j}^{\mp};q)_{\infty}
(s_i^{\pm}s_{j}^{\pm};q)_{\infty}
(q^{\frac12}x;q)_{\infty}
(q^{\frac12}s_{i}^{\pm}x;q)_{\infty}
\nonumber\\
&\times 
(1 + s_1^n+s_2^n+s_1^{-n}+s_2^{-n}).
\end{align}

In the unflavored limit, we have
\begin{align}
\langle W_1 \rangle^{SO(5)_{-4,\zeta +}}(q)&=-q^{\frac12}, \\
\langle W_2 \rangle^{SO(5)_{-4,\zeta +}}(q)&=1,\\
\langle W_3 \rangle^{SO(5)_{-4,\zeta +}}(q)&=-q^{\frac32}, \\
\langle W_4 \rangle^{SO(5)_{-4,\zeta +}}(q)&=1+q^{\frac12}+q^{\frac32}+q^{\frac52}+q^{\frac72}, \\
\langle W_5 \rangle^{SO(5)_{-4,\zeta +}}(q)&=1-q^{\frac52}-q^5, \\
\langle W_6 \rangle^{SO(5)_{-4,\zeta +}}(q)&=1, \\
\langle W_7 \rangle^{SO(5)_{-4,\zeta +}}(q)&=1-q^{\frac72}-q^7. 
\end{align}

The one-point functions of the symmetric Wilson lines are evaluated as
\begin{align}
\langle W_{(2)} \rangle^{SO(5)_{-4,\zeta +}}(q)&=0, \\
\langle W_{(3)} \rangle^{SO(5)_{-4,\zeta +}}(q)&=0, \\
\langle W_{(4)} \rangle^{SO(5)_{-4,\zeta +}}(q)&=q^{\frac72}, \\
\langle W_{(5)} \rangle^{SO(5)_{-4,\zeta +}}(q)&=-q^{5}, \\
\langle W_{(6)} \rangle^{SO(5)_{-4,\zeta +}}(q)&=0, \\
\langle W_{(7)} \rangle^{SO(5)_{-4,\zeta +}}(q)&=0. 
\end{align}

For $\chi=-$ we get
\begin{align}
\langle W_{(1)} \rangle^{SO(5)_{-4,\zeta -}}(q)&=-q^{\frac12}, \\
\langle W_{(2)} \rangle^{SO(5)_{-4,\zeta -}}(q)&=0, \\
\langle W_{(3)} \rangle^{SO(5)_{-4,\zeta -}}(q)&=0, \\
\langle W_{(4)} \rangle^{SO(5)_{-4,\zeta -}}(q)&=-q^{\frac72}, \\
\langle W_{(5)} \rangle^{SO(5)_{-4,\zeta -}}(q)&=q^{5}, \\
\langle W_{(6)} \rangle^{SO(5)_{-4,\zeta -}}(q)&=0, \\
\langle W_{(7)} \rangle^{SO(5)_{-4,\zeta -}}(q)&=0. 
\end{align}

\subsection{$SO(2n+1)_{-2n-1}, O(2n+1)_{-2n-1}$}
\label{sec_SO2n+1_-2n}
We now consider the case of $N_f=2$, again without an adjoint chiral.

\subsubsection{$SO(3)_{-3}, O(3)_{-3}$}
\label{sec_SO5_k-3}
With $n=1$ the theory is identified with the $SO(3)$ CS theory with level $k=-3$. 
The half-indices are 
\begin{align}
\mathbb{II}_{\mathcal{N}}^{SO(3)_{-3,\zeta  \chi}}(q)&=\frac12 (q)_{\infty}
\oint \frac{ds}{2\pi is}
(\chi s^{\pm};q)_{\infty}
(\chi q^{\frac12}x^{\pm};q)_{\infty}
(q^{\frac12}s^{\pm}x; q)_{\infty}
(q^{\frac12}s^{\pm}x^{-1}; q)_{\infty}. 
\end{align}

We find that the unflavored half-indices are given by
\begin{align}
\label{h_SO3_k-3}
\mathbb{II}_{\mathcal{N}}^{SO(3)_{-3,\zeta +}}(q)
&=\frac{\varphi(-q^{\frac32})}{f(-q)}
=\frac{f(-q^{\frac32},-q^{\frac32})}{f(-q)}
\nonumber\\
&=\prod_{n=1}^{\infty}
\frac{(1-q^{\frac{3n}{2}})^2}{(1-q^n)(1-q^{3n})}, \\
\label{h_SO3_k-3_chi-}
\mathbb{II}_{\mathcal{N}}^{SO(3)_{-3,\zeta -}}(q)
&=\frac{\varphi(q^{\frac32})}{f(-q)}=\frac{f(q^{\frac32},q^{\frac32})}{f(-q)}
\end{align}
with an obvious combined expression
\begin{align}
\label{h_SO3_k-3_chi-}
\mathbb{II}_{\mathcal{N}}^{SO(3)_{-3,\zeta \chi}}(q)
&=\frac{\varphi(-\chi q^{\frac32})}{f(-q)}=\frac{f(-\chi q^{\frac32}, -\chi q^{\frac32})}{f(-q)} \; .
\end{align}

By gauging the $\mathbb{Z}_2^{\mathcal{C}}$ symmetry, 
we find the half-indices of the $O(3)$ CS theory
\begin{align}
\label{h_O3_k-3}
\mathbb{II}_{\mathcal{N}}^{O(3)_{-3,\zeta +}}(q)
&=\frac{f(q^6,q^6)}{f(-q)}, \\
\label{h_O3_k-3_chi-}
\mathbb{II}_{\mathcal{N}}^{O(3)_{-3,\zeta -}}(q)
&=-q^{\frac32}\frac{f(1,q^{12})}{f(-q)}. 
\end{align}

The one-point function of the Wilson line $W_{\tiny \yng(1)}$ in the fundamental representation 
for $\chi=+$ (resp. $\chi=-$) is computed by inserting the character $1+s+s^{-1}$ (resp. $-1+s+s^{-1}$) in the integrand. 
We find that the unflavored one-point functions are given by
\begin{align}
\label{Wfund_SO3_k-3}
\langle W_{\tiny \yng(1)} \rangle^{SO(3)_{-3,\zeta +}}(q)
&=-2q^{\frac12}\frac{f(-q^{\frac12},-q^{\frac52})}{f(-q)}, \\
\langle W_{\tiny \yng(1)} \rangle^{SO(3)_{-3,\zeta -}}(q)
&=-2q^{\frac12}\frac{f(q^{\frac12},q^{\frac52})}{f(-q)}
\end{align}
or combined as
\begin{align}
\label{Wfund_SO3_k-3_chi}
\langle W_{\tiny \yng(1)} \rangle^{SO(3)_{-3,\zeta \chi}}(q)
&=-2q^{\frac12}\frac{f(- \chi q^{\frac12},- \chi q^{\frac52})}{f(-q)} \; .
\end{align}

The one-point functions of the symmetric Wilson lines are 
given in terms of the half-index (\ref{h_SO3_k-3}) and (\ref{Wfund_SO3_k-3}). 
For example, for $\chi=+$ we have
\begin{align}
\langle W_{\tiny \yng(2)} \rangle^{SO(3)_{-3,\zeta +}}(q)&=q\mathbb{II}_{\mathcal{N}}^{SO(3)_{-3,\zeta +}}(q), \\
\langle W_{\tiny \yng(3)} \rangle^{SO(3)_{-3,\zeta +}}(q)&=-q^2\mathbb{II}_{\mathcal{N}}^{SO(3)_{-3,\zeta +}}(q), \\
\langle W_{\tiny \yng(4)} \rangle^{SO(3)_{-3,\zeta +}}(q)&=-q^3\langle W_{\tiny \yng(1)} \rangle^{SO(3)_{-3,\zeta +}}(q), \\
\langle W_{\tiny \yng(5)} \rangle^{SO(3)_{-3,\zeta +}}(q)&=-q^5\mathbb{II}_{\mathcal{N}}^{SO(3)_{-3,\zeta +}}(q), \\
\langle W_{\tiny \yng(6)} \rangle^{SO(3)_{-3,\zeta +}}(q)&=q^7\mathbb{II}_{\mathcal{N}}^{SO(3)_{-3,\zeta +}}(q), \\
\langle W_{\tiny \yng(7)} \rangle^{SO(3)_{-3,\zeta +}}(q)&=q^9\langle W_{\tiny \yng(1)} \rangle^{SO(3)_{-3,\zeta +}}(q), \\
\langle W_{\tiny \yng(8)} \rangle^{SO(3)_{-3,\zeta +}}(q)&=q^{12}\mathbb{II}_{\mathcal{N}}^{SO(3)_{-3,\zeta +}}(q). 
\end{align}

For $\chi=-$ we get the same results up to some overall signs
\begin{align}
\langle W_{\tiny \yng(2)} \rangle^{SO(3)_{-3,\zeta -}}(q)&=q\mathbb{II}_{\mathcal{N}}^{SO(3)_{-3,\zeta -}}(q), \\
\langle W_{\tiny \yng(3)} \rangle^{SO(3)_{-3,\zeta -}}(q)&=q^2\mathbb{II}_{\mathcal{N}}^{SO(3)_{-3,\zeta -}}(q), \\
\langle W_{\tiny \yng(4)} \rangle^{SO(3)_{-3,\zeta -}}(q)&=q^3\langle W_{\tiny \yng(1)} \rangle^{SO(3)_{-3,\zeta -}}(q), \\
\langle W_{\tiny \yng(5)} \rangle^{SO(3)_{-3,\zeta -}}(q)&=q^5\mathbb{II}_{\mathcal{N}}^{SO(3)_{-3,\zeta -}}(q), \\
\langle W_{\tiny \yng(6)} \rangle^{SO(3)_{-3,\zeta -}}(q)&=q^7\mathbb{II}_{\mathcal{N}}^{SO(3)_{-3,\zeta -}}(q), \\
\langle W_{\tiny \yng(7)} \rangle^{SO(3)_{-3,\zeta -}}(q)&=q^9\langle W_{\tiny \yng(1)} \rangle^{SO(3)_{-3,\zeta -}}(q), \\
\langle W_{\tiny \yng(8)} \rangle^{SO(3)_{-3,\zeta -}}(q)&=q^{12}\mathbb{II}_{\mathcal{N}}^{SO(3)_{-3,\zeta -}}(q). 
\end{align}

For the $O(3)$ CS theory the one-point functions are obtained by gauging the $\mathbb{Z}_2^{\mathcal{C}}$ symmetry. 
For example, the one-point functions of the fundamental Wilson lines are given by
\begin{align}
\label{Wfund_O3_k-3}
\langle W_{\tiny \yng(1)} \rangle^{O(3)_{-2,\zeta +}}(q)
&=-2q^{\frac12} \frac{f(q^4,q^8)}{f(-q)}, \\
\label{Wfund_O3_k-3_chi-}
\langle W_{\tiny \yng(1)} \rangle^{O(3)_{-2,\zeta -}}(q)
&=2q \frac{f(q^2,q^{10})}{f(-q)}. 
\end{align}

\subsubsection{$SO(5)_{-5}, O(5)_{-5}$}
\label{sec_SO5_k-5}
For $n=2$ and $N_f=N_a=1$, we have the $SO(5)$ CS theory with level $k=-5$. 
The half-indices are evaluated from 
\begin{align}
\label{h_SO5_k-5}
&
\mathbb{II}_{\mathcal{N}}^{SO(5)_{-6,\zeta \chi}}
\nonumber\\
&=\frac{(q)_{\infty}^2}{8}
\prod_{i=1}^2 \oint \frac{ds_i}{2\pi is_i}
(\chi s_i^{\pm};q)_{\infty}
\prod_{i<j}^2 
(s_i^{\pm}s_{j}^{\mp};q)_{\infty}
(s_i^{\pm}s_{j}^{\pm};q)_{\infty}
\prod_{\alpha=1}^{2}
(\chi q^{\frac12}ax_{\alpha};q)_{\infty}
(q^{\frac12}as_{i}^{\pm}x_{\alpha};q)_{\infty}.
\end{align}
We find that the unflavored half-index can be expresses in terms of Ramanujan's general theta function 
\begin{align}
\mathbb{II}_{\mathcal{N}}^{SO(5)_{-5,\zeta \chi}}
&=\frac{\varphi(-\chi q^{\frac52})}{f(-q)}
=\frac{f(-\chi q^{\frac52},-\chi q^{\frac52})}{f(-q)}. 
\end{align}

In the unflavored limit the one-point functions of the fundamental Wilson line are given by
\begin{align}
\label{Wfund_SO5_k-5}
\langle W_{\tiny \yng(1)} \rangle^{SO(5)_{-5,\zeta \chi}}(q)
&=-2q^{\frac12}\frac{f(-\chi q^{\frac32},-\chi q^{\frac72})}{f(-q)}. 
\end{align}

The other one-point functions are given in terms of the half-index and the one-point function of the fundamental Wilson line. 
For example, 
\begin{align}
\langle W_{\tiny \yng(2)} \rangle^{SO(5)_{-5,\zeta +}}(q)&=q\mathbb{II}_{\mathcal{N}}^{SO(5)_{-5,\zeta +}}(q), \\
\langle W_{\tiny \yng(3)} \rangle^{SO(5)_{-5,\zeta +}}(q)&=0, \\
\langle W_{\tiny \yng(4)} \rangle^{SO(5)_{-5,\zeta +}}(q)&=0, \\
\langle W_{\tiny \yng(5)} \rangle^{SO(5)_{-5,\zeta +}}(q)&=-q^4\mathbb{II}_{\mathcal{N}}^{SO(5)_{-5,\zeta +}}(q), \\
\langle W_{\tiny \yng(6)} \rangle^{SO(5)_{-5,\zeta +}}(q)&=-q^5\langle W_{\tiny \yng(1)} \rangle^{SO(5)_{-5,\zeta +}}(q), \\
\langle W_{\tiny \yng(7)} \rangle^{SO(5)_{-5,\zeta +}}(q)&=-q^7\mathbb{II}_{\mathcal{N}}^{SO(5)_{-5,\zeta +}}(q), \\
\langle W_{\tiny \yng(8)} \rangle^{SO(5)_{-5,\zeta +}}(q)&=0. 
\end{align}

For $\chi=-$ we find
\begin{align}
\langle W_{\tiny \yng(2)} \rangle^{SO(5)_{-5,\zeta -}}(q)&=q\mathbb{II}_{\mathcal{N}}^{SO(5)_{-5,\zeta -}}(q), \\
\langle W_{\tiny \yng(3)} \rangle^{SO(5)_{-5,\zeta -}}(q)&=0, \\
\langle W_{\tiny \yng(4)} \rangle^{SO(5)_{-5,\zeta -}}(q)&=0, \\
\langle W_{\tiny \yng(5)} \rangle^{SO(5)_{-5,\zeta -}}(q)&=q^4\mathbb{II}_{\mathcal{N}}^{SO(5)_{-5,\zeta -}}(q), \\
\langle W_{\tiny \yng(6)} \rangle^{SO(5)_{-5,\zeta -}}(q)&=q^5\langle W_{\tiny \yng(1)} \rangle^{SO(5)_{-5,\zeta -}}(q), \\
\langle W_{\tiny \yng(7)} \rangle^{SO(5)_{-5,\zeta -}}(q)&=q^7\mathbb{II}_{\mathcal{N}}^{SO(5)_{-5,\zeta -}}(q), \\
\langle W_{\tiny \yng(8)} \rangle^{SO(5)_{-5,\zeta -}}(q)&=0. 
\end{align}

The half-indices and the line defect indices for the $O(5)$ CS theory can be obtained by gauging the $\mathbb{Z}_2^{\mathcal{C}}$ symmetry. 
We find
\begin{align}
\label{h_O5_k-5}
\mathbb{II}_{\mathcal{N}}^{O(5)_{-5,\zeta +}}
&=\frac{f(q^{10},q^{10})}{f(-q)}, \\
\label{h_O5_k-5_chi-}
\mathbb{II}_{\mathcal{N}}^{O(5)_{-5,\zeta -}}
&=-q^{\frac52}\frac{f(1,q^{20})}{f(-q)}
\end{align}
and 
\begin{align}
\label{Wfund_O5_k-5}
\langle W_{\tiny \yng(1)} \rangle^{O(5)_{-5,\zeta +}}(q)
&=-2q^{\frac12}\frac{f(q^8,q^{12})}{f(-q)}, \\
\label{Wfund_O5_k-5_chi-}
\langle W_{\tiny \yng(1)} \rangle^{O(5)_{-5,\zeta -}}(q)
&=2q^{2}\frac{f(q^2,q^{18})}{f(-q)}. 
\end{align}

\subsubsection{General case}
\label{sec_SO2n+1_-2n-1}
Now we propose the general formula of the unflavored half-index and the one-point function of the fundamental Wilson line in terms of Ramanujan's general theta function. 
We conjecture that 
\begin{align}
\label{h_SO2n+1_-2n-1}
\mathbb{II}_{\mathcal{N}}^{SO(2n+1)_{-2n-1,\zeta +}}
&=\frac{\varphi(-q^{\frac{2n+1}{2}})}{f(-q)}
=\frac{f(-q^{\frac{2n+1}{2}},-q^{\frac{2n+1}{2}})}{f(-q)}, \\
\label{h_SO2n+1_-2n-1_chi-}
\mathbb{II}_{\mathcal{N}}^{SO(2n+1)_{-2n-1,\zeta -}}
&=\frac{\varphi(q^{\frac{2n+1}{2}})}{f(-q)}
=\frac{f(q^{\frac{2n+1}{2}},q^{\frac{2n+1}{2}})}{f(-q)}, \\
\label{h_O2n+1_-2n-1_chi+}
\mathbb{II}_{\mathcal{N}}^{O(2n+1)_{-2n-1,\zeta +}}
&=\frac{f(q^{4n+2},q^{4n+2})}{f(-q)}, \\
\label{h_O2n+1_-2n-1_chi-}
\mathbb{II}_{\mathcal{N}}^{O(2n+1)_{-2n-1,\zeta -}}
&=-q^{n+\frac12}\frac{f(1,q^{8n+4})}{f(-q)} 
\end{align}
and 
\begin{align}
\label{Wfund_SO2n+1_-2n-1}
\langle W_{\tiny \yng(1)} \rangle^{SO(2n+1)_{-2n-1,\zeta+}}(q)
&=-2q^{\frac12}\frac{f(-q^{\frac{2n-1}{2}},-q^{\frac{2n+3}{2}})}{f(-q)}, \\
\label{Wfund_SO2n+1_-2n-1_chi-}
\langle W_{\tiny \yng(1)} \rangle^{SO(2n+1)_{-2n-1,\zeta-}}(q)
&=-2q^{\frac12}\frac{f(q^{\frac{2n-1}{2}},q^{\frac{2n+3}{2}})}{f(-q)}, \\
\label{Wfund_O2n+1_-2n-1_chi+}
\langle W_{\tiny \yng(1)} \rangle^{O(2n+1)_{-2n-1,\zeta+}}(q)
&=-2q^{\frac12}\frac{f(q^{4n},q^{4n+4})}{f(-q)}, \\
\label{Wfund_O2n+1_-2n-1_chi-}
\langle W_{\tiny \yng(1)} \rangle^{O(2n+1)_{-2n-1,\zeta-}}(q)
&=2q^n \frac{f(q^2,q^{8n+2})}{f(-q)}. 
\end{align}

As we have seen in the examples for $SO(3)$ and $SO(5)$ we can write combined $\chi$-dependent expressions
\begin{align}
\label{h_SO2n+1_-2n-1_chi}
\mathbb{II}_{\mathcal{N}}^{SO(2n+1)_{-2n-1,\zeta \chi}}
&=\frac{\varphi(- \chi q^{\frac{2n+1}{2}})}{f(-q)}
=\frac{f(- \chi q^{\frac{2n+1}{2}}, - \chi q^{\frac{2n+1}{2}})}{f(-q)} \; , \\
\label{Wfund_SO2n+1_-2n-1_chi}
\langle W_{\tiny \yng(1)} \rangle^{SO(2n+1)_{-2n-1,\zeta \chi}}(q)
&=-2q^{\frac12}\frac{f(- \chi q^{\frac{2n-1}{2}}, - \chi q^{\frac{2n+3}{2}})}{f(-q)} \; .
\end{align}

These results can be further combined with those for $SO(2n)$, \eqref{h_SO2n_-2n_chi} and \eqref{Wfund_SO2n_-2n_chi}, to give
\begin{align}
\label{h_SON_-N_chi}
\mathbb{II}_{\mathcal{N}}^{SO(2n+\epsilon)_{-2n-\epsilon,\zeta \chi}}
&=\frac{\varphi((-1)^{\epsilon} \chi q^{\frac{2n+\epsilon}{2}})}{f(-q)}
=\frac{f((-1)^{\epsilon} \chi q^{\frac{2n+\epsilon}{2}}, (-1)^{\epsilon} \chi q^{\frac{2n+\epsilon}{2}})}{f(-q)} \; , \\
\label{Wfund_SON_-N_chi}
\langle W_{\tiny \yng(1)} \rangle^{SO(2n+\epsilon)_{-2n-\epsilon,\zeta \chi}}(q)
&=-2q^{\frac12}\frac{f( (-1)^{\epsilon} \chi q^{\frac{2n+\epsilon-2}{2}}, (-1)^{\epsilon} \chi q^{\frac{2n+\epsilon+2}{2}})}{f(-q)} \; .
\end{align}

\subsection{$SO(2n+1)_{-4n+2}$ with adjoint}
\label{sec_SO2n+1_adj_-4n+2}
The half-index of the $SO(2n+1)$ CS theory of level $k=-4n+2$ with an adjoint chiral coincides with that of the $USp(2n)$ CS theory of level $-2n-2$ with an adjoint, in both cases without fundamental chirals, for $n=1$ and $n=2$ due to the identifications $SO(3)\simeq USp(2)$ and $SO(5)\simeq USp(4)$. However, there is no reason this should hold for higher rank cases and indeed for $n=3$ we find the half-indices differ at order $q^8$.

We also note that at least for $SO(3)$ and $SO(5)$ the half-indices do not depend on $\chi$. However, we do have $\chi$-dependence in the one-point functions of Wilson lines.
E.g., for $SO(3)$ we have
\begin{align}
    \langle s^{2k} \rangle & = q^{k^2} \frac{(q^4; q^4)_{\infty}^2}{(q^2; q^2)_{\infty}}, \\
    \langle s^{2k+1} \rangle & = -\frac{\chi}{2} q^{k(k-1)} \frac{(q^2; q^2)_{\infty}^5}{(q)_{\infty}^2 (q^4; q^4)_{\infty}^2},
\end{align}
so we can claulate for example
\begin{align}
    \langle W_1 \rangle^{SO(3)_{-4,\zeta \chi}} & = \langle \chi + s + s^{-1} \rangle
    \nonumber \\
    & = \chi \left(\frac{(q^4; q^4)_{\infty}^2}{(q^2; q^2)_{\infty}}
     - \frac{(q^2; q^2)_{\infty}^5}{(q)_{\infty}^2 (q^4; q^4)_{\infty}^2}
    \right)
\end{align}
and
\begin{align}
    \langle W_2 \rangle^{SO(3)_{-4,\zeta \chi}} & = \langle 1 + s^2 + s^{-2} \rangle
    \nonumber \\
    & = \left( 1 + 2 q^4 \right) \frac{(q^4; q^4)_{\infty}^2}{(q^2; q^2)_{\infty}}
\end{align}

\subsection{$SO(2n+1)_{-4n+2-N_f}$ with adjoint}
\label{sec_SO2n+1_adj_-4n+2-N_f}
Let us now study the CS theory with an adjoint chiral as well as $N_f$ fundamental flavors. We see that the half-indices are again given by eta-products but now involving a factor of $\eta(\frac{|k|}{2} \tau)$ where $k$ is the (odd) CS level in the case of $N_f = 1$. In all cases the weight of the eta-product is half the rank of the gauge group for $\chi = +$. For $N_f = 1$ it is reduced by $\frac{1}{2}$ for the case of $\chi = -$. Surprisingly, for $N_f = 2$ the half-index does not depend on $\chi$.

\subsubsection{$SO(3)_{-3}$ with adjoint}
\label{sec_SO3adj_-3}
For $n=1$ and $N_f=1$, we have the $SO(3)$ CS theory with level $-3$. 
The half-index reads
\begin{align}
\mathbb{II}_{\mathcal{N},D}^{SO(3)_{-3,\zeta \chi}}
&=\frac12 (q)_{\infty}
\oint \frac{ds}{2\pi is}
(\chi s^{\pm};q)_{\infty}(\chi qs^{\pm};q)_{\infty}
(\chi q^{\frac12}x;q)_{\infty}
(q^{\frac12}s^{\pm}x;q)_{\infty}. 
\end{align}
We find that the half-indices are given by the following eta-products: 
\begin{align}
\mathbb{II}_{\mathcal{N},D}^{SO(3)_{-3,\zeta +}}
&=q^{-\frac{3}{16}}
\frac{\eta(3\tau)^2}{\eta(3\tau/2)}
\nonumber\\
&=\prod_{n=1}^{\infty}
\frac{(1-q^{3n})^2}{(1-q^{\frac{3n}{2}})}, \\
\mathbb{II}_{\mathcal{N},D}^{SO(3)_{-3,\zeta -}}
&=q^{-\frac{3}{16}}
\frac{\eta(3\tau/2)\eta(6\tau)}{\eta(3\tau)}
\nonumber\\
&=\prod_{n=1}^{\infty}
\frac{(1-q^{\frac{3n}{2}}) (1-q^{6n})}{(1-q^{3n})}
\end{align}
and the combined result is
\begin{align}
\label{so3_m3_adj}
\mathbb{II}_{\mathcal{N},D}^{SO(3)_{-3,\zeta \chi}}
&=\prod_{n=1}^{\infty}
\frac{(1-q^{3n}) (1 - \chi q^{3n})}{(1 - \chi q^{\frac{3n}{2}})} \; .
\end{align}

Using the Jacobi triple product we can calculate
\begin{align}
\label{so3_m3_adj_an}
    \mathbb{II}_{\mathcal{N},D}^{SO(3)_{-3,\zeta \chi}} & =
    \frac{(\chi q^{\frac{1}{2}}; q)_{\infty} (q^6; q^6)_{\infty}^2}{(q)_{\infty} (q^{12}; q^{12})_{\infty}} \left(
    \frac{(q^4; q^4)_{\infty}^2 (q^6; q^6)_{\infty}^3}{(q^2; q^2)_{\infty} (q^3; q^3)_{\infty}^2 (q^{12}; q^{12})_{\infty}} \right.
    \nonumber \\
    & + \chi q^{\frac{1}{2}} \left.
    \frac{(q^2; q^2)_{\infty}^4}{(q)_{\infty}^2 (q^4; q^4)_{\infty}}
    \right) \; .
\end{align}
It is not possible to check this as for previous linear eta-product identities as we cannot write this involving a sum of two modular functions (up to an overall factor). However we have checked that the $q$-series expansions of \eqref{so3_m3_adj} and \eqref{so3_m3_adj_an} agree to at least order $q^{500}$ so this provides an interesting conjectured identity.

For $O(3)$ theory we find 
\begin{align}
\mathbb{II}_{\mathcal{N},D}^{O(3)_{-3,\zeta +}}
&=\sum_{m\in \mathbb{Z}}q^{3m(4m+1)}, \\
\mathbb{II}_{\mathcal{N},D}^{O(3)_{-3,\zeta -}}
&=q^{\frac32}\sum_{m\in \mathbb{Z}}q^{3m(4m-3)}. 
\end{align}

\subsubsection{$SO(5)_{-7}$ with adjoint}
\label{sec_SO5_adj-7}
For $n=2$ and $N_f=1$ the $SO(5)$ CS theory has level $-7$. 
We conjecture that the unflavored half-indices are given by the following eta-product: 
\begin{align}
\mathbb{II}_{\mathcal{N},D}^{SO(5)_{-7,\zeta +}}
&=q^{-\frac{35}{48}}\frac{\eta(7\tau)^3}{\eta(7\tau/2)}
\nonumber\\
&=\prod_{n=1}^{\infty}
\frac{(1-q^{7n})^3}{(1-q^{\frac{7n}{2}})}, \\
\mathbb{II}_{\mathcal{N},D}^{SO(5)_{-7,\zeta -}}
&=q^{-\frac{35}{48}}\eta(7\tau/2)\eta(14\tau)
\nonumber\\
&=\prod_{n=1}^{\infty}
(1-q^{\frac{7n}{2}})(1-q^{14n}), 
\end{align}
which can be combined for $\chi = \pm$ as
\begin{align}
    \mathbb{II}_{\mathcal{N},D}^{SO(5)_{-7,\zeta -}}
    & = \prod_{n=1}^{\infty}
    (1 + \chi q^{\frac{7n}{2}}) (1-q^{7n}) (1 - \chi q^{7n}). 
\end{align}

\subsubsection{$SO(3)_{-4}$ with adjoint}
\label{sec_SO3_adj-4}
For $n=1$ and $N_f=2$ we have the CS theory of level $-4$. 
The half-index is
\begin{align}
\mathbb{II}_{\mathcal{N},D}^{SO(3)_{-4,\zeta \chi}}&=\frac12 (q)_{\infty}
\oint \frac{ds}{2\pi is}
(\chi s^{\pm};q)_{\infty}
(\chi qs^{\pm};q)_{\infty}
(\chi q^{\frac12}x^{\pm})_{\infty}
(q^{\frac12}s^{\pm}x)_{\infty}
(q^{\frac12}s^{\pm}x^{-1})_{\infty}. 
\end{align}
We find that the unflavored half-index is given by
\begin{align}
\label{h_SO3adj_-4}
\mathbb{II}_{\mathcal{N},D}^{SO(3)_{-4,\zeta \chi}}
&=q^{-\frac18}
\frac{\eta(2\tau)^2}{\eta(\tau)}
\nonumber\\
&=\prod_{n=1}^{\infty}
\frac{(1-q^{2n})^2}{1-q^n}. 
\end{align}
In this case the half-index does not depend on $\chi$. 
So the unflavored half-index for the $O(3)_{-4}$ CS theory with $\chi'=+$ is the same as (\ref{h_SO3adj_-4}) and that with $\chi'=-$ vanishes. 

\subsubsection{$SO(5)_{-8}$ with adjoint}
\label{sec_SO5_adj-8}
Similarly, the unflavored index does not depend on $\chi$
for $n=2$ and $N_f=2$, i.e. the $SO(5)$ CS theory. We conjecture that the unflavored half-index is given by
\begin{align}
\mathbb{II}_{\mathcal{N},D}^{SO(5)_{-8,\zeta \chi}}&
=q^{-\frac58}
\frac{\eta(2\tau)^2\eta(8\tau)^2}{\eta(\tau)\eta(4\tau)}
\nonumber\\
&=\prod_{n=1}^{\infty}
\frac{(1-q^{2n})^2(1-q^{8n})^2}{(1-q^{n})(1-q^{4n})}. 
\end{align}

\section{$G_2$ CS theories}
\label{sec_CS_G2}
Consider 3d $\mathcal{N}=2$ gauge theory with gauge group $G_2$ and $N_f$ fundamental chirals $Q_{\alpha}$, $\alpha=1,\cdots, _f$. 
We now consider the case with Neumann boundary condition for the vector multiplet and Dirichlet for the fundamental chiral multiplets.  
The boundary conditions and charges of the field content are summarized as follows: 
\begin{align}
\label{G2_4_charges}
\begin{array}{c|c|c|c|c|c}
& \textrm{bc} & G_2 & SU(N_f) & U(1)_a & U(1)_R \\ \hline
\textrm{VM} & \mathcal{N} & {\bf Adj} & {\bf 1} & 0 & 0 \\
Q_{\alpha} & \textrm{D} & {\bf 7} & {\bf \overline{N}_f} & 1 & r
\end{array}
\end{align}

We can easily calculate the gauge and 't Hooft anomalies as follows,
\begin{align}
\label{bdy_G2_Nf_anom}
\Acal & = \underbrace{{4\Tr(s^2)} + 7r^2}_{\textrm{VM}, \; \Ncal}
 + \underbrace{\left( N_f\Tr(s^2) + \frac{7}{2}\Tr(x^2) + \frac{7}{2}N_f(a-r)^2 \right)}_{Q_{\alpha}, \; D}
  \nonumber \\
  = & (N_f + 4) \Tr(s^2) + \frac{7}{2} \Tr(x^2) + \frac{7}{2}N_f a^2 - 7N_f ar + \frac{7}{2}(N_f + 2)r^2
 \; .
\end{align}
To cancel the gauge anomaly we need to take the CS level 
\begin{align}
k&=-N_f - 4. 
\end{align}

The half-index is
\begin{align}
\label{bdy_G2_4_hindexA}
\II_{\Ncal} = & \frac{(q)_{\infty}^2}{2^2 3} \prod_{i=1}^2 \oint \frac{ds_i}{2\pi i s_i}
\left( \prod_{i \ne j}^3 (s_i s_j^{-1}; q)_{\infty} \right) \prod_{i = 1}^3 (s_i^{\pm}; q)_{\infty} 
\prod_{\alpha = 1}^{N_f} (q^{\frac12} x_{\alpha}; q)_{\infty} 
\prod_{i = 1}^3 (q^{\frac12} s_i^{\pm} x_{\alpha}; q)_{\infty} \; , 
\end{align}
where
$\prod_{i = 1}^3 s_i = \prod_{\alpha = 1}^{N_f} x_{\alpha} = 1$.

In the case with Dirichlet for the adjoint and fundamental chiral multiplets, 
the boundary conditions and charges of the field content are summarized as follows: 
\begin{align}
\label{G2_4_charges}
\begin{array}{c|c|c|c|c|c}
& \textrm{bc} & G_2 & SU(N_f) & U(1)_a & U(1)_R \\ \hline
\textrm{VM} & \mathcal{N} & {\bf Adj} & {\bf 1} & 0 & 0 \\
\Phi & \textrm{D} & {\bf Adj} & {\bf 1} & 0 & 0 \\
Q_{\alpha} & \textrm{D} & {\bf 7} & {\bf \overline{N}_f} & 1 & r
\end{array}
\end{align}

We can easily calculate the gauge and 't Hooft anomalies as follows,
\begin{align}
\label{bdy_G2_Nf_anom}
\Acal & = \underbrace{{4\Tr(s^2)} + 7r^2}_{\textrm{VM}, \; \Ncal} + \underbrace{{4\Tr(s^2)} + 7r^2}_{\Phi, \; D}
 + \underbrace{\left( N_f\Tr(s^2) + \frac{7}{2}\Tr(x^2) + \frac{7}{2}N_f(a-r)^2 \right)}_{Q_{\alpha}, \; D}
  \nonumber \\
  = & (N_f + 8) \Tr(s^2) + \frac{7}{2} \Tr(x^2) + \frac{7}{2}N_f a^2 - 7N_f ar + \frac{7}{2}(N_f + 4)r^2
 \; .
\end{align}
To cancel the gauge anomaly we need to take CS level 
\begin{align}
k&=-N_f - 8. 
\end{align}

The half-index is
\begin{align}
\label{bdy_G2_4_hindexA}
\II_{\Ncal, D} = & \frac{(q)_{\infty}^4}{2^2 3} \prod_{i=1}^2 \oint \frac{ds_i}{2\pi i s_i}
\left( \prod_{i \ne j}^3 (s_i s_j^{-1}; q)_{\infty} (q s_i s_j^{-1}; q)_{\infty} \right) \prod_{i = 1}^3 (s_i^{\pm}; q)_{\infty} (q s_i^{\pm}; q)_{\infty} 
\nonumber \\
 & \times \prod_{\alpha = 1}^{N_f} (q^{1 - r/2} a^{-1} x_{\alpha}; q)_{\infty} \prod_{i = 1}^3 (q^{1 - r/2} s_i^{\pm} a^{-1} x_{\alpha}; q)_{\infty}
\end{align}
where
$\prod_{i = 1}^3 s_i = \prod_{\alpha = 1}^{N_f} x_{\alpha} = 1$.

The character of the representation of $G_2$ with highest weight $a\omega_1+b\omega_2$ 
where $\omega_1$ and $\omega_2$ are the two fundamental weights 
and $a$ and $b$ are non-negative integers is \cite{MR1153249}
\begin{align}
\chi_{a,b}^{\mathfrak{g}_2}(s)
&=\frac{s_{(a+2b+1,a+b+1)}-s_{(a+2b+1,b)}}
{s_{\tiny \yng(1,1)}-s_{\tiny \yng(1)}}. 
\end{align}
For example, the character of the fundamental representation with $(a,b)$ $=$ $(1,0)$ is
\begin{align}
\chi_{1,0}^{\mathfrak{g}_2}(s)
&=1+s_1+s_2+s_1^{-1}+s_2^{-1}+s_1s_2+s_1^{-1}s_2^{-1}. 
\end{align}

\subsection{$G_{2_{-4}}$}
\label{sec_G2_-4}
For the pure $G_2$ CS theory with level $k=-4$ obeying the Neumann boundary condition, 
there are no BPS local operators living at the boundary so that the half-index is trivial. 
However, in the presence of the BPS line defects, one can find the BPS local operators living at the junction of the lines and the boundary 
which are detectable from the non-trivial line defect half-indices. 
For example, we find the following non-vanishing $\langle W_{a,b} \rangle^{G_{2_{-4}}}(q)$ for small values of $a$ and $b$
\begin{align}
\langle W_{0,1} \rangle^{G_{2_{-4}}}(q)&=-q, \\
\langle W_{3,0} \rangle^{G_{2_{-4}}}(q)&=q^2, \\
\langle W_{4,0} \rangle^{G_{2_{-4}}}(q)&=-q^3, \\
\langle W_{0,4} \rangle^{G_{2_{-4}}}(q)&=q^7, \\
\langle W_{7,0} \rangle^{G_{2_{-4}}}(q)&=-q^7. 
\end{align}

\subsection{$G_{2_{-5}}$}
\label{sec_G2_-5}
Consider the case with $N_f=1$ where the $G_2$ CS theory has level $k=-5$. 
We find that the unflavored half-index precisely agrees with 
\begin{align}
\label{h_G2_k-5}
\II_{\Ncal}^{G_{2_{-5}}}(q) = \frac{f(-q^{\frac32},-q)}{f(-q)}. 
\end{align}

The one-point function of the fundamental Wilson line is evaluated as
\begin{align}
\langle W_{1,0} \rangle^{G_{2_{-5}}}(q)
&=\frac{(q)_{\infty}^2}{2^2 3}
\oint 
\prod_{i=1}^{2}
\frac{ds_i}{2\pi is_i}
\prod_{i \neq j}
(s_is_j^{-1};q)_{\infty}
\prod_{i=1}^3(s_i^{\pm};q)_{\infty}
(q^{\frac12}x;q)_{\infty}
(q^{\frac12}s_i^{\pm}x;q)_{\infty}
\nonumber\\
&\times 
(1+s_1+s_2+s_1^{-1}+s_2^{-1}+s_1s_2+s_1^{-1}s_2^{-1})
\end{align}
We find that in the unflavored limit it is given by
\begin{align}
\label{W10_G2_k-5}
\langle W_{1,0} \rangle^{G_{2_{-5}}}(q)
&=-q^{\frac12}\frac{f(-q^{\frac12},-q^2)}{f(-q)}. 
\end{align}

The other one-point functions can be described by the half-index (\ref{h_G2_k-5}) and the one-point function (\ref{W10_G2_k-5}). 
For example, 
\begin{align}
\label{W_G2_k-5}
\langle W_{0,1} \rangle^{G_{2_{-5}}}(q)&=0, \\
\langle W_{2,0} \rangle^{G_{2_{-5}}}(q)&=0, \\
\langle W_{1,1} \rangle^{G_{2_{-5}}}(q)&=-q\langle W_{1,0} \rangle^{G_{2_{-5}}}(q), \\
\langle W_{0,2} \rangle^{G_{2_{-5}}}(q)&=-q^{2}\mathbb{II}_{\mathcal{N}}^{G_{2_{-5}}}(q), \\
\langle W_{3,0} \rangle^{G_{2_{-5}}}(q)&=0, \\
\langle W_{2,1} \rangle^{G_{2_{-5}}}(q)&=0, \\
\langle W_{1,2} \rangle^{G_{2_{-5}}}(q)&=0, \\
\langle W_{0,3} \rangle^{G_{2_{-5}}}(q)&=0, \\
\langle W_{4,0} \rangle^{G_{2_{-5}}}(q)&=q^2\langle W_{1,0} \rangle^{G_{2_{-5}}}(q), \\
\langle W_{3,1} \rangle^{G_{2_{-5}}}(q)&=q^{3}\mathbb{II}_{\mathcal{N}}^{G_{2_{-5}}}(q), \\
\langle W_{2,2} \rangle^{G_{2_{-5}}}(q)&=0, \\
\langle W_{1,3} \rangle^{G_{2_{-5}}}(q)&=0, \\
\langle W_{0,4} \rangle^{G_{2_{-5}}}(q)&=0. 
\end{align}

\subsection{$G_{2_{-8}}$ with adjoint}
\label{sec_G2_-8_adj}

With an adjoint chiral but no fundamentals we have calculated the half-index to order $q^{15}$ and it is consistent with an eta-product expression
\begin{align}
    q^{-\frac{1}{6}} \frac{\eta(4 \tau) \eta(6 \tau) \eta(8 \tau)}{\eta(2 \tau) \eta(12 \tau)} \; .
\end{align}
We note that only even powers of $q$ are seen in the expansion.

However, since we have only calculated up to order $q^{15}$ there may be higher order terms. Indeed, in previous results we always saw the level appearing (the least common multiple of the $d_i$ in the notation of \eqref{eta_prod}). Hence we would conjecture that the result (assuming it is an eta-product) would include also a factor of $\eta(24\tau)$, and of course we cannot exclude even higher terms. If we further conjecture that the pattern holds of the weight being half the rank of the gauge group, it is natural to conjecture the result
\begin{align}
    \II_{\Ncal, D}^{G_{2_{-8}}} & \sim q^{-\frac{7}{6}} \frac{\eta(4 \tau) \eta(6 \tau) \eta(8 \tau) \eta(24 \tau)}{\eta(2 \tau) \eta(12 \tau)} \; .
\end{align}

\appendix
\section{Modular Forms}
\label{app_modular}
Under a modular transformation $\tau \to \frac{a \tau + b}{c \tau + d}$ we use the definition that a modular form of weight $k$ is a function $f(\tau)$ which transforms as $f(\tau) \to \lambda(a, b, c, d) (c \tau + d)^k f(\tau)$.\footnote{An alternative definition is stricter, requiring that $\lambda(a, b, c, d) = 1$.}
With this definition the Dedekind eta function $\eta(\tau)$ is a modular form of weight $\frac{1}{2}$ obeying 
\begin{align}
\eta(\tau+1)&=e^{\frac{\pi i}{12}}\eta(\tau),\\
\eta(-1/\tau)&=\sqrt{\tau/i} \eta(\tau). 
\end{align}

The weight of an eta-product $\prod_i \eta(d_i \tau)^{m_i}$ is defined to be $\frac{1}{2} \sum_i m_i$. 
The eta-products can be cusp forms or modular forms 
if they satisfy certain conditions. 
If we define the level, $N$, to be the lowest common multiple of the $d_i$ then such an eta-product is a modular form of weight $k$ on the congruence subgroup $\Gamma_0(N)$ $\subset$ $SL(2,\mathbb{Z})$ that is defined as the matrices which are upper triangular mod $N$ 
\begin{align}
\Gamma_0(N)&:=\left\{
\left(
\begin{matrix}
a&b\\
c&d\\
\end{matrix}
\right)
\in SL(2,\mathbb{Z}): 
c\equiv 0 \mod N
\right\}
\end{align}
if and only if the following conditions all hold, as proved by Newman \cite{MR91352,MR107629} 
\begin{itemize}
    \item $\frac{1}{2} \sum_i m_i = k$
    \item $\sum_i d_i m_i = 0 \mod 24$
    \item $\sum_i \frac{N}{d_i} m_i = 0 \mod 24$
    \item $\prod_i d_i^{|m_i|}$ is a square integer.
\end{itemize}

There exist linear identities for eta-products.
In the case that the eta-products are modular functions (modular forms of weight zero) on some $\Gamma_0(N)$ there is an algorithmic way to prove such identities. This has been formulated and automated in Maple packages provided by Garvan \cite{garvan2019tutorialmapleetapackage}. We use this later to prove some interesting identities which allow us in some examples to reexpress a sum of eta-products, derived by analytically evaluating a half-index, as a single eta-product.

\subsection*{Acknowledgements}
The work of T.O. was supported by the Startup Funding no.\ 4007012317 of the Southeast University. 
DJS was supported in part by the STFC Consolidated grant ST/T000708/1.

\appendix

\bibliographystyle{utphys}
\bibliography{ref}

\end{document}